\documentclass[11pt,british]{article}
\usepackage{libertineRoman}
\usepackage{biolinum}
\usepackage{libertineMono}
\usepackage[T1]{fontenc}
\usepackage[latin9]{inputenc}
\usepackage{geometry}
\usepackage{color}
\usepackage{refstyle}
\usepackage{float}
\usepackage{amsmath}
\usepackage{amsthm}
\usepackage{amssymb}
\usepackage{cancel}
\usepackage{stmaryrd}
\usepackage{graphicx}
\PassOptionsToPackage{normalem}{ulem}
\usepackage{ulem}
\usepackage{titling}
\usepackage{needspace}

\makeatletter

\AtBeginDocument{\providecommand\Defref[1]{\ref{Def:#1}}}
\AtBeginDocument{\providecommand\Eqref[1]{\ref{Eq:#1}}}
\AtBeginDocument{\providecommand\Remref[1]{\ref{Rem:#1}}}
\AtBeginDocument{\providecommand\Thmref[1]{\ref{Thm:#1}}}
\AtBeginDocument{\providecommand\Figref[1]{\ref{Fig:#1}}}
\AtBeginDocument{\providecommand\Propref[1]{\ref{Prop:#1}}}
\AtBeginDocument{\providecommand\Lemref[1]{\ref{Lem:#1}}}
\AtBeginDocument{\providecommand\Claimref[1]{\ref{Claim:#1}}}
\AtBeginDocument{}
\AtBeginDocument{\providecommand\Subsecref[1]{\ref{Subsec:#1}}}
\AtBeginDocument{\providecommand\lemref[1]{\ref{lem:#1}}}
\RS@ifundefined{subsecref}
  {\newref{subsec}{name = \RSsectxt}}
  {}
\RS@ifundefined{thmref}
  {\def\RSthmtxt{theorem~}\newref{thm}{name = \RSthmtxt}}
  {}
\RS@ifundefined{lemref}
  {\def\RSlemtxt{lemma~}\newref{lem}{name = \RSlemtxt}}
  {}

\theoremstyle{plain}
\newtheorem{thm}{\protect\theoremname}
\theoremstyle{definition}
\newtheorem{defn}[thm]{\protect\definitionname}
\theoremstyle{remark}
\newtheorem{rem}[thm]{\protect\remarkname}
\theoremstyle{plain}
\newtheorem{prop}[thm]{\protect\propositionname}
\theoremstyle{remark}
\newtheorem{notation}[thm]{\protect\notationname}
\theoremstyle{plain}
\newtheorem{lem}[thm]{\protect\lemmaname}
\theoremstyle{remark}
\newtheorem{claim}[thm]{\protect\claimname}

\usepackage{hyperref}

\hypersetup{
    colorlinks=true,
    linkcolor=blue,
    filecolor=magenta,      
    urlcolor=cyan,
	citecolor=blue,
  hidelinks,
backref=true,
pagebackref=true,
hyperindex=true,
breaklinks=true,
colorlinks=true,%
urlcolor=blue,
bookmarks=true,
bookmarksopen=false}

\newref{thm}{name=theorem~,Name=Theorem~,names=theorems~,Names=Theorems~}
\RS@ifundefined{defref}{\newref{def}{name=definition~,Name=Definition~,names=definitions~,Names=Definitions~}}{}
\RS@ifundefined{corref}{\newref{cor}{name=corollary~,Name=Corollary~,names=corollaries~,Names=Corollaries~}}{}
\newref{lem}{name=lemma~,Name=Lemma~,names=lemmas~,Names=Lemmas~}
\RS@ifundefined{claimref}{\newref{claim}{name=claim~,Name=Claim~,names=claims~,Names=Claims~}}{}
\newref{sec}{name=section~,Name=Section~,names=sections~,Names=Sections~}
\newref{subsec}{name=section~,Name=Section~,names=sections~,Names=Sections~}
\RS@ifundefined{propref}{\newref{prop}{name=proposition~,Name=Proposition~,names=propositions~,Names=Propositions~}}{}
\RS@ifundefined{conjref}{\newref{conj}{name=conjecture~,Name=Conjecture~,names=conjectures~,Names=Conjectures~}}{}
\RS@ifundefined{assuref}{\newref{assu}{name=assumption~,Name=Assumption~,names=assumptions~,Names=Assumptions~}}{}
\RS@ifundefined{remref}{\newref{rem}{name=remark~,Name=Remark~,names=remarks~,Names=Remarks~}}{}
\RS@ifundefined{algref}{\newref{alg}{name=algorithm~,Name=Algorithm~,names=algorithms~,Names=Algorithms~}}{}
\RS@ifundefined{noteref}{\newref{note}{name=note~,Name=Note~,names=notes~,Names=Notes~}}{}
\RS@ifundefined{notaref}{\newref{nota}{name=notation~,Name=Notation~,names=notations~,Names=Notations~}}{}
\RS@ifundefined{exaref}{\newref{exa}{name=example~,Name=Example~,names=examples~,Names=Examples~}}{}
\RS@ifundefined{tabref}{\newref{tab}{name=table~,Name=Table~,names=tables~,Names=Tables~}}{}
\RS@ifundefined{factref}{\newref{fact}{name=fact~,Name=Fact~,names=facts~,Names=Facts~}}{}
\RS@ifundefined{partref}{\newref{part}{name=part~,Name=Part~,names=parts~,Names=Parts~}}{}
\RS@ifundefined{probref}{\newref{prob}{name=problem~,Name=Problem~,names=problems~,Names=Problems~}}{}
\RS@ifundefined{eqref}{\newref{eq}{name=equation~,Name=Equation~,names=equations~,Names=Equations~}}{}
\RS@ifundefined{enumref}{\newref{enum}{name=item~,Name=Item~,names=items~,Names=Items~}}{}

\usepackage{color}
\definecolor{purple}{RGB}{0,0,0} %

\usepackage{tikzsymbols}

\usepackage[bottom]{footmisc}
\usepackage{physics}

\usepackage{tikz}
\usepackage{pgfplots}
\pgfplotsset{compat=1.18}

\usepackage{color,graphicx}

\newcommand{\kb}[1]{| #1 \rangle \langle #1 |}

\newcommand{\pg}[2]{\left \llbracket \; #1 \;, \; #2 \; \right\rrbracket} 
\newcommand{\nobarfrac}{\genfrac{}{}{0pt}{}}

\newcommand{\lwcf}{{$\Lambda$-penWCF}}

\newcommand{\half}{\frac{1}{2}}
\newcommand{\err}{\mathsf{err}}

\newcommand{\spacec}{\mathsf{sc}}
\newcommand{\roundc}{\mathsf{rc}}

\newcommand{\biasMochon}{\epsilon_\text{Moch}}

\newcommand{\approxError}{\varepsilon_{\mathsf{approx}}}
\newcommand{\supp}{\mathsf{supp}}

\usepackage{tabto}

\usepackage{tcolorbox}
\tcbuselibrary{theorems} 

\newtcbtheorem[]{mytheo}{Theorem}%
{colback=gray!10,colframe=black!50,fonttitle=\bfseries}{th}

\newtcbtheorem[]{mytheomr}{Main Result}
{colback=gray!10,colframe=black!50,fonttitle=\bfseries}{th}

\newtcbtheorem[]{mydef}{Definition}%
{colback=gray!10,colframe=black!50,fonttitle=\bfseries}{th} 

\newtcbtheorem[]{myq}{Question}%
{colback=gray!10,colframe=black!50,fonttitle=\bfseries}{th}

\usepackage{mdframed}

\newenvironment{theorembox}
  {\vspace{1em} \begin{mdframed}[
      backgroundcolor=gray!20,
      linewidth=0pt,
      nobreak=true,
      innerleftmargin=10pt,
      innerrightmargin=10pt,
      innertopmargin=2pt,
      innerbottommargin=10pt
    ]\begin{thm}}
  {\end{thm}\end{mdframed} \vspace{1em}}
\ifdefined\showcaptionsetup
 \PassOptionsToPackage{caption=false}{subfig}
\fi
\usepackage{subfig}
\makeatother

\usepackage{babel}

\providecommand{\claimname}{Claim}
\providecommand{\definitionname}{Definition}
\providecommand{\lemmaname}{Lemma}
\providecommand{\notationname}{Notation}
\providecommand{\propositionname}{Proposition}
\providecommand{\remarkname}{Remark}
\providecommand{\theoremname}{Theorem}

\begin{document}
\pagenumbering{roman}   %
\title{Cheat-Penalised Quantum Weak Coin-Flipping}
\author{
Atul Singh \textsc{Arora},\thanks{Centre for Quantum Science and Technology (CQST), IIIT Hyderabad\newline Joint Center for Quantum Information and Computer Science (QuICS)\newline
Department of Computer Science, University of Maryland, College Park \newline
 \href{mailto:atul.singh.arora@gmail.com}{atul.singh.arora@gmail.com}} 
~~Carl A. \textsc{Miller},\thanks{Joint Center for Quantum Information and Computer Science (QuICS)\newline 
University of Maryland, College Park\newline 
\href{mailto:camiller@umd.edu}{camiller@umd.edu}}
~~Mauro E.S. \textsc{Morales},\thanks{Joint Center for Quantum Information and Computer Science (QuICS)\newline
Department of Computer Science, University of Maryland, College Park\newline 
\href{mailto:mauroms@umd.edu}{mauroms@umd.edu}}
~~Jamie \textsc{Sikora}\thanks{Virginia Tech Center for Quantum Information Science and Engineering (VTQ)\newline 
Department of Computer Science, Virginia Tech \newline
\href{mailto:sikora@vt.edu}{sikora@vt.edu}\newline 
\emph{The authors are listed in alphabetical order.}} 
}
\date{October 3, 2025} 
\maketitle
\begin{abstract}

Coin-flipping is a fundamental task in two-party cryptography where two remote mistrustful parties wish to generate a shared uniformly random bit. While quantum protocols promising near-perfect security exist for \emph{weak} coin-flipping---when the parties want opposing outcomes---it has been shown that they must be inefficient in terms of their round complexity, and it is an open question of how space efficient they can be. In this work, we consider a variant called \emph{cheat-penalised} weak coin-flipping in which if a party gets caught cheating, they lose $\Lambda$ points (compared to $0$ in the standard definition). We find that already for a small cheating penalty, the landscape of coin-flipping changes dramatically. For example, with $\Lambda=0.01$, we exhibit a protocol where neither Alice nor Bob can bias the result in their favour beyond $1/2 + 10^{-8}$, which uses $24$ qubits and $10^{16}$ rounds of communication (provably $10^{7}$ times better than any weak coin-flipping protocol with matching security). For the same space requirements, we demonstrate how one can choose between lowering how much a malicious party can bias the result (down to $1/2 + 10^{-10}$) and reducing the rounds of communication (down to $25,180$), depending on what is preferred. To find these protocols, we make two technical contributions. First, we extend the point game-protocol correspondence introduced by Kitaev and Mochon, to incorporate: (i) approximate point games, (ii) the cheat-penalised setting, and (iii) round and space complexity. Second, we give the first (to the best of our knowledge) numerical algorithm for constructing (approximate) point games that correspond to high security and low complexity. Our results open up the possibility of having secure and practical quantum protocols for multiparty computation.

\end{abstract}
\global\long\def\poly{{\rm \mathsf{poly}}}%
\global\long\def\tr{\mathsf{tr}}%
\global\long\def\perm#1#2{\!_{#1}P_{#2}}%
\global\long\def\comb#1#2{\,{}_{#1}C_{#2}}%
\global\long\def\td{{\rm TD}}%
\global\long\def\f{\mathcal{F}}%
\global\long\def\g{\mathcal{G}}%
\global\long\def\negl{\mathsf{negl}}%
\global\long\def\BQP{\mathsf{BQP}}%
\global\long\def\BPP{\mathsf{BPP}}%
\global\long\def\TD{{\rm TD}}%
\global\long\def\calD{\mathcal{D}}%
\global\long\def\calL{\mathcal{L}}%
\global\long\def\calM{\mathcal{M}}%

\global\long\def\NP{\mathsf{NP}}%

\global\long\def\gen{\mathsf{Gen}}%

\global\long\def\verify{\mathsf{Verify}}%

\global\long\def\prove{\mathsf{Prove}}%

\global\long\def\pred{\mathsf{pred}}%

\global\long\def\pk{\mathsf{pk}}%

\global\long\def\sk{\mathsf{sk}}%

\global\long\def\bbF{\mathbb{F}}%

\global\long\def\bit{\{0,1\}}%

\global\long\def\G{\mathsf{G}}%

\global\long\def\Call{C^{{\rm all}}}%

\global\long\def\commit{\mathsf{commit}}%

\global\long\def\QTM{\mathsf{QTM}}%

\global\long\def\QIM{\mathsf{QIM}}%

\global\long\def\val{\mathsf{val}}%

\global\long\def\PPT{\mathsf{PPT}}%

\global\long\def\hc{\mathsf{h.c.}}%

\global\long\def\ksampBQP#1{#1\text{-}\mathsf{sampBQP}}%

\global\long\def\sampBQP{\mathsf{sampBQP}}%

\global\long\def\sampBPP{\mathsf{sampBPP}}%

\global\long\def\points#1{\left\llbracket #1\right\rrbracket }%

\global\long\def\spectrum{\mathsf{spectrum}}%

\global\long\def\Prob{\mathsf{Prob}}%

\global\long\def\Rnonneg{[0,\infty)}%

\global\long\def\Proj{\mathsf{Proj}}%

\global\long\def\spn{\mathsf{span}}%

\global\long\def\SWP{\mathsf{SWP}}%

\global\long\def\supp{\mathsf{supp}}%

\global\long\def\sideal{s_{\mathsf{ideal}}}%

\global\long\def\eideal{e_{\mathsf{ideal}}}%

\global\long\def\serror{s_{\mathsf{error}}}%

\global\long\def\eerror{e_{\mathsf{error}}}%

\global\long\def\erest{e_{\mathsf{rest}}}%

\global\long\def\deltacreate{\delta_{\mathsf{create}}}%

\global\long\def\deltaabsorb{\delta_{\mathsf{absorb}}}%

\global\long\def\deltacatalyst{\delta_{\mathsf{clyst}}}%

\global\long\def\deltasfix{\delta_{\mathsf{s\text{-}fix}}}%

\global\long\def\ncatalyst{n_{\mathsf{catalyst}}}%

\global\long\def\maxcoordinate{\mathsf{max\text{-}coordinate}}%

\global\long\def\mincoordinate{\mathsf{min\text{-}coordinate}}%

\newpage{}

\tableofcontents{}

\newpage{}
\pagenumbering{arabic}

\newcommand{\note}[1]{\textcolor{red}{[Jamie: #1]}}

\section{Introduction}

Can two parties who are communicating at a distance, flip a coin in such a way that both are assured that the coin-flip was fair?  In other words, is there an interactive protocol that allows two parties who do not trust one another to create a shared random bit? 
Classically, this task is impossible in the unconditional setting,\footnote{For any classical coin-flipping strategy, von Neumann's mini-max theorem implies that one of the parties will have a way to force a particular outcome.  See \cite{miller2022mathematics}.} although it becomes possible if computational hardness assumptions are made (e.g., \cite{moran2009optimally}).  In the quantum setting, however, there are secure protocols for coin-flipping that do not make computational hardness assumptions and rely only on minimal physical assumptions \cite{aharonov2000quantum, spekkens2001degrees, SR02, Ambainis04b, mochon2004quantum, Mochon05, Mochon07}.  Coin-flipping is a prime example of the unique capabilities of quantum mechanics for cryptographic tasks, and it was one of the original problems that started the field of quantum cryptography \cite{BB84}.

Quantum coin-flipping is part of the larger landscape of \textit{two-party cryptography}, in which two persons who do not trust one another try to accomplish a cooperative task while simultaneously guarding against cheating by the other person. 
Other two-party primitives, such as bit commitment and oblivious transfer, 
automatically imply an ability to perform two-party coin-flipping.  The task of coin-flipping is thus a natural watermark for measuring the power and the limitations of quantum mechanics in two-party cryptography.

The body of work in two-party cryptography offers many results that are beautiful from a theoretical standpoint but daunting from an experimental standpoint. Secure two-party computation, bit commitment, oblivious transfer, and strong coin-flipping have all been proved to be impossible to perform \cite{lo1997insecurity, LO1998177, ambainis2004multiparty,lo1997quantum, Mayers1997}---the best that one can achieve for those tasks are protocols with a fixed constant bias. 
For example, Kitaev proved that for strong coin-flipping---a coin-flipping problem where players are oblivious to the preference of the other---one of the parties can force a particular outcome with probability $\frac{1}{\sqrt{2}}$, meaning that the bias (i.e., the gap between $\frac{1}{2}$ and the probability of successful cheating) is at least 
$\frac{1}{\sqrt{2}} - \frac{1}{2}$. 
In 2007, Mochon \cite{Mochon07} offered a ray of hope by proving that \textit{weak} coin-flipping---coin-flipping in which the desired outcome for each party is known to all a priori---is possible with bias arbitrarily close to zero. 

{However, Mochon's constructions are rather involved and complex. Specifically, the round complexity (the number of rounds of communication between the parties) required to achieve bias $\epsilon$ is very high.  The best known upper bound on the number of communication rounds in Mochon's construction is $(1/\epsilon)^{O(1/\epsilon)}$ \cite{ACG+14}. In fact, the situation was much worse because he used a non-constructive step in his proof, making even the circuit description of his protocols elusive. This was resolved in subsequent works that further improved upon and enriched Mochon's work \cite{ACG+14, chailloux2013optimal, chailloux2017physical, ganz2017quantum, Arora2019,Arora2019b,arora2024protocolsquantumweakcoin}. Yet, none of them managed to improve the communication complexity. In 2020, Miller \cite{miller2020impossibility} proved that any weak coin-flipping protocol with bias $\epsilon$ must have at least $\exp ( \Omega ( 1/\sqrt{\epsilon} ))$ round complexity.  Consequently, there is no efficient protocol---i.e., no protocol in which the time-complexity is polynomial in $(1/\epsilon)$---for the original quantum coin-flipping problem. Unlike round complexity, no space-complexity lower bounds (i.e., lower bounds on the number qubits needed) are known for weak coin-flipping protocols. Therefore, even though Mochon's constructions are known to achieve bias $\epsilon$ with $O(\log(1/\epsilon))$ qubits \cite{ACG+14}, this is not known to be optimal. It is worth noting that Mochon's so-called \emph{Dip-Dip-Boom} protocol approaches bias $1/6$ using only two qutrits and a qubit. In fact, Dip-Dip-Boom may be seen as the first protocol in Mochon's family of protocols, parametrised by $k$, that approach bias $\biasMochon(k):=1/(4k+2)$. Interestingly, beyond $k=1$, no protocol is known to approach bias $\biasMochon(k)$ with \emph{constant} space complexity.}

Two party-cryptography therefore seems challenged from the start.  And yet, two-party cryptography was identified in \cite{wehner2018quantum} (along with QKD and position verification) as one of three potential first-stage applications of  quantum networks, and many experimental works have been done on the topic (see \cite{bozzio2024quantum} for a summary).  There is hope for two-party cryptography if certain theoretical obstacles can be overcome.

The simplest way to circumvent an impossibility proof such as \cite{miller2020impossibility} is to change the mathematical model on which the proof is based.  Fortunately, in the case of coin-flipping, existing literature~\cite{ambainis2004multiparty,Mochon07} gives us one natural and effective way of doing so---namely, by modifying the rules for coin-flipping to include a penalty $\Lambda \geq 0$ for cheating.

In this extended model, Alice and Bob each have three possible outputs: $0$ (Alice wins the coin toss), $1$ (Bob wins the coin toss), and $\bot$ (indicates the other party has been caught cheating). If Alice wins, she gains one point (Bob gains nothing); if Bob wins, he gains one point (Alice gains nothing); and if either party is caught cheating, they lose $\Lambda$ points. 
This model, which we call \emph{$\Lambda$-penalty weak coin-flipping}, or {\bf \lwcf} for short, makes a distinction between a party honestly losing the exchange (and receiving a score of $0$) and a party getting caught cheating (and receiving a score of $-\Lambda$). This distinction is natural when considered as part of a broader scheme where parties engage in multiple interactions, introducing disincentive for malicious behaviour.
There are a few key differences between the previously studied version of weak coin-flipping and the cheat-penalised setting. 
Specifically, in the cheat-penalised setting, each party wishes to maximise their respective \emph{expected reward}, not just the probability that they will win. 
Thus, we want protocols where neither Alice nor Bob can cheat and obtain an expected reward more than $1/2 + \epsilon$, for a preferably small value of $\epsilon > 0$. 
In this work, we not only want the bias $\epsilon$ %
to be small, but also the round complexity $\roundc$ and the space complexity $\spacec$. This motivates the following questions. 
\begin{tcolorbox}[colback=gray!10, colframe=black!50]
\emph{To what extent can introducing a cheat penalty improve the efficiency of weak coin-flipping protocols? How large must the cheat penalty be to achieve this?}
\end{tcolorbox}
Ambainis, Buhrman, Dodis and Roehrig~\cite{ambainis2004multiparty} considered the cheat-penalised setting for multipartite coin-flipping. In particular, their results show that for $\Lambda \ge 4$, one can construct (bipartite) cheat-penalised weak coin-flipping protocols with constant space and round complexity, where the bias $\epsilon$ vanishes as $\Lambda\to \infty$. Mochon~\cite{Mochon07}, in a brief detour, gave a cheat-penalised protocol (a variant of Dip-Dip-Boom) for any $\Lambda\ge 0$ that uses constant space. He asserted, heuristically, that its bias also vanishes in the limit of both the round complexity and the cheating parameter $\Lambda$ going to $\infty$. Besides these two works, we are unaware of any results in this direction.

\begin{figure}[H]
\centering
\includegraphics[width=0.6\paperwidth]{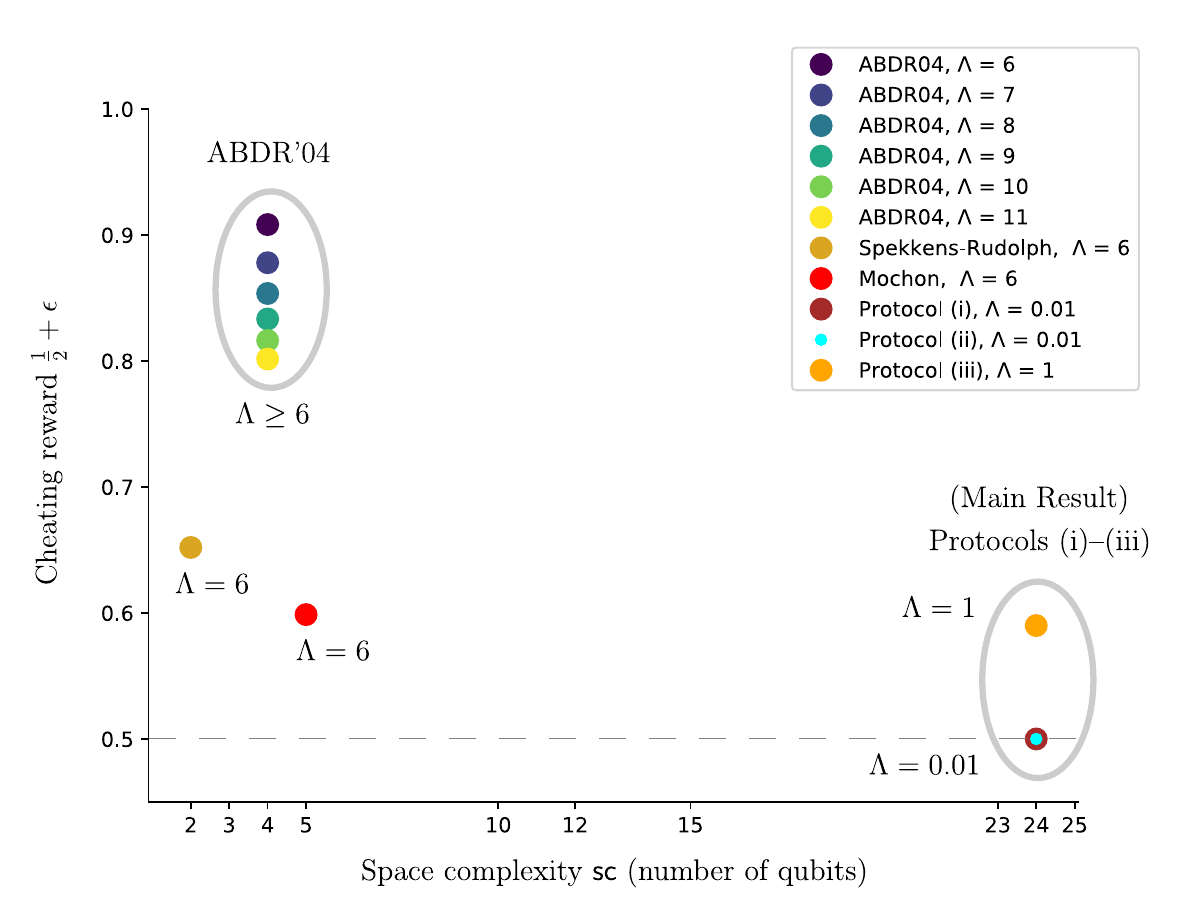}
\caption{Comparison of cheat-penalised weak coin flipping protocols in terms of the greatest expected reward, i.e. $\frac{1}{2} + \epsilon$, and the number of qubits used in the protocol. We compare our Protocols (i)--(iii) to that of Ambainis, Buhrman, Dodis and Roehrig~\cite{ambainis2004multiparty}, the Spekkens-Rudolph protocol~\cite{SR02} (which we extend to the cheat-penalty setting), and Mochon's Dip-Dip-Boom cheat-penalised version~\cite{Mochon07} (we rigorously derive and extend his heuristic bounds on the bias). Protocols~(i)--(iii) are constructed using abstract objects we call cheat-penalised Time-Independent Point Games (penTIPGs). Protocols~(i) and (ii) have $\roundc=10^{16}$ and $\roundc=10^{18}$ respectively and are based on penTIPG 3 which is illustrated in~\Figref{TIPG4_plot} (not to scale). Protocol (iii) has round complexity $\roundc=25,180$ and is based on penTIPG 3. All these penTIPGs and related details appear in~\Secref{other_protocols}.} %
\label{fig:reward-qubits}
\end{figure}

By extending the formalism of point games (which we briefly recall below) to the cheat-penalised setting and developing a new numerical algorithm for constructing approximate point games, we obtain protocols that %
significantly outperform prior constructions in the following precise sense.
\begin{mytheomr*}{The existence of more efficient, secure cheat-penalised weak coin-flipping} %
There exist weak coin-flipping protocols with cheat penalty $\Lambda=0.01$, space complexity (number of qubits) $\spacec=24$ and the following trade-offs between the bias $\epsilon$ and the round complexity $\roundc$.
\begin{description} %
    \item[{(i) More round efficient than WCF.}] \tabto{5.6cm} Bias $\epsilon=10^{-8}$, round complexity $\roundc=10^{16}$ \\ \tabto{5.6cm}($\roundc$ is still $10^7$ times better than\\ 
    \tabto{5.6cm}~any possible WCF protocol with a matching bias)
    \item[{(ii) Constant space with low bias.}] \tabto{5.6cm} Bias $\epsilon=10^{-10}$, round complexity $\roundc=10^{18}$, $24$ qubits\\ \tabto{5.6cm} ($\roundc$ is still $10^5$ times better)
\end{description}
To obtain a better trade-off, we use $\Lambda=1$ below.
\begin{description}
    \item[{(iii) Potentially amenable to experiments.}] \tabto{5.6cm} Bias $\epsilon=0.09$, round complexity $\roundc=25,180$\\ \tabto{5.6cm} ($\epsilon < \biasMochon(2) = 1/10$) %
\end{description} %
\end{mytheomr*} 
In all the protocols above, Alice and Bob hold 8 qubits each and they exchange messages using a 8 qubit register. The comparison of Protocol (i) above with (non-cheat-penalised) WCF protocols is based on Miller's work~\cite{miller2020impossibility} that show that any WCF protocol with bias $\epsilon=10^{-8}$, must have round complexity at least $\roundc=10^{19}$ (these explicit numbers were computed later in~\cite{Alnawakhtha2025}). Protocol (ii) shows that with 24 qubits, one can get very close to zero bias---i.e., we have a constant space protocol that can go as low as $\epsilon=10^{-10}$. The only remotely comparable protocol in the (non-cheat-penalised) coin-flipping setting is Mochon's Dip-Dip-Boom that uses two qutrits and a qubit, to approach bias $\epsilon=1/6$ in the limit of infinite round complexity. Protocol (ii) strongly suggests that in the cheat-penalised setting, there may be a constant space protocol that has bias arbitrarily close to zero. Protocol (iii) may be a reasonable compromise between the bias and the round complexity. Crucially, all our protocols are such that the message register can be discarded after every two rounds (i.e., after two messages have been exchanged).\footnote{We clarify that we count one message from one party to the other as constituting one round. In other works, sometimes one sees the convention where two messages are treated as one round of communication.} Consequently, one only needs to keep the local quantum memories of the parties coherent throughout the execution of the protocol. These features suggest that Protocol (iii)  may \emph{potentially} be amenable to experiments.
Figure~\ref{fig:reward-qubits} compares known protocols and our results, in terms of the bias and space complexity while \Figref{reward-messages} (in the main text) compares bias and round complexity. 

Our work makes several technical contributions in order to achieve this main result. To describe these, we briefly introduce some of the key concepts we use. %

\paragraph{Point Games.}  
Kitaev and Mochon~\cite{Mochon07} introduced various so-called \emph{point games} to design and analyse coin-flipping protocols. For a full introduction to this formalism we refer the reader to \cite{Mochon07,ACG+14}. Here, we focus on two variants: \emph{time-independent point games} ({\bf TIPGs}) and \emph{time-dependent point games} ({\bf TDPGs}). %
A TIPG is specified by two bivariate functions $(h,v)$ where each encodes a finite set of weighted configurations of points on a two-dimensional plane. %
We use the notation $p \llbracket x, y \rrbracket$ to indicate that the point $(x,y)$ has weight $p$ where $p\in \mathbb{R}$. Slightly more formally, $\points{x,y}$ is a bivariate function that is zero everywhere except on input $(x,y)$ where it evaluates to one.
A TIPG $(h,v)$ for (non-penalised) weak coin-flipping has bias $\epsilon$ if it satisfies
\begin{equation} 
h + v = 
         \underbrace{\points{\half + \epsilon, \half + \epsilon}}_{\text{final config.}}
        - \underbrace{\left(\frac{1}{2} \llbracket 0, 1 \rrbracket 
        + \frac{1}{2} \llbracket 1, 0 \rrbracket\right)}_{\text{initial config.}} 
\end{equation} 
together with extra ``validity conditions'' that we are suppressing for now. The functions $h$ and $v$ encode a protocol for weak coin-flipping, i.e., for any such TIPG, there exist weak coin-flipping protocols that approach bias $\epsilon$ (up to arbitrary precision). 
As alluded to earlier, Mochon~\cite{Mochon07} constructed a family of TIPGs parametrised by $k$ that approach bias $\epsilon=1/(4k+2)$, thereby establishing the existence of weak coin-flipping with vanishing bias, $\epsilon\to0$.
However, in 2020, Miller showed that there is a relationship between the round complexity of weak coin-flipping protocols and the norms of $h$ and $v$ in any point game, leading to the conclusion that 
any weak coin-flipping protocol must have at least $\exp(\Omega(\epsilon^{-1/2}))$ rounds of communication. This renders them necessarily inefficient as $\epsilon \to 0$. A TDPG provides an alternative way of specifying a weak coin-flipping protocol. Instead of being defined by two functions, it is specified by a sequence of bivariate functions $(p_0,p_1,\dots, p_n)$ such that  $p_0=\half\points{0,1}+\half\points{1,0}$, $p_n=\points{\half + \epsilon, \half + \epsilon}$ and each pair of functions $(p_i,p_{i+1})$ fulfil a validity condition, analogous to the one imposed in the TIPG formalism. While TIPGs encode protocols in a time-independent fashion---capturing only the start and end conditions with an implicit validity constraint---TDPGs represent the protocol as a sequence of valid transitions across rounds. This representation is therefore more directly connected to the round complexity of the protocol.

\paragraph{Contribution 1: Approximate cheat-penalised TIPGs and how to find them.}  
We revisit the construction of point games in~\cite{Mochon07}, and define the notion of a \emph{cheat-penalised TIPG}, which we denote by {\bf penTIPG}. In essence, here one translates both the ``initial'' and ``final configurations'' by $\Lambda$ (along both axes).
While Mochon already informally considered such penTIPGs, his focus was establishing security for weak coin-flipping (without penalty) and not efficiency. Consequently, he did not study such games in further detail.
In our work, we are concerned with \emph{efficiency} as well, so our starting point is finding penTIPGs with small norm which can potentially give more efficient protocols.
It does not take long to realise that a brute force search is hopeless, as the search space includes the space of all bivariate functions $(h,v)$. 
 Our first contribution is a numerical algorithm to find such point games. Not only does our algorithm yield solutions, it does so even for small values of $\Lambda$. However, these solutions are \emph{approximate} and we therefore need to slightly relax the definition of penTIPG as follows.
\begin{mydef}{$(\Lambda,\approxError)$-penTIPG---Approximate cheat-penalised TIPGs (\Defref{penTIPGformal} simplified)}{theoexample}
    We say $(h,v)$ is a $(\Lambda,\approxError)$-penTIPG with bias $\epsilon$ if it satisfies 
    \begin{equation}\label{eq:penTIPG}
        \Bigg\rVert h+v - 
                \Bigg(\underbrace{\points{\Lambda + \half + \epsilon, \Lambda + \half + \epsilon }}_{\text{final config.}} - 
                \underbrace{\left(  
                    \half \points{\Lambda+1,\Lambda} +
                    \half \points{\Lambda,\Lambda + 1}
                \right)}_{\text{initial config.}} \Bigg)  \Bigg\lVert_1\leq \approxError
    \end{equation}
in addition to the ``validity conditions'' mentioned previously. %
By the \emph{number of points} of $(h,v)$, we mean the number of input pairs that are assigned non-zero weight by $h$ and $v$, i.e., $|\supp\{h,v\}|$. By the \emph{norm of $(h,v)$} we mean $\max\{\|h\|_1,\|v\|_1\}$.
\end{mydef} 
We obtain various solutions but we highlight the one that results in a cheat-penalised WCF protocol with final bias $\epsilon=10^{-8}$, where the cheat penalty is $\Lambda=0.01$. As for efficiency, it uses 24 qubits and has round complexity more than \emph{seven orders of magnitude} lower compared to Miller and Alnawaktha's~\cite{miller2020impossibility,Alnawakhtha2025} lower bound for the analogous WCF protocol. 

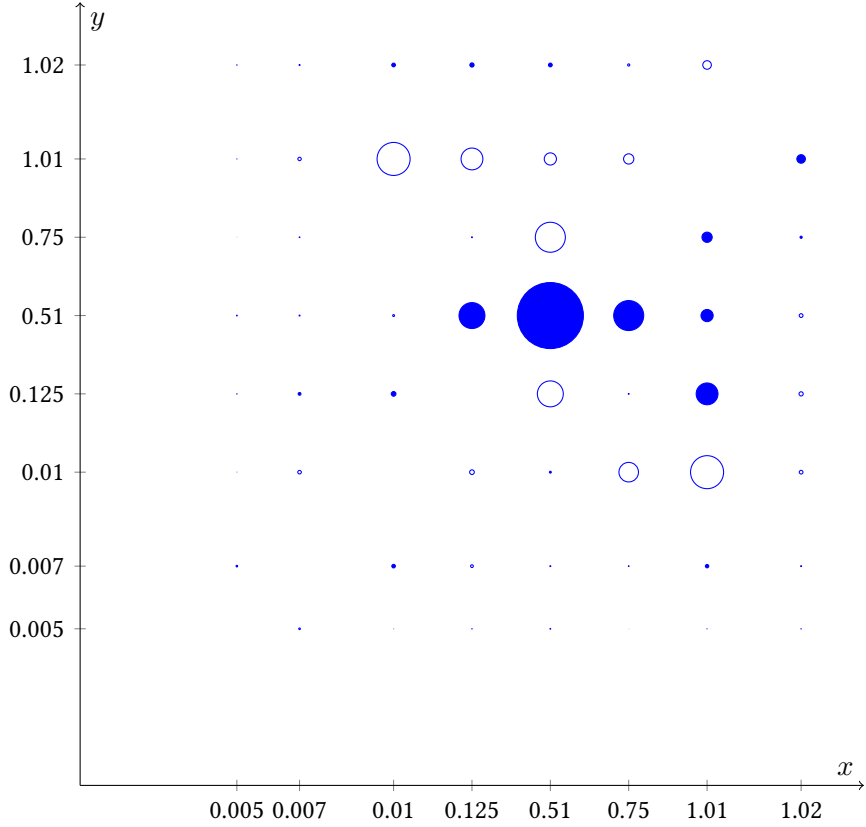
\begin{figure}
    \begin{center}
        \begin{tikzpicture}
\begin{axis}[
    axis lines=middle,
    xlabel={$x$},
    ylabel={$y$},
    xtick={0.005, 0.007, 0.01, 0.0125, 0.015000000000005, 0.0175, 0.02, 0.023},
    ytick={0.005, 0.007, 0.01, 0.0125, 0.015000000000005, 0.0175, 0.02, 0.023},
    xticklabels={0.005, 0.007, 0.01, 0.125, 0.51, 0.75, 1.01, 1.02},
    yticklabels={0.005, 0.007, 0.01, 0.125, 0.51, 0.75, 1.01, 1.02},
    scaled ticks=false,
    grid=none,
    xmin=0, xmax=0.025,
    ymin=0, ymax=0.025,
    ticklabel style={font=\small},
    axis line style={->},
    width=12cm,
    height=12cm
]

\pgfmathsetmacro{\scale}{25} %

\addplot[mark=*, mark size=\scale*abs(8.32667e-18), blue, fill=blue] coordinates {(0.005,0.005)};
\addplot[mark=*, mark size=\scale*abs(-0.012549), draw=blue, fill=white] coordinates {(0.007,0.005)};
\addplot[mark=*, mark size=\scale*abs(0.000867), draw=blue, fill=white] coordinates {(0.01,0.005)};
\addplot[mark=*, mark size=\scale*abs(0.002990), blue, fill=blue] coordinates {(0.0125,0.005)};
\addplot[mark=*, mark size=\scale*abs(-0.005655), blue, fill=blue] coordinates {(0.015000000000005,0.005)};
\addplot[mark=*, mark size=\scale*abs(0.000098), draw=blue, fill=white] coordinates {(0.0175,0.005)};
\addplot[mark=*, mark size=\scale*abs(0.001934), blue, fill=blue] coordinates {(0.02,0.005)};
\addplot[mark=*, mark size=\scale*abs(0.002993), draw=blue, fill=white] coordinates {(0.023,0.005)};

\addplot[mark=*, mark size=\scale*abs(0.012549), blue, fill=blue] coordinates {(0.005,0.007)};
\addplot[mark=*, mark size=\scale*abs(6.93889e-18), blue, fill=blue] coordinates {(0.007,0.007)};
\addplot[mark=*, mark size=\scale*abs(-0.027156), blue, fill=blue] coordinates {(0.01,0.007)};
\addplot[mark=*, mark size=\scale*abs(-0.021163), draw=blue, fill=white] coordinates {(0.0125,0.007)};
\addplot[mark=*, mark size=\scale*abs(0.007004), draw=blue, fill=white] coordinates {(0.015000000000005,0.007)};
\addplot[mark=*, mark size=\scale*abs(0.005266), blue, fill=blue] coordinates {(0.0175,0.007)};
\addplot[mark=*, mark size=\scale*abs(0.025788), blue, fill=blue] coordinates {(0.02,0.007)};
\addplot[mark=*, mark size=\scale*abs(-0.007110), draw=blue, fill=white] coordinates {(0.023,0.007)};

\addplot[mark=*, mark size=\scale*abs(-0.000867), blue, fill=blue] coordinates {(0.005,0.01)};
\addplot[mark=*, mark size=\scale*abs(0.027156), draw=blue, fill=white] coordinates {(0.007,0.01)};
\addplot[mark=*, mark size=\scale*abs(2.77556e-18), blue, fill=blue] coordinates {(0.01,0.01)};
\addplot[mark=*, mark size=\scale*abs(-0.035587), draw=blue, fill=white] coordinates {(0.0125,0.01)};
\addplot[mark=*, mark size=\scale*abs(-0.014383), blue, fill=blue] coordinates {(0.015000000000005,0.01)};
\addplot[mark=*, mark size=\scale*abs(-0.147784), draw=blue, fill=white] coordinates {(0.0175,0.01)};
\addplot[mark=*, mark size=\scale*abs(-0.25), draw=blue, fill=white] coordinates {(0.02,0.01)};
\addplot[mark=*, mark size=\scale*abs(-0.028092), draw=blue, fill=white] coordinates {(0.023,0.01)};

\addplot[mark=*, mark size=\scale*abs(-0.002990), draw=blue, fill=white] coordinates {(0.005,0.0125)};
\addplot[mark=*, mark size=\scale*abs(0.021163), blue, fill=blue] coordinates {(0.007,0.0125)};
\addplot[mark=*, mark size=\scale*abs(0.035587), blue, fill=blue] coordinates {(0.01,0.0125)};
\addplot[mark=*, mark size=\scale*abs(2.77556e-18), blue, fill=blue] coordinates {(0.0125,0.0125)};
\addplot[mark=*, mark size=\scale*abs(-0.196252), draw=blue, fill=white] coordinates {(0.015000000000005,0.0125)};
\addplot[mark=*, mark size=\scale*abs(0.006301), blue, fill=blue] coordinates {(0.0175,0.0125)};
\addplot[mark=*, mark size=\scale*abs(0.165097), blue, fill=blue] coordinates {(0.02,0.0125)};
\addplot[mark=*, mark size=\scale*abs(-0.031862), draw=blue, fill=white] coordinates {(0.023,0.0125)};

\addplot[mark=*, mark size=\scale*abs(0.005655), draw=blue, fill=white] coordinates {(0.005,0.015000000000005)};
\addplot[mark=*, mark size=\scale*abs(-0.007004), blue, fill=blue] coordinates {(0.007,0.015000000000005)};
\addplot[mark=*, mark size=\scale*abs(0.014383), draw=blue, fill=white] coordinates {(0.01,0.015000000000005)};
\addplot[mark=*, mark size=\scale*abs(0.196252), blue, fill=blue] coordinates {(0.0125,0.015000000000005)};
\addplot[mark=*, mark size=\scale*abs(0.5), blue, fill=blue] coordinates {(0.015000000000005,0.015000000000005)};
\addplot[mark=*, mark size=\scale*abs(0.227074), blue, fill=blue] coordinates {(0.0175,0.015000000000005)};
\addplot[mark=*, mark size=\scale*abs(0.092193), blue, fill=blue] coordinates {(0.02,0.015000000000005)};
\addplot[mark=*, mark size=\scale*abs(-0.028695), draw=blue, fill=white] coordinates {(0.023,0.015000000000005)};

\addplot[mark=*, mark size=\scale*abs(-0.000098), blue, fill=blue] coordinates {(0.005,0.0175)};
\addplot[mark=*, mark size=\scale*abs(-0.005266), draw=blue, fill=white] coordinates {(0.007,0.0175)};
\addplot[mark=*, mark size=\scale*abs(0.147784), blue, fill=blue] coordinates {(0.01,0.175)};
\addplot[mark=*, mark size=\scale*abs(-0.006301), draw=blue, fill=white] coordinates {(0.0125,0.0175)};
\addplot[mark=*, mark size=\scale*abs(-0.227074), draw=blue, fill=white] coordinates {(0.015000000000005,0.0175)};
\addplot[mark=*, mark size=\scale*abs(4.16334e-18), blue, fill=blue] coordinates {(0.0175,0.0175)};
\addplot[mark=*, mark size=\scale*abs(0.077654), blue, fill=blue] coordinates {(0.02,0.0175)};
\addplot[mark=*, mark size=\scale*abs(0.016646), blue, fill=blue] coordinates {(0.023,0.0175)};

\addplot[mark=*, mark size=\scale*abs(-0.001934), draw=blue, fill=white] coordinates {(0.005,0.02)};
\addplot[mark=*, mark size=\scale*abs(-0.025788), draw=blue, fill=white] coordinates {(0.007,0.02)};
\addplot[mark=*, mark size=\scale*abs(-0.25), draw=blue, fill=white] coordinates {(0.01,0.02)};
\addplot[mark=*, mark size=\scale*abs(-0.165097), draw=blue, fill=white] coordinates {(0.0125,0.02)};
\addplot[mark=*, mark size=\scale*abs(-0.092193), draw=blue, fill=white] coordinates {(0.015000000000005,0.02)};
\addplot[mark=*, mark size=\scale*abs(-0.077654), draw=blue, fill=white] coordinates {(0.0175,0.02)};
\addplot[mark=*, mark size=\scale*abs(3.98986e-18), blue, fill=blue] coordinates {(0.02,0.02)};
\addplot[mark=*, mark size=\scale*abs(0.064828), blue, fill=blue] coordinates {(0.023,0.02)};

\addplot[mark=*, mark size=\scale*abs(-0.002993), blue, fill=blue] coordinates {(0.005,0.023)};
\addplot[mark=*, mark size=\scale*abs( 0.007110), blue, fill=blue] coordinates {(0.007,0.023)};
\addplot[mark=*, mark size=\scale*abs(0.028092), blue, fill=blue] coordinates {(0.01,0.023)};
\addplot[mark=*, mark size=\scale*abs(0.031862), blue, fill=blue] coordinates {(0.0125,0.023)};
\addplot[mark=*, mark size=\scale*abs(0.028695), blue, fill=blue] coordinates {(0.015000000000005,0.023)};
\addplot[mark=*, mark size=\scale*abs(-0.016646), draw=blue, fill=white] coordinates {(0.0175,0.023)};
\addplot[mark=*, mark size=\scale*abs( -0.064828 ), draw=blue, fill=white] coordinates {(0.02,0.023)};
\addplot[mark=*, mark size=\scale*abs(8.67362e-18), blue, fill=blue] coordinates {(0.023,0.023)};
\end{axis}
\end{tikzpicture}
    \end{center}
    \caption{\small Graphical depiction (axes not to scale) of the parameters that entirely specify our bias $\epsilon=10^{-10}$ protocol---except for the parameter that controls the trade-off with the round complexity. More precisely, the graph shows the function $h$ of a time-independent point game $(h,v)$ with cheat penalty $\Lambda=0.01$ and approximation error $\approxError=10^{-18}$, referenced above as a $(\Lambda,\approxError)$-penTIPG. The filled circles correspond to positive weights and unfilled circles to negative weights while the radius indicates the magnitude of the weight.}\label{fig:TIPG4_plot}
\end{figure}

\begin{mytheo}{Existence of approximate cheat-penalised time-independent point games with small norm (see \Secref{tipgsforpenwcf})}{theoexample}
For $\Lambda=0.01$ and $\approxError=10^{-18}$, there exists a $(\Lambda,\approxError)$-penTIPG with bias $\epsilon=10^{-14}$ (see Figure~\ref{fig:TIPG4_plot}). Furthermore, it has norm $1.05$, and uses at most $64$ points.
\end{mytheo}
Although we leave a rigorous performance analysis of our algorithm for future work, we note that once our algorithm finds a solution, the validity of the functions $h$ and $v$ can be easily checked.
The code, together with other penTIPG solutions, are on GitHub~\cite{penWCFcode}. 
It is important to emphasise that this result, on its own, does not resolve the central question of achieving efficient and secure weak coin-flipping. We need to tackle the reduction from, approximate penTIPG to protocols. This turns out to be subtle and requires new ideas. We proceed in two steps which we summarise next.

\paragraph{Contribution 2: From approximate penTIPGs to protocols.}
We first map the approximate penTIPGs to (exact) penTDPGs and then map the penTDPGs to protocols, for which we also give round and space complexities. %

\paragraph{Mapping approximate time-independent point games to (exact) time-dependent point games.}  
First, we convert an \emph{approximate} cheat-penalised TIPG (penTIPG) into an \emph{(exact) cheat-penalised time-dependent point game} ({penTDPG}). We defer its description to the Technical Overview (see \Subsecref{point}) %
but note that it serves a similar purpose here as TDPGs do in the (non-penalised) WCF setting. Unlike a $\Lambda$-penTIPG, a {\bf $\Lambda$-penTDPG} with bias $\epsilon$ is specified by a sequence of $n$ configurations with positive weights $(p_0,p_1,p_2,\dots,p_n)$ starting with $p_0$ which is the ``initial configuration'' and $p_n$ which is the ``final configuration'' as in Equation~\ref{eq:penTIPG}. The intermediate configurations are required to satisfy ``validity conditions'' similar to those of TIPGs. As before, each change in configuration, intuitively, corresponds to one round of interaction.

\begin{mytheo}{Mapping approximate penTIPGs to (exact) penTDPGs (\Thmref{LambdaPenTIPGtoLambdaPenTDPG} simplified)}{theoexample}
    Let $(h,v)$ be a $(\Lambda,\approxError)$-penTIPG with bias $\epsilon$. Then, one can construct an (exact) $\Lambda$-penTDPG $(p_0,\dots, p_n)$ with bias $\epsilon + \err$ where the trade-off between $n$ and $\err$ can be tuned using two parameters (details suppressed). However, using these parameters, $\err$ can only be made to approach zero if $\approxError=0$. Finally, the number of points in the (exact) penEBM point game $(p_0, \dots, p_n)$ is at most the number of points in the approximate penTIPG $(h,v)$, i.e., $\max_i\{|\supp(p_i)|\} \le |\supp\{h,v\}|$.
\end{mytheo} 

We emphasise that the above deals with \emph{approximate} point games and that such a reduction is new; it has not been studied in the non-cheat-penalised setting and could find application in other coin-flipping or quantum multiparty frameworks. Note that even if the bias $\epsilon$ is small, the error term $\err$ could dominate, thereby making the final bias of the protocol higher.

\paragraph{Mapping time-dependent point games to protocols.}  
In Miller~\cite{miller2020impossibility}, it was shown that an efficient WCF protocol leads to a point game with small norm (which he shows cannot exist).  
What we want is the converse in the cheat-penalised setting: point games with small norm lead to efficient protocols---together with explicit bounds on the resources required.
While the former is a simple application of the triangle inequality, the latter turns out to be more delicate and involved (even given the point game analysis for non-cheat-penalised WCF). As a result, we get the following theorem.

\begin{mytheo}{Mapping time-dependent point games to protocols (\Thmref{LambdaPenTDPG-implies-Lambda-penWCF} simplified)}{theoexample}  
Let $(p_0,\dots,p_n)$ be a $\Lambda$-penTDPG with bias $\epsilon$. Then there exists a $\Lambda$-penWCF protocol with the same bias, $\epsilon$, that has round-complexity $\roundc=2n$ and space-complexity (number of qubits)\\ $\spacec=3\cdot \lceil \log_2 (2 \mu + 1) \rceil$ where $\mu$ is the greatest number of points in a configuration.
\end{mytheo}

Combining these, we obtain the main result we stated earlier.

\paragraph{Significance.} 
Weak coin-flipping stands out as one of the few cryptographic primitives that admits \emph{unconditional security} in the quantum multiparty setting---a rare property in this field.  
This alone makes it an essential object of study, and our work shows that it can now be done more efficiently, giving hope that these can be implemented \emph{and used} on real hardware. 
Moreover, optimal weak coin-flipping protocols are known to underpin a broad class of other cryptographic constructions. In particular, the only known optimal quantum protocols for strong coin-flipping~\cite{CK09}, bit commitment~\cite{CK11}, and oblivious transfer~\cite{chailloux2013optimal} all critically rely on having access to optimal weak coin-flipping subroutines. As a consequence, the inefficiency of weak coin-flipping directly limits the practicality of these higher-level protocols. 
By revisiting the question of efficient and secure weak coin-flipping, we not only address an open problem but potentially unlock new avenues for secure quantum multiparty computation.

\paragraph{Organisation.}
The remainder of this article is organised as follows. Section~\ref{sec:TechOverview} provides a technical overview of our results. Sections~\ref{sec:SDP-penWCF} through~\ref{sec:Lambda-TIPG} present our first main contribution: extending the point game formalism to the cheat-penalised setting. Sections~\ref{sec:alg} through~\ref{sec:comparisonPenWCF} then describe our second main contribution: a numerical algorithm for constructing point games, together with an analysis of the resulting protocols.

For the first contribution, we introduce $\Lambda$-penalty weak coin-flipping (\lwcf) protocols and the point game formalism associated with them.
We present the formulation of {\lwcf} protocols as semidefinite programs (SDPs) in Section \ref{sec:SDP-penWCF}. Then, we present the formalism of EBM point games in the cheat penalise setting in Section \ref{sec:Lambda-EBM}. Sections \ref{sec:Lambda-TDPG} and \ref{sec:Lambda-TIPG} introduce time-dependent point games and time-independent point games respectively. We show how all these formulations are equivalent to {\lwcf} protocols.

For the second contribution, Section~\ref{sec:alg} presents our numerical algorithm along with the intuition underlying its design. Section~\ref{sec:tipgsforpenwcf} illustrates some of the TIPGs obtained using this algorithm. Section~\ref{sec:other_protocols} discusses cheat-penalised versions of existing WCF protocols from the literature. Finally, Section~\ref{sec:comparisonPenWCF} compares the protocols obtained via our numerical approach with previously known constructions.

\newpage
\section{Technical Overview\label{sec:TechOverview}}

In this section, we begin by presenting Kitaev's formalism for two-party protocols for (non-cheat-penalised) weak coin-flipping and its cheat-penalised variant. We then introduce the point game---explaining first time-dependent point games and subsequently time-independent point games---for both the non-cheat-penalised and cheat-penalised cases. We conclude by describing our numerical algorithm to find good point games. 
\vspace{1em}

\subsection{Problem Statement}
\subsubsection*{Weak coin-flipping (non-cheat-penalised)}

Let $n$ be an even positive integer. An $n$-message \emph{weak coin flipping protocol} (\textbf{WCF}) may be described as follows.  Suppose that two parties, Alice and Bob, wish to flip a coin with the outcome being either $0$ (``heads'') or $1$ (``tails''). %
We assume that the parties have opposite preferences: Alice ``wins'' if the outcome is $0$ and Bob ``wins'' when the outcome is $1$.
Alice possesses a local (quantum) memory register $\mathcal{A}$, and Bob possesses a local (quantum) memory register $\mathcal{B}$ and there is a third quantum message register $\mathcal{M}$ that is passed between the two parties. At the end of the protocol, each party outputs a bit representing what they believe to be the outcome of the coin flip. The protocol proceeds as follows.

\begin{enumerate}
    \item Before round $1$, the state of $(\mathcal{A}, \mathcal{B}, \mathcal{M})$ is a pure tripartite product state.

    \item Alice possesses $\mathcal{M}$ on odd rounds, and Bob possesses $\mathcal{M}$ on even rounds.

    \item For $i$ odd, on the $i$th round Alice first applies a binary measurement to $(\mathcal{A},  \mathcal{M} )$ to determine whether Bob has cheated.  If Alice detects cheating, she aborts the protocol and outputs $0$ (her desired outcome).  If she does not detect cheating, she applies a unitary operator to $(\mathcal{A},  \mathcal{M} )$ and then sends $\mathcal{M}$ to Bob.

    \item For $i$ even, Bob  performs an analogous (possibly different) operation on the registers $(\mathcal{B}, \mathcal{M})$ and then sends $\mathcal{M}$ back to Alice. 
    \item After the $n$th round, Alice performs a binary measurement on $\mathcal{A}$ to obtain her output bit $a$, and Bob performs a binary measurement on $\mathcal{B}$ to obtain his output bit $b$.
    
\end{enumerate} 

Each player has a prescribed (honest) behaviour during round $i$, which is represented mathematically by binary measurements $\{ E_i, \mathbb{I} - E_i \}$ and unitary operators $U_i$ (both on $\mathcal{A} \otimes \mathcal{M}$ or $\mathcal{B} \otimes \mathcal{M}$ depending on whether $i$ is even or odd) and final projection operators $\{ \Pi_A^{(0)}, \Pi_A^{(1)} \}$ and $\{ \Pi_B^{(0)}, \Pi_B^{(1)} \}$ (on $\mathcal{A}$ and $\mathcal{B}$ respectively).  
These operators define the coin flipping protocol. We assume that these operators are chosen so that if both players behave honestly, then the protocol behaves correctly, i.e., 
\begin{eqnarray}
\label{eq:intro_correctness}
    \Pr( a = b = 1) = \Pr( a = b = 0 ) = \frac{1}{2}.
\end{eqnarray} 
This condition enforces that the outputs of Alice and Bob are the same and generated uniformly by the protocol. 

A malicious party may deviate arbitrarily\footnote{No bounds on their computational abilities are assumed.} (except that it cannot influence registers held by the honest party) from the prescribed protocol to bias the output towards their preferred value. Denote by $P^*_A$ the greatest probability $\Pr(b=0)$ with which a malicious Alice interacting with an honest Bob, can make him output $b=0$. Denote by $P^*_B$ the analogous quantity where Bob is malicious and interacts with an honest Alice. Given a WCF protocol, one can cast both $P^*_A$ and $P^*_B$ as semidefinite programs. The bias of the protocol is given by 
\begin{eqnarray}    
    \epsilon & := & \max \{ P^*_A, P^*_B \} - \frac{1}{2} \; \in \; [0, 1/2]
\end{eqnarray}
which is the maximum probability one of the parties can force their desired outcome above $1/2$, the honest outcome probability. 

\subsubsection*{Cheat-penalised weak coin-flipping}

The structure of the problem changes significantly if we (following \cite{ambainis2004multiparty}) introduce a third parameter $\Lambda$, the \emph{cheat penalty}. Suppose that we allow Alice and Bob to output, in addition to $0$ and $1$, an abort symbol $\bot$ that indicates cheating has been detected. Intuitively, we still want outcomes $0,1$ to determine who ``wins'' but the outcome $\bot$ means that the malicious party not only ``loses'' but is penalised. To make this quantitative, on outcomes $0$ and $1$, we assign $1$ point to the winner and $0$ to the loser, while on outcome $\bot$, we penalise the cheating party by subtracting $\Lambda$ points. We assume that each party wants to maximise their expected reward. This discussion leaves room for confusion---who is the ``we'' in the description above, what happens if both parties output $\bot$ and so on. To clarify that, we need to be slightly more formal. We first modify the protocol to accommodate three outcomes.

\begin{itemize}
    \item For steps $3-4$, if Alice or Bob detects cheating, they abort the protocol and return the symbol $\bot$ (instead of just declaring themselves the winner). 

    \item On step $5$, Alice performs a ternary measurement $\{ \Pi_A^{(0)}, \Pi_A^{(1)},
    \Pi_A^{(\bot)} \}$ and returns the result.  Bob likewise applies $\{ \Pi_B^{(0)}, \Pi_B^{(1)},
    \Pi_B^{(\bot)} \}$ and returns the result.
\end{itemize}

We enforce the same correctness condition (as in \Eqref{intro_correctness}). As a consequence, we note that when both parties are honest, the outcome $\bot$ never occurs, and one of the parties receives 1 point, each with equal probability.

As for security, we consider the setting where one party is malicious %
and the other honest (i.e., follows the protocol).
\begin{itemize}
    \item Alice is honest, Bob is malicious. 
    \begin{itemize}
        \item If Alice outputs $1$ (i.e., ``Bob wins''), Bob gets 1 point.
        \item If Alice outputs $0$ (i.e., ``Alice wins''), Bob gets 0 points.
        \item If Alice outputs $\bot$ (i.e., ``cheating detected''), Bob loses $\Lambda$ points.
    \end{itemize}
    \item Bob is honest, Alice is malicious.
    \begin{itemize}
        \item If Bob outputs $0$ (i.e., ``Alice wins''), Alice gets 1 point.
        \item If Bob outputs $1$ (i.e., ``Bob wins''), Alice gets 0 points.
        \item If Bob outputs $\bot$ (i.e., ``cheating detected''), Alice loses $\Lambda$ points.
    \end{itemize}
\end{itemize} 

Notice that this clarifies the issues raised earlier---the goal for a malicious party is to convince an honest party to give them as many points as possible. Denote by $R'_A$ the supremum of the average scores achieved by a malicious Alice when interacting with an honest Bob. Similarly, denote by $R'_B$ the analogous quantity for malicious Bob interacting with an honest Alice. Then, the bias of the protocol is given by 
\begin{eqnarray}
    \epsilon &:= &\max\{R'_A,R'_B\} - \frac{1}{2} \; \in \; [0, 1/2].
\end{eqnarray}

It is also elementary to note that for $\Lambda=0$, we reduce to (non-cheat-penalised) WCF. To see this, notice that the reward for malicious Bob is the same when Alice outputs $0$ or if she outputs $\bot$. One can argue similarly for malicious Alice and honest Bob. Thus, the distinction between the outcome $\bot$ and the bit corresponding to ``losing'', vanishes, bringing us back to (non-cheat-penalised) WCF where instead of $\bot$ the honest party declares themselves to be the winner. 

\subsubsection*{Cheat-penalised weak coin-flipping, reformulated}
It turns out that, for technical reasons which we explain shortly, it is easier to work with a ``translated'' scoring convention where the ``cheating detection'' outcome $\bot$ corresponds to getting $0$ points. More precisely, we add $\Lambda$ points globally, to get the following convention. 

\begin{itemize}
    \item Alice is honest, Bob is malicious. 
    \begin{itemize}
        \item If Alice outputs $1$ (i.e., ``Bob wins''), Bob gets $\Lambda + 1$ points.
        \item If Alice outputs $0$ (i.e., ``Alice wins''), Bob gets $\Lambda$ points.
        \item If Alice outputs $\bot$ (i.e., ``cheating detected''), Bob gets $0$ points.
    \end{itemize}
    \item Bob is honest, Alice is malicious.
    \begin{itemize}
        \item If Bob outputs $0$ (i.e., ``Alice wins''), Alice gets $\Lambda+1$ points.
        \item If Bob outputs $1$ (i.e., ``Bob wins''), Alice gets $\Lambda$ points.
        \item If Bob outputs $\bot$ (i.e., ``cheating detected''), Alice gets $0$ points.
    \end{itemize}
\end{itemize} 

Denoting by $R^*_A$ (resp. $R^*_B$) the greatest average scores a malicious Alice (resp. Bob) gets against an honest Bob (resp. Alice), we can rewrite the bias as 
\begin{eqnarray}
    \epsilon &= &\max\{R^*_A,R^*_B\} - \Lambda - \frac{1}{2} \; \in \; [0, 1/2].
\end{eqnarray}

Just as one can cast $P^*_A$ and $P^*_B$ from (non-cheat-penalised) WCF as a semidefinite program (SDP), one can also cast $R^*_A$ and $R^*_B$ as an SDP. Having non-negative reward helps not only because one can still work with positive semidefinite matrices easily, but also because the way the ``cheat detection projectors'' $\{E_i,\mathbb{I} - E_i\}$ appear in the SDP remains largely unchanged---in the (non-cheat-penalised) setting, these projectors project onto a subspace that never contributes to the support of the objective being maximised, and with this convention, it continues to be the case. Henceforth, we stick with the $\Lambda$-translated scoring convention to compute the rewards.

\subsection{The point game formalism}
\label{subsec:point}

We now return to (non-cheat-penalised) WCF and describe the point game formalism Kitaev and Mochon~\cite{Mochon07} introduced to build secure protocols, along with its generalisation to the cheat-penalised setting. We begin by presenting time-dependent point games and then simplify them further to time-independent point games. 

To appreciate the origin of point games, it is instructive to recall the connection between SDPs and WCF. As we already mentioned, the cheating probabilities $P^*_A$ and $P^*_B$ for Alice and Bob can each be cast as an SDP. Using SDP duality, one can establish upper bounds on $P^*_A$ and $P^*_B$ by finding a feasible solution to the dual SDP. The key insight of Kitaev and Mochon was to combine the spectra of these dual feasible solutions with the honest state of the protocol, and distil them into ``configurations of points on a plane'' subject to specific rules that correspond to the constraints of the dual SDP.
Given the structure of this dual SDP (as imposed by WCF), one can go beyond analysing a specific protocol and consider such constraints more generically. Viewed this way, they can be shown to have an alternate characterisation in terms of operator monotone functions. Since operator monotone functions have been studied separately and admit a simple characterisation themselves, one can use this characterisation to simplify the constraints on ``configurations'' into what are called ``validity conditions''. 
These sequences of ``configurations'' are what we have been calling \emph{time-dependent point games}. Since these ``configurations'' still correspond to the protocol, there are as many of them as there are interactions in the protocol---and hence the adjective ``time dependent''. These are then simplified further by, abstractly, merging all of Alice's steps into one step, and Bob's steps into another step to obtain a much simpler object called a \emph{time-independent point game} that is specified using only two ``moves'' (captures the transition from one configuration to the next). Going from protocols to time-dependent point games and finally time-independent point games is relatively straightforward---in hindsight---but going from time-independent point games all the way back to protocol, is highly non-trivial~\cite{Mochon07,ACG+14}. %

\subsubsection{Time-Dependent Point Games}
\begin{figure}
\begin{center}
\includegraphics[scale=0.63]{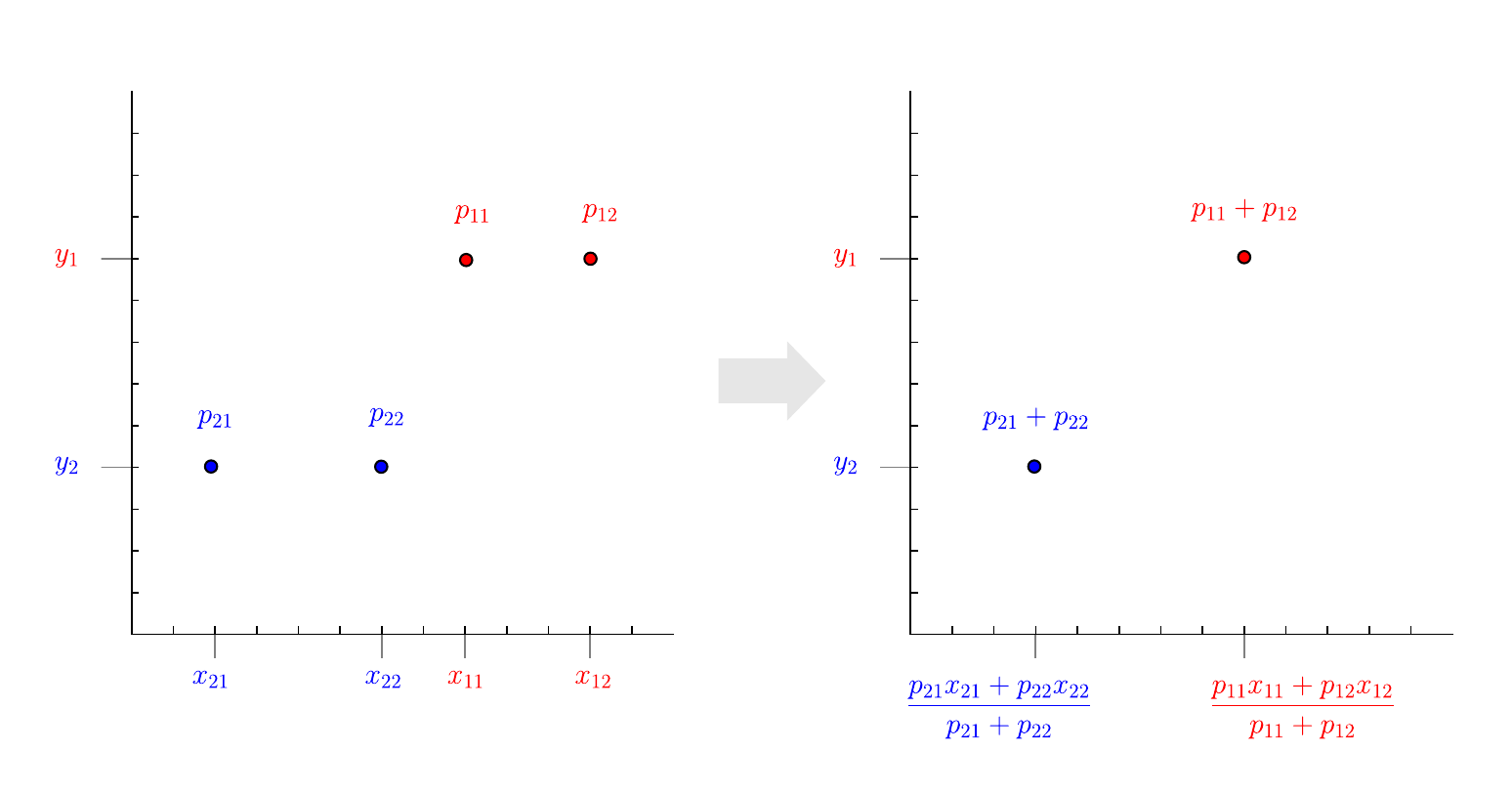}
\end{center}
\caption{An example of a horizontally valid
transition. The red points along $y=y_1$ are \emph{merged} into single point at the average $x$-coordinate of the initial two red points. Similarly, blue points along $y=y_2$ are \emph{merged} into a single point at the average $x$-coordinate of the two initial blue points. Such \emph{merge} operations are known to always satisfy the validity conditions in Equations~\ref{valid1} and~\ref{valid2}. %
}  
\label{fig:hvalid}
\end{figure}

Fortunately, the description of these point games is straightforward and that suffices for our purposes here. %
We start by defining the term \emph{one-dimensional configuration} (later we use \emph{frame} and configuration interchangeably) to mean a function $f \colon \mathbb{R}_{\geq 0} \to \mathbb{R}_{\geq 0}$ that has finite support (i.e., $f(x) = 0$ for all but a finite number of values of $x$). Given two such configurations $f_1$ and $f_2$, we say that the transition $f_1 \to f_2$ is \emph{valid} if:
\begin{eqnarray}
    \label{valid1}
    \sum_x f_1(x) = \sum_x f_2(x)
\end{eqnarray}
and for all $\lambda > 0$,
\begin{eqnarray}
    \label{valid2}
    \sum_x f_1(x)\left( \frac{x \lambda}{x + \lambda} \right) \leq \sum_x f_2(x)\left( \frac{x \lambda}{x + \lambda} \right).
\end{eqnarray}
A \emph{two-dimensional configuration} is a function $p \colon \mathbb{R}_{\geq 0} \times \mathbb{R}_{\geq 0} \to \mathbb{R}_{\geq 0}$ (i.e., a non-negative function on the upper right-hand quadrant of a Cartesian coordinate system) that has finite support.  (If we use the word ``configuration'' or ``frame'' by itself, we mean a two-dimensional configuration.) A transition $p_1 \to p_2$ is \emph{horizontally valid} if it is valid on all horizontal lines in the Cartesian coordinate system, and it is \emph{vertically valid} if it is valid on all vertical lines in the Cartesian coordinate system.  An example of a horizontally valid move is shown in Figure~\ref{fig:hvalid}. 

We continue using the $\points{x,y}$ notation: for any pair of non-negative real numbers $(x,y)$, we denote by $\llbracket x, y \rrbracket$ the function from $\mathbb{R}_{\geq 0} \times \mathbb{R}_{\geq 0}$ to $\mathbb{R}_{\geq 0}$ that maps $(x,y)$ to $1$ and all other points to $0$. With this, we define the start-configuration as $s:=\frac{1}{2}\points{0,1} + \frac{1}{2}\points{1,0}$ and the end-configuration as $e:=\points{\beta,\alpha}$. Now, a (valid) \emph{time-dependent point game} (\textbf{TDPG}) with final point $(\beta,\alpha)$ is a sequence of two-dimensional configurations $(p_0:=s,p_1,p_2,\dots, p_{n-1},p_n:=e)$ such that $p_i \to p_{i+1}$ is horizontally valid for all even $i$ and vertically valid for all odd $i$. 

In~\cite{Mochon07}, it was proved that time-dependent point games are equivalent to weak coin-flipping protocols in the following sense.

\Needspace{6\baselineskip} 
\begin{thm}[WCF protocol--TDPG equivalence (informal;~\cite{Mochon07})]The following holds.
    \begin{description}
        \item[\textbf{WCF $\implies$ TDPG.}] Given any $\delta>0$, and a weak coin-flipping protocol with cheating probabilities $P^*_A$ and $P^*_B$, using $n$ rounds of interaction, there exists a time-dependent point game $(p_0,\dots, p_n)$ with final point $\beta,\alpha$ where $\beta = P^*_B +\delta$ and $\alpha = P^*_A + \delta$.
        \item[\textbf{WCF $\impliedby$ TDPG.}] Given any $\delta>0$ and any time-dependent point game $(p_0,\dots, p_n)$ with final point $(\beta,\alpha)$, there exists a weak coin-flipping protocol with cheating probabilities $P^*_A = \alpha + \delta$ and $P^*_B = \beta + \delta$.
    \end{description}
\end{thm}

These were established in~\cite{Mochon07,ACG+14} and one of our main technical contributions is to adapt to these to the cheat-penalised settings. We make a few small remarks about these results since our focus here is efficiency. 
\begin{itemize}
    \item[-] 
    We emphasise that the ``TDPG $\implies$ WCF'' in these works relied on a \emph{non-constructive} step. As a consequence, for a given a TDPG, it was known that a protocol exists but there was no way to find the unitaries corresponding to this protocol---besides doing a brute-force search. This was later remedied in~\cite{Arora2019} who gave a general numerical algorithm, they called the EMA algorithm, for finding the required unitaries corresponding to any valid transition. We mention this here because below, when we consider the cheat-penalised setting, the start-configuration and end-configurations change. However, the notion of valid transitions stays unchanged and therefore one can still use the EMA algorithm to construct \emph{explicit unitaries} for these protocols in the cheat-penalised setting. 
    \item[-] The original proofs in~\cite{Mochon07,ACG+14} produced protocols with round-complexity $n$ (the length of the TDPG) where the message register had to kept coherent throughout the execution of the protocol. The proof in~\cite{Arora2019} produced protocols where the message register can be discarded after every interaction---however, the round complexity of this protocol was far from optimal. We omit the details but briefly, one could only make transitions line by line (and not ``frame by frame'') leading to inefficiencies. When we consider the cheat-penalised setting below, we show how to get the best of both approaches.\footnote{In fact, this improvement also applies to (non-cheat-penalised) weak coin-flipping.}
    \item[-] In~\cite{Arora2019} it was also noted that one can have $\delta=0$ in the ``TDGP $\implies$ WCF'' step by using projectors after the unitaries (instead of before, as in~\cite{ACG+14}) appropriately. We therefore use that convention here. In the cheat-penalised, with the translated scoring system, we are able to preserve the benefits of placing the projectors after. 
\end{itemize}

Fix any $\Lambda\ge 0$. A \emph{$\Lambda$-penalty time-dependent point game}, denoted by \textbf{$\Lambda$-penTDPG}, with final point $(\beta,\alpha)$ is a sequence of configurations $(p_0,\dots, p_n)$ such that $p_0 = \frac{1}{2} \points{\Lambda,\Lambda+1} + \frac{1}{2} \points{\Lambda+1,\Lambda}$, $p_n=\points{\beta,\alpha}$ and the transitions $p_i \to p_{i+1}$ are valid (exactly as in the (non-cheat-penalised) WCF setting; see \Defref{LambdaPointGameWithValidFunctions}). Skipping the ``easy direction''\footnote{In fact, we realised that one of the steps in~\cite{ACG+14} that is used to establish this ``easy direction'' for (non-cheat-penalised) WCF---specifically, the proof of Proposition 9---contains an error. The statement, fortunately, is still correct. Very briefly, the issue is that they implicitly assume that the spectrum of a matrix remains unchanged under certain kinds of projections. This is easily fixed by deferring all projectors to the end. See~\Propref{penWCF_implies_penEBM} which also holds for $\Lambda=0$.} for brevity, below we state a stronger variant of \Thmref{inf_TDPGequivTIPG} that we prove holds in the cheat-penalised setting. In fact, this result also applies to (non-cheat-penalised) WCF when $\Lambda=0$.

\begin{theorembox}[$\Lambda$-penTDPG $\implies$ $\Lambda$-penWCF protocol (\Thmref{LambdaPenTDPG-implies-Lambda-penWCF} simplified)]
Fix any $\Lambda\ge0$ (also works for $\Lambda=0$). Let $(p_{0},\dots, p_{n})$ be a $\Lambda$-penalty Time-Dependent Point Game that has $(\beta,\alpha)$
as the final point. Then, there exists a cheat-penalised WCF protocol with penalty $\Lambda$
whose ``cheating rewards'' are bounded as $R_{A}^{*}\le\alpha$ and $R_{B}^{*}\le\beta$. Furthermore, it uses $2n$ rounds of communication and $3\cdot\log\lceil\left(\max_{j}\left(2\mu_j\right)+1\right)\rceil$
qubits where $\mu_j= |\supp(p_j) |$ is the number of points with non-zero weight in the configuration $p_j$.
\end{theorembox}

To prove the theorem, we overcome the aforementioned challenges in \Subsecref{penEBMtopenWCF} and \Subsecref{penTDPGimpliespenWCF}. In fact, we show the equivalence by going through an intermediate object called ``Expressible-by-Matrices'' (EBM) point games. We defer these details to \Secref{Lambda-EBM}. 

\begin{figure}
\hspace{-0.5cm}\includegraphics[scale=0.6]{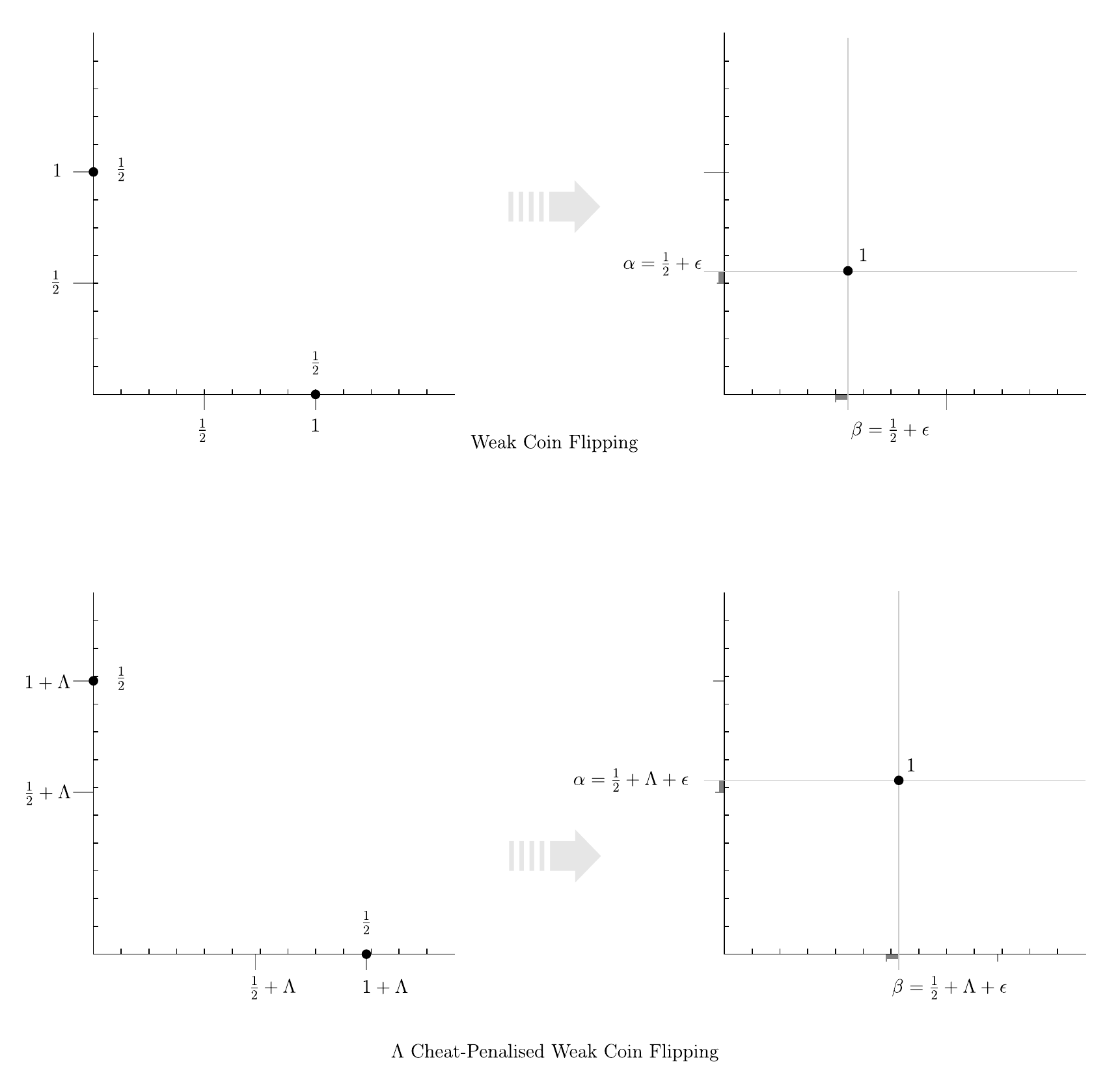}
\caption{The point game problems corresponding to ordinary weak coin-flipping (top) and $\Lambda$-penalty weak coin-flipping (bottom).  The points in the initial configurations each have weight $1/2$, and the points in the final configurations each have weight $1$.}  
\label{fig:targets}
\end{figure}

\subsubsection{Time-Independent Point Games}

Moving on, as we already suggested, one can simplify time-dependent point games even further by considering time-independent point games.  %
To this end, we first consider the (non-cheat-penalised) WCF setting and introduce some notation. Let us say that a \emph{one-dimensional move} is a (not necessarily non-negative) function from $\mathbb{R}_{\geq 0}$ to $\mathbb{R}$ with finite support, and that a \emph{two-dimensional move} is a function from $\mathbb{R}_{\geq 0}^2$ to $\mathbb{R}$ with finite support.  A one-dimensional move $f$ is \emph{valid} if
\begin{eqnarray}
\sum_{x \in \mathbb{R}_{\geq 0}} f(x) & = & 0     
\end{eqnarray}
and 
\begin{eqnarray}
\sum_{x \in \mathbb{R}_{\geq 0}} f(x)
\left( \frac{x \lambda }{x + \lambda} \right) & \geq & 0 \quad \forall \lambda \ge 0.   
\end{eqnarray}
A two-dimensional move is horizontally valid if its rows are valid, and vertically valid if its columns are valid. A \emph{time-independent point game} (\textbf{TIPG}) with final point $(\beta,\alpha)$ is a pair $(h,v)$ of two-dimensional moves such that $h$ is horizontally valid and $v$ is vertically valid such that $h+v=e-s$ (where recall $s:=1/2\points{0,1} + 1/2\points{1,0}$ and $e=\points{\beta,\alpha}$). 

It is straightforward to observe that given a time-dependent point game $(p_0, p_1, \ldots, p_n)$ from $s$ to $e$, one can construct a time-independent point game from $s$ to $e$ as
\begin{eqnarray}
h & := & - p_0 + p_1 - p_2 + p_3 + \ldots + p_{n-1} \\
v & := & - p_1 + p_2 - p_3 + p_4 - \ldots + p_n.
\end{eqnarray}
The converse also holds~\cite{Mochon07,ACG+14} although establishing it takes much more work---and it is closely related to the round complexity. 

\begin{thm}[TDPG--TIPG equivalence (informal~\cite{Mochon07,ACG+14})] The following holds.\label{thm:inf_TDPGequivTIPG}
    \begin{description}
        \item[\textbf{TDPG $\implies$ TIPG.}] Given a time-dependent point game (TDPG) with final point $(\beta,\alpha)$ there is a time-independent point game (TIPG) with final point $(\beta,\alpha)$.
        \item[\textbf{TDPG $\impliedby$ TIPG.}] For every $\delta>0$, given a time-independent point game (TIPG) with final point $(\beta,\alpha)$ there is a time-dependent point game TDPG $(p_0,\dots, p_n)$ with final point $(\beta+\delta,\alpha,+\delta)$ where $n$ grows as $\delta$ shrinks.
    \end{description}
\end{thm}

At a high level, the key difficulty in converting a TIPG into a TDPG is that there is no way to ``start the TIPG''. Specifically, the starting configuration for a TDPG is always $s$. However, to use the move $h$ as a transition in the TDPG, one must ensure that there are points at $h^-$ already present in the previous configuration. Here $h^-$ is the negative part of $h$ and encodes all the points with negative weight. To remedy this, Mochon introduces the notion of a \emph{catalyst state}. The idea is to take some small amount of weight from $s$ and deposit it as per $h^-$. This process also creates a few extra points with high coordinates. However, since the total weight of these points is small, it does not influence the final point of the TDPG too much. With a scaled down version of $h^-$ present, one can apply a scaled down version of the $h$ move, followed by the $v$, repeatedly until most of the weight has accumulated at the point $(\beta,\alpha)$. Finally, all remaining points are ``absorbed'' into this $(\beta,\alpha$ point to obtain the final point $(\beta+\delta,\alpha+\delta)$ where $\delta$ depends on a variety of parameters. Intuitively, smaller the $\delta$, smaller the weight of the catalyst state---and thus greater the number of transitions in the TDPG. 

We briefly remark on some of the challenges that arise when one introduces cheat penalty and considers cheat-penalised TDPGs (penTDPGs) and TIPGs (penTIPGs) that we describe shortly. %

\begin{itemize}
    \item[-] 
    For the cheat-penalised version, like we explained in the introduction, we need to allow the TIPG to be approximate---otherwise we cannot use results from our numerical algorithm (explained in \Subsecref{algForGoodTIPG} below). This means that not only does one need to prepare the catalyst, but one also needs to deposit weight on the points that arise because of this approximation. While the weight on the catalyst in the non-cheat-penalised setting is essentially a free parameter, it is unclear if this is still the case in view of the approximation error. 
    \item[-] 
    In (non-cheat-penalised) TDPGs and TIPGs, no point has either coordinate below $0$. In the cheat-penalised setting, everything gets translated by $(\Lambda,\Lambda)$. One might therefore suspect that in this setting, no point in the point game can have either coordinate below $\Lambda$. However, this is not the case---and as we explain later, numerical evidence suggests that being able to use points with coordinates below $\Lambda$ seems to be crucial to obtain improvements. Why is this pertinent to the connection between cheat-penalised TDPGs and TIPGs? It turns out that Mochon's original construction for catalyst state, when applied to the cheat-penalised setting, cannot handle points that have coordinates lower than $\Lambda$. We use a slightly different route to circumvent this issue (but assume $\Lambda>0$ to simplify the analysis slightly; note that $\Lambda=0$ works for the previous results).
    \item[-]
    Finally, the analysis in~\cite{Mochon07} is not concerned with resource usage while the analysis in~\cite{ACG+14} is only considers asymptotic dependence of the round complexity on the bias. This is not enough to compute concrete numbers for specific protocols. We need to compute explicit formulae relating the bias and round complexity, in terms of various parameters (including approximation error)---while preserving properties that control the space-complexity. %
\end{itemize}

Cumulatively, when deriving the analogous result (i.e., cheat-penalised TIPG $\implies$ cheat-penalised TDPG), %
we need to be much more careful about our parametrisation and even so, we find that $\delta$ cannot be arbitrarily small. This, in turn, means that the approximation factor limits the bias, regardless of how many rounds of communication are allowed. 

In the technical overview, we have not yet detailed penTIPGs. Unsurprisingly---at least at this point---a \emph{$\Lambda$-penalty, time-independent point game with approximation error $\approxError$}, or more briefly, \textbf{$(\Lambda,\approxError)$-penTIPG} is a pair of moves $(h,v)$ where $h$ is horizontally valid, $v$ is vertically valid and they satisfy $\norm{h+v - (e-s)}_1\le \approxError$ for $e:=\points{\beta,\alpha}$, and $s:=1/2\points{\Lambda+1,\Lambda} + 1/2\points{\Lambda,\Lambda+1}$. %
We show that these approximate time-independent point games can also be converted into time-dependent point games with a slightly increased final coordinate. This involves various parameters so we first state the result and specify the parameters afterwards.

\begin{theorembox}[$\Lambda,\approxError$-pen TIPG $\implies$ $\Lambda$-pen
TDPG (\Thmref{LambdaPenTIPGtoLambdaPenTDPG} reformulated)]
    Fix any $\Lambda>0$ (cannot have $\Lambda=0$). 
    Given a symmetric $(\Lambda,\approxError)$-penTIPG $(h,v)$, i.e.,  
            $h=v^{T}$, with
            final point $(\beta,\alpha)$,
            choose any \[c\in\Big(0,\frac{m_{\mathsf{min}}^{2}}{(\Lambda+1)\cdot\Lambda}\Big)\] where $m_{\mathsf{min}}\ge 0$ is the minimum coordinate in $(h,v)$. The choice of $c$ specifies a $\delta_{\min}<1$.
    Then, for every $\delta\in(\delta_{\min},1)$ there is a $\Lambda$-penalty
TDPG $(p_{0},\dots, p_{n})$ with final point $(\beta+\mathsf{err},\alpha+\mathsf{err})$
where \[ \mathsf{err}=\sqrt{\delta\cdot(m_{\mathsf{max}}-\alpha)(m_{\mathsf{max}}-\beta)},\]
$n=10+2/\eta_{\mathsf{clyst}}$. Here $m_{\mathsf{max}}$ is closely related with the maximum coordinate in $(h,v)$ and the catalyst state and $\eta_{clyst}$ is proportional to the weight assigned to the catalyst state (controlled by $\delta$) but inversely proportional to $\norm{h^-}$. Explicit formulae for these are listed below. The number of points in $p_{i}$ is at most the number of points in $(h,v)$, i.e., $|\supp(p_{i})|\le |\supp\{ h \} \cup \supp\{ v \} |$ for all $i\in \{0,\dots n\}$.
\end{theorembox}

To describe the parameters exactly, we observe that there exist $\varepsilon_{1},\varepsilon_{2}\le\approxError$ such that
negative part of $h+v$ equals $\left(1-\varepsilon_{1}\right)s+\varepsilon_{1}\cdot\serror$
and the positive part of $h+v$ equals $(1-\varepsilon_{2})e+\varepsilon_{2}\cdot\eerror$ where
$\left\Vert \serror\right\Vert _{1},\left\Vert \eerror\right\Vert _{1}=1$. Recall that $c$ was already fixed above (in the theorem). Using these, we have:
\begin{itemize}
\item $\delta_{\min}:=(1-\varepsilon_{2})\cdot\frac{c'\varepsilon_{1}}{1+c'\varepsilon_{1}}+\varepsilon_{2}$,
with $c':=c^{-1}-1$,
\item $m_{\mathsf{max}}:=\max\left\{ \maxcoordinate(h),\tilde{m}_{\mathsf{max}}\right\} $ with 
\item $\tilde{m}_{\mathsf{max}}:=\min\left\{ \left| (1-w)\left(\frac{1}{(\Lambda+1)}-\frac{w}{m_{\mathsf{min}}}\right)^{-1} \right|:w\in\{w^{\pm}\}\right\} $
for 
\[
w^{\pm}=\frac{\sqrt{8c\Lambda^{2}(\Lambda+1)^{2}+m_{\mathsf{min}}^{2}(8c\Lambda(\Lambda+1)+1)-8c\Lambda\left(2\Lambda^{2}+3\Lambda+1\right)m_{\mathsf{min}}}\pm m_{\mathsf{min}}}{2(\Lambda+1)(m_{\mathsf{min}}-\Lambda)},
\]
and finally,
\item $\eta_{\mathsf{clyst}}:=\frac{\deltacatalyst}{\left\Vert h^{-}\right\Vert }\cdot\frac{c(1-\varepsilon_{1})+\varepsilon_{1}}{(1-\deltacatalyst)}$
with $\deltacatalyst=1-\left((1-\varepsilon_{1})+\varepsilon_{1}/c\right)\cdot\frac{1-\delta}{1-\varepsilon_{2}}.$
\end{itemize}
\vspace{1em}

In \Secref{Lambda-TIPG}, we prove the theorem by overcoming the challenges listed above it. We note that our result relating time-independent point games and time-dependent point games (in the cheat-penalised setting) assumes the point game is symmetric (i.e., $h=v^T$) and $\Lambda>0$ because otherwise some of the analysis gets more involved. We leave the relaxation of these assumptions to future work.

\subsection{Good TIPGs and how to find them}\label{subsec:algForGoodTIPG}
    Even using all the point game machinery discussed so far, and its extension to the cheat-penalised setting, it is unclear how to design good point games. Any point game for weak coin-flipping can be immediately lifted to the cheat-penalised setting---simply because the constraints on ``valid move'' get weakened as one moves farther away from the origin. Indeed, this was noted by Mochon~\cite{Mochon07} and he gave a heuristic calculation for how it might apply to his Dip-Dip-Boom protocol. This protocol for (non-cheat-penalised) weak coin-flipping has bias $1/6$. Mochon's informal calculations suggest that in the cheat-penalised setting, its bias vanishes as the cheating penalty grows. 
    
    \subsubsection{Analytic TIPGs}
    To find analytic TIPGS, we start by looking at the Spekkens-Rudolph protocol~\cite{SR02} (the simplest instance of Dip-Dip-Boom) in the cheat-penalised setting in Section~\ref{sec:SR-protocol}. We then look at the Dip-Dip-Boom protocol. We provide a careful derivation of the fact that the bias vanishes as the cheating penalty grows in Section~\ref{sec:DipBoom}. While Mochon's analysis established vanishing bias to order $O(1/\Lambda)$, we extend this result by deriving the higher order terms as well. 
    Besides Mochon's analysis, we also look at cheat-penalised versions of other protocols in the literature. 
    While a promising start, it was unclear how or whether one can build on these to have efficient protocols with low bias and small penalty. For instance, Mochon's family of TIPGs that approach zero bias---of which Dip-Dip-Boom is the simplest instance---are insensitive to the penalty parameter. More concretely, the polynomial-based technique that Mochon uses to assign weights to the points in his TIPGs, depends only on the relative positions of the points under consideration---and therefore remain unchanged under translations (i.e., the cheating penalty parameter).
    On the other hand, doing a brute-force numerical search quickly becomes untenable as we explain below. Numerical searches were attempted already for (non-cheat-penalised) weak coin-flipping~\cite{NST14} with no luck until the question was resolved via other techniques~\cite{Mochon07,Arora2019,Arora2019b}. %

    \subsubsection{Numerical Algorithm for finding TIPGs}
    To overcome these challenges, we introduce a numerical algorithm that harnesses the structure underlying point games and works generically to solve \textbf{(time-independent) point game problems}---given a start-configuration $s$ and an end-configuration $e$, find horizontally valid and vertically valid moves $(h,v)$ such that $$h+v=e-s.$$ 

    We begin with the concept of a bivariate \emph{profile function} from \cite{miller2020impossibility}.  For any two-dimensional move $p$, we define an associated profile function $\hat{p} \colon \mathbb{R}^2 \to \mathbb{R}$. Deferring a formal definition to \Subsecref{setup}, informally, for $\lambda,\gamma >0$, we have
    \begin{eqnarray}
        \hat{p} ( \lambda, \gamma ) & := & \sum_{x,y \geq 0} \left( \frac{\lambda x}{\lambda + x} \right) \left(\frac{\gamma y}{\gamma + y } \right) p(x,y).
    \end{eqnarray}
    The definition is essentially a restatement of the conditions used to define valid moves (see \Subsecref{point}). Any horizontally or vertically valid function must have a non-negative profile function. As a consequence, a point game problem $(s,e)$ %
    can only be solved if $\hat{e} \geq \hat{s}$. 

    In \cite{miller2020impossibility}, the profile function was used to prove the impossibility of constructing weak coin-flipping protocols with low communication complexity.  In Section~\ref{sec:alg}, we show how the same concept can be used to \emph{find} good point games. %
    Below, we provide a four-step description of the algorithm, together with the intuition behind it. %

    \paragraph{1. Discretisation} Recall that we want to find a pair $(h,v)$ that solves the \emph{point game problem} for a given $s$ and $e$. A brute-force search will have to look through the set of all possible pairs $(h,v)$ which is an infinite-dimensional space that is described by an infinite number of linear constraints. Concretely, $h$ and $v$ must satisfy the validity conditions from Equations \ref{valid1} and \ref{valid2}. Instead, we can choose a discretisation of the plane for the coordinates $x$ and $y$, together with a maximum and a minimum coordinate. This defines a finite set $S\subseteq \mathbb{R}_{\geq 0}$ which includes the coordinates appearing in the start-configuration $s$ and end-configuration $e$.
    Since our search will require the use of the profile functions of $h$ and $v$, we also discretise the parameter space for the profile functions, obtaining a finite set $T\subseteq \mathbb{R}_{\geq 0}$. Finally, we fix a threshold parameter $\delta \in \mathbb{R}_{>0}$ that will be used in the next step. In summary, the appropriate choice of the sets $S$ and $T$ together with the parameter $\delta$ is the first step of the algorithm. %

    \paragraph{2. Profile Matching} %
    We search for a move $h$ that is horizontally valid (when restricted to $T$), such that the Euclidean norm 
    $$\norm{(\hat{h}+\hat{v})-(\hat{e}-\hat{s})}$$
    is small and $v=h^\top$. Since $T$ is finite, the profile functions $\hat{h}, \hat{v}, \hat{e}$, and $\hat{s}$ can all be represented as finite-dimensional vectors. %

    We restrict the search for $(h,v)$ to a smaller subspace $V$ defined as follows. Consider the matrix $M$ mapping moves with coordinates in $S$ to profiles with coordinates in $T$. Let $V$ be the span of right singular vectors of $M$ with singular values greater than the threshold parameter $\delta$ (recall $\delta$ was fixed in the previous step). Restricting the search to $V$ ensures that we %
    only retain those components of a move that contribute the most to their profile function.
    This not only reduces the complexity of the search but also, crucially, keeps the norm of $(h,v)$ small.

    \paragraph{3. Configuration Matching} 
    Denote by $(h,v)$ the pair of moves produced by the previous step. This is such that the profile $\hat{h}+\hat{v}$ is close to $\hat{e}-\hat{s}$. That, in turn, means that $(e-s) - (h+v) = t$ has a small profile. %
    It turns out that this fact helps with constructing approximately valid moves. 
        \begin{defn}
        A function $g:\mathbb{R}_{\geq 0}\to \mathbb{R}$ is \emph{$\eta$-approximately valid} with $\eta>0$, if
        \begin{align}
            \sum_x g(x) + \eta &\geq 0,\\
            \sum_x g(x) \left(\frac{\lambda x}{x+\lambda} \right) + \eta & \geq 0.
        \end{align}
        The definition extends naturally to approximately horizontally and vertically valid  moves.
        \end{defn}

        Our goal in this step, is to find moves $p$ and $q$ such that $(h+p)+ (v+q) = e-s$ and $p$ and $q$  are approximately horizontally and vertically valid, respectively. 
        To find such a pair $(p,q)$, we decompose $t$ into two parts where one part is approximately horizontally valid while the other is approximately vertically valid. It is not too hard to see that such a decomposition is guaranteed because $t$ has a small profile. In fact, given that $h=v^{\top}$, one can also ensure that $p=q^{\top}$. Finally, while adding $p$ and $q$ increases the norm of the moves, this is unavoidable if one wants to match $e-s$ exactly.

    \paragraph{4. Validity Enforcement} 
    So far, we have approximately valid functions $(h',v')$ where $h'=h+p$ and $v'=v+q$ such that $h'+v'=e-s$. In this final step, we want the opposite trade-off: we want $h'+v'$ to be approximately equal to $e-s$ while ensuring $h'$ and $v'$ are exactly valid. 
    To achieve this, we project $h'$ onto the closest (exactly) valid vector $h_*$. This turns out to be a quadratic program that can be solved efficiently. Since $h_*$ was close to $h'$ (and $v'=h^{\prime \top}$), it follows that $h_* + v_*$ is close to $e-s$ as desired.

    \vspace{2em}
    The algorithm above results in approximate penTIPGs and in the previous section we saw how to map precisely such penTIPGs into cheat-penalised WCF protocols.
    By running this algorithm numerically, we find several TIPGS with good parameters. We describe two of these in Theorem~\ref{thm:final-TIPGs} below. These penTIPGs were used to construct the protocols described in the Main Result box. 
    See \Secref{list_TIPGs} for an explicit description of these TIPGs and our GitHub page~\cite{penWCFcode} for a Mathematica implementation of our algorithm.

\begin{theorembox}[Existence of $(\Lambda,\approxError)$-penTIPG (from~\Secref{list_TIPGs})]\label{thm:final-TIPGs}
    There exist $(\Lambda,\approxError)$-penTIPGs with starting point $s=\tfrac12 \points{\Lambda,\Lambda+1}+\tfrac12 \points{\Lambda+1,\Lambda}$, final point $e=\points{\Lambda + \tfrac12 + \epsilon, \Lambda + \tfrac12 + \epsilon}$ and a grid of $64$ points, with the following parameters:
    \begin{itemize}
        \item $\Lambda=1$, $\approxError = 10^{-18}$ and $\epsilon=5\times10^{-3}$. 
        \item $\Lambda=0.01$, $\approxError= 10^{-18}$ and $\epsilon=10^{-14}$. 
    
    \end{itemize} 
\end{theorembox}

\newpage{}

\section{The Primal and Dual Formulation of penWCF}\label{sec:SDP-penWCF}
We start by defining what it means to be a $\Lambda$-penalty weak coin flipping protocol formally. 
\begin{defn}[$\Lambda$-penWCF protocol with bias $\epsilon$]
 \label{def:penWCF}For $n$ even, an $n$-message $\Lambda$-penalty
weak coin-flipping protocol ($\Lambda$-penWCF protocol) between
two parties, Alice and Bob, is described by:
\begin{itemize}
\item three Hilbert spaces ${\cal A},{\cal B}$ corresponding to Alice's
and Bob's private workspaces and a message space ${\cal M}$;
\item an initial product state $\left|\psi_{0}\right\rangle :=\left|\psi_{A,0}\right\rangle \otimes\left|\psi_{M,0}\right\rangle \otimes\left|\psi_{B,0}\right\rangle \in{\cal A}\otimes{\cal M}\otimes{\cal B}$;
\item a set of $n$ unitaries $\{U_{1},\dots, U_{n}\}$ acting on ${\cal A}\otimes{\cal B}\otimes{\cal C}$
with $U_{i}=U_{A,i}\otimes\mathbb{I}_{{\cal B}}$ for odd $i$ and
$U_{i}=\mathbb{I}_{{\cal A}}\otimes U_{B,i}$ for even $i$;
\item a set of honest states $\left\{ \left|\psi_{0}\right\rangle ,\dots,\left|\psi_{n}\right\rangle \right\} $
defined by $\left|\psi_{i}\right\rangle :=U_{i}U_{i-1}\cdots U_{1}\left|\psi_{0}\right\rangle $;
\item a set of $n$ projectors $\{E_{1},\dots, E_{n}\}$ acting on ${\cal A}\otimes{\cal M}\otimes{\cal B}$
with $E_{i}=E_{A,i}\otimes\mathbb{I}_{{\cal B}}$ for $i$ odd, and
$E_{i}=\mathbb{I}_{{\cal A}}\otimes E_{B,i}$ for $i$ even, such
that $E_{i}\left|\psi_{i}\right\rangle =\left|\psi_{i}\right\rangle $;
\item two final POVMs $\left\{ \Pi_{A}^{(0)},\Pi_{A}^{(1)},\Pi_{A}^{(\bot)}\right\} $
acting on ${\cal A}$ and $\{\Pi_{B}^{(0)},\Pi_{B}^{(1)},\Pi_{B}^{(\bot)}\}$
acting on ${\cal B}$ where the superscripts $(0),(1)$ denote outcome
$0,1$, respectively, while the superscript $(\bot)$ denotes the outcome
$\bot$ which denotes ``cheat detection''. 
\end{itemize}
The penWCF protocol proceeds as follows: 
\begin{itemize}
\item In the beginning, Alice holds $\left|\psi_{A,0}\right\rangle \otimes $ $\left|\psi_{M,0}\right\rangle $
and Bob holds $\left|\psi_{B,0}\right\rangle $.
\item For $i=1$ to $n$:
\begin{itemize}
\item If $i$ is odd, Alice measures the incoming state with the POVM $\{E_{i-1},\mathbb{I}-E_{i-1}\}$.
On the first outcome, she applies $U_{i}$ and sends the message qubits
to Bob; on the second outcome, she ends the protocol by outputting
$\bot$ (i.e., Alice declares Bob cheated).
\item If $i$ is even, Bob measures the incoming state with the POVM $\{E_{i-1},\mathbb{I}-E_{i-1}\}$.
On the first outcome, he applies $U_{i}$ and sends the message qubits
to Alice; on the second outcome, he ends the protocol by outputting
$\bot$ (i.e., Bob declares himself to be the winner).
\item Alice and Bob measure their part of the state with the final POVM
and output the outcome of their measurements. 
\end{itemize}
\item Points allocation:
\begin{itemize}
\item If the output is $0$, Alice gets $\Lambda+1$ points, Bob gets $\Lambda$
points. 
\item If the output is $1$, Bob gets $\Lambda+1$ points, Alice gets $\Lambda$
points. 
\item If Alice (resp. Bob) outputs $\bot$, then Bob (resp. Alice) gets
$0$ points while Alice (resp. Bob) gets $\Lambda$ points. 
\end{itemize}
\end{itemize}
We require that a $\Lambda$-penWCF protocol satisfy the following
properties: 
\begin{itemize}
\item Correctness: When both players are honest, Alice and Bob never output
$\bot$ and their outcomes are identical, i.e. $\Pi_{{\cal A}}^{(0)}\otimes\mathbb{I}_{{\cal M}}\otimes\Pi_{{\cal B}}^{(1)}\left|\psi_{n}\right\rangle =\Pi_{{\cal A}}^{(1)}\otimes\mathbb{I}_{{\cal M}}\otimes\Pi_{{\cal B}}^{(1)}\left|\psi_{n}\right\rangle =0$ and for all $x\in\{0,1\}$, $\Pi_{{\cal A}}^{(\perp)}\otimes\mathbb{I}_{{\cal M}}\otimes\Pi_{{\cal B}}^{(x)}\left|\psi_{n}\right\rangle =\Pi_{{\cal A}}^{(x)}\otimes\mathbb{I}_{{\cal M}}\otimes\Pi_{{\cal B}}^{(\perp)}\left|\psi_{n}\right\rangle =0$.
\item Balanced: When both players are honest, they each win with probability
$1/2$:
\[
P_{A}=\norm{\Pi_{{\cal A}}^{(0)}\otimes\mathbb{I}_{{\cal M}}\otimes\Pi_{{\cal B}}^{(0)}\left|\psi_{n}\right\rangle }^{2}=\frac{1}{2}
\]
 and 
\[
P_{B}=\norm{\Pi_{A}^{(1)}\otimes\mathbb{I}_{{\cal M}}\otimes\Pi_{{\cal B}}^{(1)}\left|\psi_{n}\right\rangle }^{2}=\frac{1}{2}.
\]
\item $\epsilon$ bias: 
\begin{itemize}
\item Notation:
\begin{itemize}
\item Denote by $R_{A}^{*}$ (resp. $R_{B}^{*}$) the expected number of
points Alice (resp. Bob) obtains, maximised over all cheating strategies
for Alice (resp. Bob) and call it ``cheating rewards'' for Alice
(resp. Bob).
\item Define $P_{A}^{*}:=R_{A}^{*}-\Lambda$ and similarly $P_{B}^{*}:=R_{B}^{*}-\Lambda$
and call them ``cheating probabilities'' for Alice and Bob respectively.
\end{itemize}
\item We say the protocol has bias $\epsilon$ if $\max\{P_{A}^{*},P_{B}^{*}\}\le\frac{1}{2}+\epsilon$. 
\end{itemize}
\end{itemize}
\end{defn}

\begin{rem}[$P_{A}^{*},P_{B}^{*}$ can be viewed as probabilities]
 From correctness, it follows that $P_{A}^{*},P_{B}^{*}\ge0$ are
both non-negative. Specifically, this is because correctness guarantees
that on an honest execution, $\bot$ never occurs which means that
an honest strategy already yields $\Lambda$ points and that, together
with the definitions of $R_{A}^{*},P_{A}^{*},R_{B}^{*},P_{B}^{*}$,
establishes $P_{A}^{*},P_{B}^{*}\ge0$. From the definition of the
problem, it is also clear that a malicious party can gain at most
$\Lambda+1$ points, thus, $P_{A}^{*},P_{B}^{*}\le1$. Thus, we can
view formally $P_{A}^{*},P_{B}^{*}$ as ``cheating probabilities''. 
\end{rem}

\begin{rem}[$\Lambda$-penWCF without intermediate projectors]
 One can defer the intermediate measurements in \Defref{penWCF}
above to the very end, without changing the security of the protocol.
We add projectors for convenience later. \label{rem:noIntermediateMeasurement} 
\end{rem}

\begin{figure}[H]
\centering
\includegraphics[width=0.4\paperwidth]{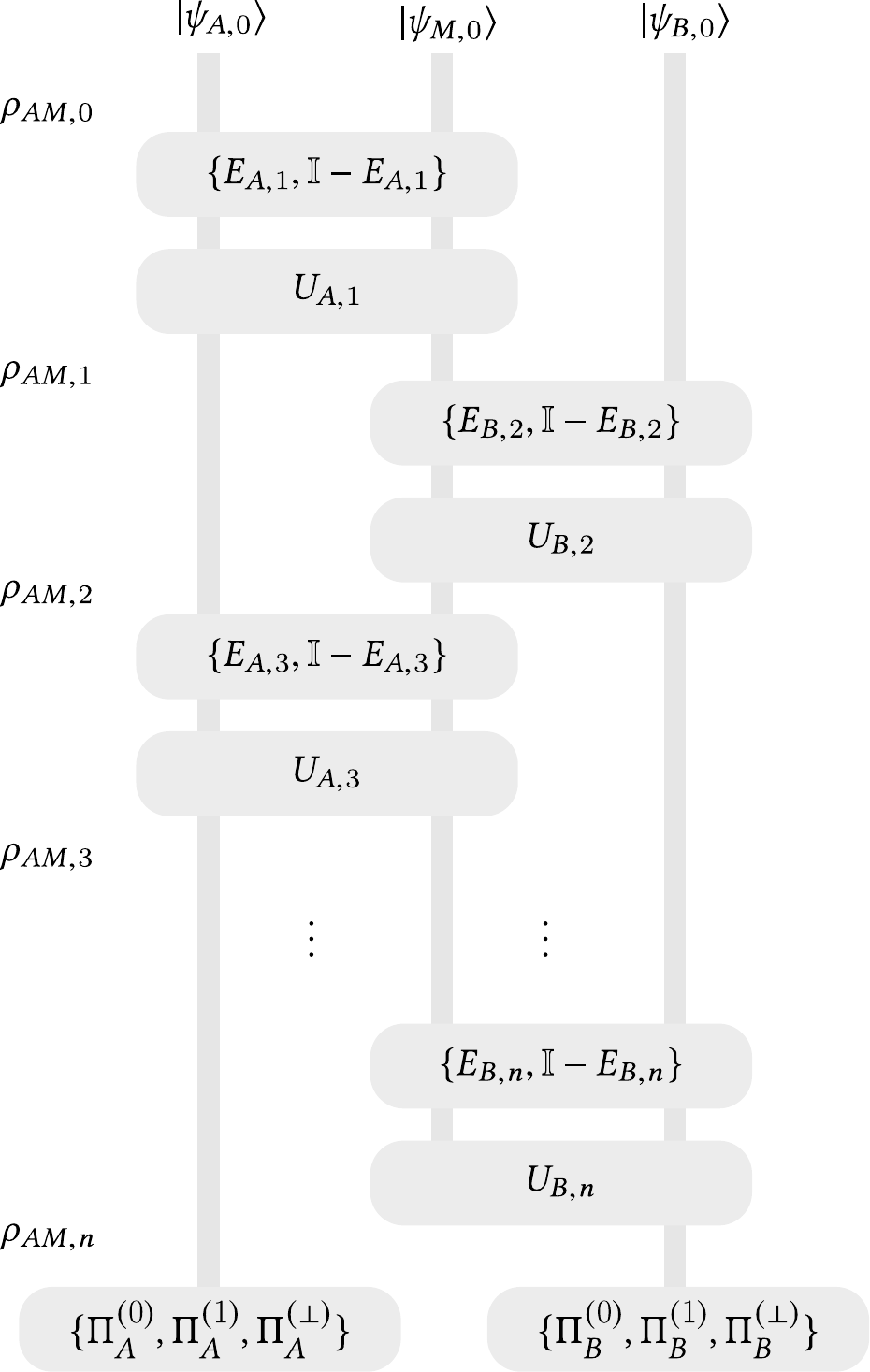}
\caption{Schematic of a cheat-penalised weak coin-flipping protocol.}%
\end{figure}

We now cast $R_{A}^{*}$ and $R_{B}^{*}$ as SDPs. This formulation enables us to relate point games to protocols for \lwcf.

\begin{thm}[Primal SDP]
 It holds that\label{thm:primalSDP}\\
\begin{align*}
R_{B}^{*}=\max\Big[(\Lambda+1)\cdot\tr(\Pi_{A}^{(1)}\otimes\mathbb{I}_{M}\rho_{AM,n}) & +\Lambda\cdot\tr(\Pi_{A}^{(0)}\otimes\mathbb{I}_{M}\rho_{AM,n})\\
 & \cancel{+0\cdot\tr(\Pi_{A}^{(\bot)}\otimes\mathbb{I}_{M}\rho_{AM,n})}\Big]
\end{align*}
 over all $\rho_{AM,i}$ satisfying the constraints
\begin{itemize}
\item $\tr_{M}(\rho_{AM,0})=\tr_{MB}(\left|\psi_{0}\right\rangle \left\langle \psi_{0}\right|)=\left|\psi_{A,0}\right\rangle \left\langle \psi_{A,0}\right|$
\item for $i$ odd, $\tr_{M}(\rho_{AM,i})=\tr_{M}(U_{i} E_{i} \rho_{AM,i-1}E_{i} U_{i}^{\dagger})$
\item for $i$ even, $\tr_{M}(\rho_{AM,i})=\tr_{M}(\rho_{AM,i-1})$.
\end{itemize}
Similarly, it holds that

\begin{align*}
R_{A}^{*}=\max\Big[(\Lambda+1)\cdot\tr(\Pi_{B}^{(0)}\otimes\mathbb{I}_{M}\rho_{BM,n}) & +\Lambda\cdot\tr(\Pi_{B}^{(1)}\otimes\mathbb{I}_{M}\rho_{BM,n})\\
 & \cancel{+0\cdot\tr(\Pi_{B}^{(\perp)}\otimes\mathbb{I}_{M}\rho_{BM,n})}\Big]
\end{align*}
over all $\rho_{BM,i}$ satisfying the following constraints
\begin{itemize}
\item $\tr_{M}(\rho_{MB,0})=\tr_{AM}(\left|\psi_{0}\right\rangle \left\langle \psi_{0}\right|)=\left|\psi_{B,0}\right\rangle \left\langle \psi_{B,0}\right|$
\item for $i$ even, $\tr_{M}(\rho_{MB,i})=\tr_{M}(U_{i}E_{i}\rho_{MB,i-1}E_{i} U_{i}^{\dagger})$
\item for $i$ odd, $\tr_{M}(\rho_{MB,i})=\tr_{M}(\rho_{MB,i-1})$.
\end{itemize}
\end{thm}

\begin{proof}
This is essentially the same as WCF (as in~\cite{ACG+14}) except that we changed the objective
function to reflect the points a player gains on average (and also, we apply projections before instead of after).
\end{proof}

\begin{rem}
Note here that using the translated game already helps simplify the
notation. For instance, since the outcome corresponding to $\mathbb{I}-E_{i}$
results in $0$ points being awarded to the malicious party, one need not
include the support of $U_{i}\rho_{AM,i-1}U_{i}^{\dagger}$ on $\mathbb{I}-E_{i}$
when considering the primal SDP for Bob, for instance. 
\end{rem}
Next, we consider the dual SDP.
\begin{thm}[Dual SDP]
It holds that \label{thm:dualSDP}\\
$R_{B}^{*}=\min[\tr(Z_{A,0}\left|\psi_{A,0}\right\rangle \left\langle \psi_{A,0}\right|)]$
over all dual variables $Z_{A,i}$ under the following dual constraints
\begin{enumerate}
\item $\forall i$ $Z_{A,i}\ge0$
\item for $i$ odd, $Z_{A,i-1}\otimes\mathbb{I}_{M}\ge E_{A,i-1}U_{A,i}^{\dagger}(Z_{A,i}\otimes\mathbb{I}_{M})U_{A,i}E_{A,i-1}$
\item for $i$ even, $Z_{A,i-1}=Z_{A,i}$
\item $Z_{A,n}=\Lambda\cdot\Pi_{A}^{(0)}+(\Lambda+1)\cdot\Pi_{A}^{(1)}$.
\end{enumerate}
$R_{A}^{*}=\min[\tr(Z_{B,0}\left|\psi_{B,0}\right\rangle \left\langle \psi_{B,0}\right|)]$
over all dual variables $Z_{B,i}$ satisfying the following dual constraints
\begin{enumerate}
\item $\forall i$, $Z_{B,i}\ge0$
\item for $i$ even, $\mathbb{I}_{M}\otimes Z_{B,i-1}\ge E_{B,i-1}U_{B,i}^{\dagger}(\mathbb{I}_{M}\otimes Z_{B,i})U_{B,i}E_{B,i-1}$,
\item for $i$ odd, $Z_{B,i-1}=Z_{B,i}$
\item $Z_{B,n}=\Lambda\cdot\Pi_{B}^{(1)}+(\Lambda+1)\cdot\Pi_{B}^{(0)}$.
\end{enumerate}
The collection of matrices $(Z_{A,0},\dots Z_{A,n})$ and $(Z_{B,0},\dots Z_{B,n})$
satisfying the constraints above, and also the following constraint,
are defined to be \emph{dual feasible points}.\\

5. $\left|\psi_{A,0}\right\rangle $ is an eigenvector of $Z_{A,0}$
with eigenvalue $\beta>0$ and $\left|\psi_{B,0}\right\rangle $ is
an eigenvector of $Z_{B,0}$ with eigenvalue $\alpha>0$.

It holds that 
\begin{equation}
R_{B}^{*}=\inf\alpha \text{ and }R_{A}^{*}=\inf\beta\label{eq:Ralphainf}
\end{equation}
 where the infimum is over all dual feasible points (with $\alpha,\beta$
are as in condition 5 above).
\end{thm}

\begin{proof}
Using the standard weak duality argument, one can check that indeed
the dual is as stated (see for instance Appendix B of Mochon's WCF
paper)---except for condition 5.

We focus on proving \Eqref{Ralphainf}. We show $R_{B}^{*}=\inf\alpha$
(the proof for $R_{A}^{*}=\inf\beta$ is analogous). Suppose we are
given an optimal solution $(Z_{A,i})_{i=1}^{n}$ to the dual SDP satisfying
the first four constraint. This, in particular, entails that $R_{B}^{*}=\left\langle \psi_{A,0}\right|Z_{A,0}\left|\psi_{A,0}\right\rangle $.
If $\left|\psi_{A,0}\right\rangle $ were an eigenvector of $Z_{A,0}$
then there is nothing left to prove. Let us assume the contrary. If
we can construct another operator $Z'_{A,0}$ such that (1) $Z'_{A,0}\ge Z_{A,0}$
and (2) $Z'_{A,0}\left|\psi_{A,0}\right\rangle =(R_{B}^{*}+\epsilon)\left|\psi_{A,0}\right\rangle $
where $\epsilon>0$ can be made arbitrarily small, then we would have
proven \Eqref{Ralphainf}. This is because we could use the matrices
$(Z'_{A,0},Z_{A,1}\dots Z_{A,n})$ as a feasible point to the dual
SDP whose objective $\tr(Z'_{A,0}\left|\psi_{A,0}\right\rangle \left\langle \psi_{A,0}\right|)$
equals $R_{B}^{*}$ in the limit $\epsilon\to0$. 

We drop the subscript $A,0$ henceforth for clarity, viz. we use $(Z,Z',\left|\psi\right\rangle )$
instead of $(Z_{A,0},Z'_{A,0},\left|\psi_{A,0}\right\rangle )$. We
show that for any $\epsilon>0$, there is a $\val\in\mathbb{R}_{\ge 0}$
such that 
\begin{equation}
Z':=\left(\left\langle \psi\right|Z\left|\psi\right\rangle +\epsilon\right)\left|\psi\right\rangle \left\langle \psi\right|+\val(\mathbb{I}-\left|\psi\right\rangle \left\langle \psi\right|)\label{eq:ZprimeForm}
\end{equation}
 satisfies the aforementioned requirements (1) and (2). In fact, requirement
(2) is satisfied by construction. We impose requirement (1) to obtain
$\val$ as follows: we require that for all normalised vectors $\left|\phi\right\rangle \in{\cal A}$,
it holds that 
\[
\left\langle \phi\right|\left(Z'-Z\right)\left|\phi\right\rangle =\left[\left(R_{B}^{*}+\epsilon\right)\left|\left\langle \phi|\psi\right\rangle \right|^{2}+\val\cdot\left(1-\left|\left\langle \phi|\psi\right\rangle \right|^{2}\right)\right]-\left\langle \phi\right|Z\left|\phi\right\rangle \ge0.
\]
One can write $\left|\phi\right\rangle =a\left|\psi\right\rangle +\tilde{a}\left|\psi^{\perp}\right\rangle $
where $a'$ is real (the phase can be absorbed in $\left|\psi^{\perp}\right\rangle $)
satisfying $|a|^{2}+\tilde{a}^{2}=1$ and $\left\langle \psi|\psi^{\perp}\right\rangle =0$.
Using this, one has 
\[
\left\langle \phi\right|Z\left|\phi\right\rangle =|a|^{2}\cancelto{R_{B}^{*}}{\left\langle \psi\right|Z\left|\psi\right\rangle }+\tilde{a}^{2}\left\langle \psi^{\perp}\right|Z\left|\psi^{\perp}\right\rangle +\tilde{a}\left(a\left\langle \psi^{\perp}\right|Z\left|\psi\right\rangle +\hc{}\right).
\]
Substituting this in the previous equation, one obtains
\begin{align*}
\left\langle \phi\right|\left(Z'-Z\right)\left|\phi\right\rangle  & =\left(R_{B}^{*}+\epsilon\right)\left|a\right|^{2}+\val\cdot\cancelto{\tilde{a}^{2}}{(1-|a|^{2})}\\
 & \quad-|a|^{2}R_{B}^{*}-\tilde{a}^{2}\left\langle \psi^{\perp}\right|Z\left|\psi^{\perp}\right\rangle -\tilde{a}\left(a\left\langle \psi^{\perp}\right|Z\left|\psi\right\rangle +\hc{}\right)\\
 & =\tilde{a}^{2}\left(\val-\left\langle \psi^{\perp}\right|Z\left|\psi^{\perp}\right\rangle \right)-\tilde{a}\left(a\left\langle \psi^{\perp}\right|Z\left|\psi\right\rangle +\hc{}\right)+|a|^{2}\epsilon\ge0.
\end{align*}
If $\left|\phi\right\rangle $ is such that $a=0$, then one can simply
pick any $\val\ge\left\Vert Z\right\Vert $ (where $\left\Vert Z\right\Vert $
is the operator norm of $Z$, i.e. the highest eigenvalue of $Z$
in this case). But this may not be enough otherwise. When $a\neq0$,
one can view the expression above as a quadratic in $\tilde{a}$.
We take $\val\ge\left\Vert Z\right\Vert $ so that the coefficient
of $\tilde{a}^{2}$ is positive. Now, if we can ensure that the quadratic
has no roots then the inequality is guaranteed to hold. To this end,
we require that the discriminant is negative. This yields 
\begin{align*}
 &  & \left(a\left\langle \psi^{\perp}\right|Z\left|\psi\right\rangle +\hc{}\right)^{2}-4|a|^{2}\epsilon\left(\val-\left\langle \psi^{\perp}\right|Z\left|\psi^{\perp}\right\rangle \right) & \le0\\
\iff &  & \frac{1}{4|a|^{2}\epsilon}\left(a\left\langle \psi^{\perp}\right|Z\left|\psi\right\rangle +\hc{}\right)^{2}+\left\langle \psi^{\perp}\right|Z\left|\psi^{\perp}\right\rangle  & \le\val.
\end{align*}
Using the fact that $z+\hc=2\mathfrak{R}(z)\le2\left|z\right|$ for
any complex number $z$, we have that 
\[
\frac{1}{4|a|^{2}\epsilon}\left(a\left\langle \psi^{\perp}\right|Z\left|\psi\right\rangle +\hc{}\right)^{2}\le\frac{\cancel{4|a|^{2}}}{\cancel{4|a|^{2}}\epsilon}\left\Vert Z\right\Vert 
\]
which means that it suffices to require that $\val$ satisfies 
\[
\frac{\left\Vert Z\right\Vert }{\epsilon}\le\frac{\left\Vert Z\right\Vert }{\epsilon}+\left\langle \psi^{\perp}\right|Z\left|\psi^{\perp}\right\rangle \le\val
\]
to ensure the determinant is negative. Using $\left\langle \psi^{\perp}\right|Z\left|\psi^{\perp}\right\rangle \le\left\Vert Z\right\Vert $,
we conclude that it suffices to set $\val=\max\left\{ \left(\frac{1}{\epsilon}+1\right)\left\Vert Z\right\Vert ,2\left\Vert Z\right\Vert \right\} $. 

One can proceed analogously for the $R_{A}^{*}$ case.
\end{proof}

\section{$\Lambda$-Penalty EBM Point Games}\label{sec:Lambda-EBM}
Following \cite{ACG+14}, we define EBM point games as a step towards time-dependent point games.
\begin{defn}[Prob]
 Consider $Z\ge0$ and let $\Pi^{[z]}$ represent the projector on
the eigenspace of eigenvalue $z\in\spectrum(Z)$. We have $Z=\sum_{z}z\Pi^{[z]}$.
Let $\left|\psi\right\rangle $ be a (not necessarily normalised)
vector. We define the function with finite support $\Prob[Z,\psi]:=\Rnonneg\to\Rnonneg$
as 
\[
\Prob[Z,\psi](z)=\begin{cases}
\left\langle \psi\right|\Pi^{[z]}\left|\psi\right\rangle  & \text{if \ensuremath{z\in\spectrum(Z)}}\\
0 & \text{else}.
\end{cases}
\]
If $Z=Z_{A}\otimes\mathbb{I}_{M}\otimes Z_{B}$, using the same notation,
we define the bivariate function with finite support $\Prob[Z_{A},Z_{B},\psi]:\Rnonneg\times\Rnonneg\to\Rnonneg$
as 
\[
\Prob[Z_{A},Z_{B},\psi](z_{A},z_{B}):=\begin{cases}
\left\langle \psi\right|\Pi^{[z_{A}]}\otimes\mathbb{I}_{M}\otimes\Pi^{[z_{B}]}\left|\psi\right\rangle  & \text{if \ensuremath{(z_{A},z_{B})\in\spectrum(Z_{A})\times\spectrum(Z_{B})}}\\
0 & \text{else}.
\end{cases}
\]
\end{defn}

\begin{defn}[EBM line transition]
\label{def:EBMlinetransition} Let $g,h:\Rnonneg\to\Rnonneg$ be
two functions with finite supports. The line transition $g\to h$
is \emph{expressible by matrices (EBM)} if there exist two matrices
$0\le G\le H$ and a (not necessarily normalised) vector $\left|\psi\right\rangle $
such that $g=\Prob[G,\left|\psi\right\rangle ]$ and $h=\Prob[H,\left|\psi\right\rangle ]$.
We say $g\to h$ above is \emph{EBM with spectrum in $[a,b]$} if
$\spectrum(G)\cup\spectrum(H)\subseteq[a,b]$.
\end{defn}

\begin{defn}[EBM transition]
 Let $p,q:\Rnonneg\times\Rnonneg\to\Rnonneg$ be two functions with
finite supports. The transition $p\to q$ is an: 
\begin{itemize}
\item EBM horizontal transition if for all $y\in\Rnonneg$, $p(\cdot,y)\to q(\cdot,y)$
is an EBM line transition, and
\item EBM vertical transition if for all $x\in\Rnonneg$, $p(x,\cdot)\to q(x,\cdot)$
is an EBM line transition. 
\end{itemize}
\end{defn}

The previous definitions were the same as in the standard WCF setting. Now, we introduce the cheat-penalised version of the EBM point games. The only change is in the starting and final frames. 
\begin{defn}[$\Lambda$-penEBM point game]
\label{def:EBMpointGame} A $\Lambda$-penalty EBM point game is a sequence of functions $(p_{0},\dots, p_{n})$ with
finite support such that 
\begin{itemize}
\item $p_{0}=\frac{1}{2}\points{\Lambda,\Lambda+1}+\frac{1}{2}\points{\Lambda+1,\Lambda}$,
\item for all even $i$, $p_{i}\rightarrow p_{i+1}$ is an EBM vertical
transition,
\item for all odd $i$, $p_{i}\to p_{i+1}$ is an EBM horizontal transition,
\item $p_{n}=1\points{\beta,\alpha}$ for some $\alpha,\beta\in[\Lambda,\Lambda+1]$.
We call $\points{\beta,\alpha}$ the final point of the EBM point
game.
\end{itemize}
\end{defn}
We sometimes refer to the functions $p_{i}$ as frames or configurations. Next, we show the relation between {\lwcf} protocols and $\Lambda$-penEBM point games.

\subsection{$\Lambda$-penWCF protocol implies $\Lambda$-penEBM point game}
\begin{prop}[$\Lambda$-penWCF $\implies$ $\Lambda$-penEBM point game]
 \label{prop:penWCF_implies_penEBM}Given 
\begin{itemize}
\item a $\Lambda$-penWCF protocol with cheating rewards $R_{A}^{*}$ and
$R_{B}^{*}$ (see \Defref{penWCF}), and
\item any positive number $\delta>0$, 
\end{itemize}
there exists a $\Lambda$-penEBM point game with final point $\points{R_{A}^{*}+\delta,R_{B}^{*}+\delta}$.
\end{prop}

\begin{rem}
    The proof of this proposition closely follows that of the analogous statement for (standard) WCF, although we note and correct a minor error in the argument presented in \cite{ACG+14}. The main difference from the standard case is in the boundary conditions.
\end{rem}
\begin{proof}
Assume we are given a $\Lambda$-penWCF protocol (with no intermediate
measurements as justified in \Remref{noIntermediateMeasurement}),
together with the dual certificates $\left\{ Z_{A,i}\right\} _{i=1}^{n}$
and $\left\{ Z_{B,i}\right\} _{i=1}^{n}$ (see \Thmref{dualSDP})
witnessing cheating points $R_{A}^{*}$ and $R_{B}^{*}$. Using the
proof of the last part of \Thmref{dualSDP}, we can use $Z'_{A,0}$
(resp. $Z'_{B,0}$) as in \Eqref{ZprimeForm} instead of $Z_{A,0}$
(resp. $Z_{B,0}$) which admits $\left|\psi_{A,0}\right\rangle $
(resp. $\left|\psi_{B,0}\right\rangle $) as eigenvectors with eigenvalues
$\beta=R_{B}^{*}+\delta$ (resp. $\alpha=R_{A}^{*}+\delta$). 

We use the ``reversed time'' convention: $Z_{A}^{(i)}:=Z_{A,n-i},Z_{B}^{(i)}:=Z_{B,n-i}$
and $\left|\psi^{(i)}\right\rangle :=\left|\psi_{n-i}\right\rangle $
for $i\in\{0,\dots,n\}$ (note that the notion conflicts slightly;
we already used $\Pi_{A}^{(0/1/\bot)}$ (resp. $\Pi_{B}^{(0/1/\bot)}$)
to denote Alice's (resp. Bob's) final POVM measurements). We use this
notation also for $U_{A,i}$ and $U_{B,i}$.

With the notation in place, we start the proof by defining the first
and last frames. We continue writing the projectors $E_{A,i},E_{B,i}$
for illustrating where they make a difference.

\textbf{First frame.} Define $p_{0}:=\Prob[Z_{A}^{(0)},Z_{B}^{(0)},\left|\psi^{(0)}\right\rangle ]$.
Consider $Z_{A}^{(0)}\otimes\mathbb{I}_{M}\otimes Z_{B}^{(0)}$ and
the state $\left|\psi^{(0)}\right\rangle $. Recall (see \Thmref{dualSDP})
that $Z_{A}^{(0)}=\Lambda\cdot\Pi_{A}^{(0)}+(\Lambda+1)\cdot\Pi_{A}^{(1)}$
and $Z_{B}^{(0)}=\Lambda\cdot\Pi_{B}^{(1)}+(\Lambda+1)\cdot\Pi_{B}^{(0)}$.
Recall also that by definition of a $\Lambda$-penWCF protocol (see
\Defref{penWCF}), it holds that
\begin{align*}
\tr\left(\Pi_{A}^{(1)}\otimes\mathbb{I}_{M}\otimes\Pi_{B}^{(1)}\left|\psi^{(0)}\right\rangle \left\langle \psi^{(0)}\right|\right)=\tr\left(\Pi_{A}^{(0)}\otimes\mathbb{I}_{M}\otimes\Pi_{B}^{(0)}\left|\psi^{(0)}\right\rangle \left\langle \psi^{(0)}\right|\right) & =\frac{1}{2} &  & \text{(Balanced)}\\
\tr\left(\Pi_{A}^{(a)}\otimes\mathbb{I}_{M}\otimes\Pi_{B}^{(b)}\left|\psi^{(0)}\right\rangle \left\langle \psi^{(0)}\right|\right) & =0 & \forall(a,b)\in S_{\text{invalid}}\quad & \text{(Correctness)}
\end{align*}
where $S_{{\rm invalid}}:=\{(0,1),(1,0),(\bot,0),(\bot,1),(0,\bot),(1,\bot),(\bot,\bot)\}$
captures events where the honest parties either output abort or disagree
on the output. One can thus write 
\begin{align*}
p_{0} & =\Prob\left[(\Lambda+1)\cdot\Pi_{A}^{(1)}+\Lambda\cdot\Pi_{A}^{(0)}+0\cdot\Pi_{A}^{(\bot)},(\Lambda+1)\cdot\Pi_{B}^{(0)}+\Lambda\cdot\Pi_{B}^{(1)}+0\cdot\Pi_{B}^{(\bot)},\left|\psi^{(0)}\right\rangle \right].
\end{align*}
Note that one can also write
\begin{align*}
p_0 = \sum_{(x,y)\in\{\Lambda+1,\Lambda,0\}\times\{\Lambda+1,\Lambda,0\}}\mathsf{coeff}_{xy}\points{x,y}
\end{align*}
for some choice of $\mathsf{coeff}_{xy}\in \mathbb{R}$ for each $x,y$. Further, we have
\begin{align*}
 p_0 & =\tr\left(\Pi_{A}^{(1)}\otimes\mathbb{I}_{M}\otimes\Pi_{B}^{(1)}\cdot\psi^{(0)}\right)\points{\Lambda+1,\Lambda}+\\
 & \quad\tr\left(\Pi_{A}^{(0)}\otimes\mathbb{I}_{M}\otimes\Pi_{B}^{(0)}\cdot\psi^{(0)}\right)\points{\Lambda,\Lambda+1} & \text{(other \ensuremath{\mathsf{coeff}_{xy}=0}; Correctness)}\\
 & =\frac{1}{2}\cdot\points{\Lambda+1,\Lambda}+\frac{1}{2}\cdot\points{\Lambda,\Lambda+1} & \text{(Balanced)}
\end{align*}
where we use $\psi^{(0)}$ to denote $\left|\psi^{(0)}\right\rangle \left\langle \psi^{(0)}\right|$. 

\textbf{Last frame.} Now, define $p_{n}:=\Prob\left[Z_{A}^{(n)},Z_{B}^{(n)},\left|\psi^{(n)}\right\rangle \right]$.
Consider $Z_{A}^{(n)}\otimes\mathbb{I}_{M}\otimes Z_{B}^{(n)}$ and
$\left|\psi^{(n)}\right\rangle $. Recall that $\left|\psi^{(n)}\right\rangle =\left|\psi_{A,0}\right\rangle \otimes\left|\psi_{M,0}\right\rangle \otimes\left|\psi_{B,0}\right\rangle $
(see \Defref{penWCF}) and as stated at the beginning, we can take
$Z_{A}^{(n)}=\beta\left|\psi_{A}^{(n)}\right\rangle \left\langle \psi_{A}^{(n)}\right|+\val_{A}\left(\mathbb{I}-\left|\psi_{A}^{(n)}\right\rangle \left\langle \psi_{A}^{(n)}\right|\right)$,
$Z_{B}^{(n)}=\alpha\left|\psi_{B}^{(n)}\right\rangle \left\langle \psi_{B}^{(n)}\right|+\val_{B}\left(\mathbb{I}-\left|\psi_{B}^{(n)}\right\rangle \left\langle \psi_{B}^{(n)}\right|\right)$
(using \Eqref{ZprimeForm}). Since $\left|\psi_{A}^{(n)}\right\rangle $
and $\left|\psi_{B}^{(n)}\right\rangle $ are normalised vectors,
together these imply that $p_{n}=1\cdot\points{\beta,\alpha}$.

\textbf{Intermediate frames.} We now define the remaining frames $p_{i}$
as 
\begin{equation}
p_{i}:=\Prob\left[Z_{A}^{(i)},Z_{B}^{(i)},\left|\psi^{(i)}\right\rangle \right]=\sum_{(x,y)\in\spectrum(Z_{A}^{(i)})\times\spectrum(Z_{B}^{(i)})}\tr\left[\Proj(Z_{A}^{(i)},x)\otimes\mathbb{I}_{M}\otimes\Proj(Z_{B}^{(i)},y)\cdot\psi^{(i)}\right]\points{x,y}\label{eq:Prob_i_intermediate}
\end{equation}
where $\Proj(Z_{A}^{(i)},x)$ is a projector on the $x$-eigenvalue
subspace of $Z_{A}^{(i)}$ ($\Proj(Z_{B}^{(i)},y)$ is defined analogously)
and $\psi^{(i)}$ denotes $\left|\psi^{(i)}\right\rangle \left\langle \psi^{(i)}\right|$.
We now show that $p_{i}\to p_{i+1}$ is an EBM transition (either
horizontal or vertical, depending on $i$). This part of the proof
is exactly the same as in the standard WCF case. Suppose $i$ is such
that $Z_{B}^{(i+1)}=Z_{B}^{(i)}$ while 
\begin{equation}
Z_{A}^{(i+1)}\otimes\mathbb{I}_{M}\ge U_{A}^{(i)\dagger}\left(Z_{A}^{(i)}\otimes\mathbb{I}_{M}\right)U_{A}^{(i)}\label{eq:Z_le_UEZ_bound}
\end{equation}
(this depends on whether $n-i$ is odd or even; in the other case,
$Z_{A}^{(i+1)}=Z_{A}^{(i)}$ while $Z_{B}^{(i+1)}$ will satisfy an
inequality). It also holds that $\left|\psi^{(i+1)}\right\rangle =U^{(i)\dagger}\left|\psi^{(i)}\right\rangle $.
We note that 
\begin{align}
p_{i}(\cdot,y) & =\sum_{x\in\spectrum(Z_{A}^{(i)})}\tr[\Proj(Z_{A}^{(i)},x)\otimes\mathbb{I}_{M}\otimes\Proj(Z_{B}^{(i)},y)\cdot\psi^{(i)}]\points x\nonumber \\
 & =\Prob\left[Z_{A}^{(i)}\otimes\mathbb{I}_{M}\otimes\Proj(Z_{B}^{(i)},y),\left|\psi^{(i)}\right\rangle \right] & \text{using the Definition of Prob}\nonumber \\
 & =\Prob\left[U_{A}^{(i)\dagger}\left(Z_{A}^{(i)}\otimes\mathbb{I}_{M}\right)U_{A}^{(i)}\otimes\Proj(Z_{B}^{(i)},y),U^{(i)\dagger}\left|\psi^{(i)}\right\rangle \right] & \text{\ensuremath{\because} the unitaries cancel and}\label{eq:addingProjectorsNotGood}\\
 &  & \ensuremath{\spectrum(Z)=\spectrum(U^{\dagger}ZU)}\nonumber \\
 & =\Prob[G_{y},\left|\psi^{(i+1)}\right\rangle ]\nonumber 
\end{align}
where $G_{y}:=U_{A}^{(i)\dagger}\left(Z_{A}^{(i)}\otimes\mathbb{I}_{M}\right)U_{A}^{(i)}\otimes\Proj(Z_{B}^{(i+1)},y)$
because $Z_{A}^{(i+1)}=Z_{A}^{(i)}$.\footnote{We point out that \Eqref{addingProjectorsNotGood} would not hold
if there were a projector because $\spectrum(Z)\neq\spectrum(EZE)$
for an arbitrary projector $E$ (even if it is constrained to be such
that $E\left|\psi\right\rangle =\left|\psi\right\rangle $. Previous
work~\cite{ACG+14} implicitly assume this in their proof making
it slightly incorrect---as one can proceed like we did here by absorbing
all projectors to the very end. } Similarly, one can write 
\begin{align*}
p_{i+1}(\cdot,y) & =\Prob\left[Z_{A}^{(i+1)}\otimes\mathbb{I}_{M}\otimes\Proj(Z_{B}^{(i+1)},y),\left|\psi^{(i+1)}\right\rangle \right]\\
 & =\Prob\left[H_{y},\left|\psi^{(i+1)}\right\rangle \right]
\end{align*}
where $H_{y}:=Z_{A}^{(i+1)}\otimes\mathbb{I}_{M}\otimes\Proj(Z_{B}^{(i+1)},y)$.
Now clearly, for each $y$, using \Eqref{Z_le_UEZ_bound}, we have
that $H_{y}\ge G_{y}$ and so $p_{i}(\cdot,y)\to p_{i+1}(\cdot,y)$
is an EBM transition for each $y$. 

\end{proof}

It turns out that, just as for standard WCF, the converse of also
holds. 

\subsection{$\Lambda$-penEBM point game implies $\Lambda$-penWCF protocol\protect\label{subsec:Lambda-penEBM-point-game}}
\begin{thm}
\label{thm:penEBMtopenWCF}Given a $\Lambda$-penEBM point game with
final point $\points{\beta,\alpha}$, there exists a $\Lambda$-penWCF
protocol with $R_{A}^{*}\le\beta$ and $R_{B}^{*}\le\alpha$. 
\end{thm}

The proof of this is also very similar to the proof for the standard
WCF case. Here, we improve the proof from \cite{Arora2019} by
making the resulting protocol more round efficient. To this end,
we use the following characterisation of EBM transitions.
\begin{notation}
We often consider (EBM line) transitions from $g\to h$, from $n_{g}>0$
points to $n_{h}>0$ points (here subscript $g$ and $h$ are not
indices---but behaving more like the symbol $\prime$ behaves in
$g'$). We write them as $g=\sum_{i=1}^{n_{g}}p_{g_{i}}\points{x_{g_{i}}}$
and $h=\sum_{i=1}^{n_{h}}p_{h_{i}}\points{x_{h_{i}}}$ where (again
subscript $g$ and $h$ are not indices) $p_{g_{i}}>0,p_{h_{i}}>0$
are the weights, and $x_{g_{i}}\ge0,x_{h_{i}}\ge0$ are distinct coordinates
(i.e. $x_{g_{i}}\neq x_{g_{i'}}$ for $i\neq i'$, and similarly $x_{h_{i}}\neq x_{h_{i'}}$
for $i\neq i'$). 
\end{notation}

\begin{lem}
\label{lem:EBM-line-characterisation}Let $g\to h$ be an EBM line
transition with spectrum in $[a,b]$ and suppose $g$ and $h$ have
disjoint support. Then, there exists a unitary $U$, and diagonal
matrices $X_{h},X_{g}$ (with no multiplicities except possibly those
of $a$ and $b$) of size at most $m\times m$ for $m:=n_{g}+n_{h}-1$
such that 
\[
\underbrace{\left[\begin{array}{cccccc}
x_{h_{1}}\\
 & \ddots\\
 &  & x_{h_{n_{h}}}\\
 &  &  & b\\
 &  &  &  & \ddots\\
 &  &  &  &  & b
\end{array}\right]}_{:=X_{h}}\ge U\underbrace{\left[\begin{array}{cccccc}
x_{g_{1}}\\
 & \ddots\\
 &  & x_{g_{n_{g}}}\\
 &  &  & a\\
 &  &  &  & \ddots\\
 &  &  &  &  & a
\end{array}\right]}_{:=X_{g}}U^{\dagger}
\]
and 
\[
\left[\begin{array}{c}
\sqrt{p_{h_{1}}}\\
\vdots\\
\sqrt{p_{h_{n_{h}}}}\\
0\\
\vdots\\
0
\end{array}\right]=U\left[\begin{array}{c}
\sqrt{p_{g_{1}}}\\
\vdots\\
\sqrt{p_{g_{n_{g}}}}\\
0\\
\vdots\\
0
\end{array}\right]=:\left|\psi\right\rangle 
\]
or more briefly, $X_{h}\ge UX_{g}U^{\dagger}$ and $\sum_{i=1}^{n_{h}}\sqrt{p_{h_{i}}}\left|i\right\rangle =U\left(\sum_{i=1}^{n_{g}}\sqrt{p_{g_{i}}}\left|i\right\rangle \right)=:\left|\psi\right\rangle .$ 

This is also true if $g$ and $h$ have common support, i.e., if $g=\sum_{i=1}^{n_{k}}p_{k_{i}}\points{x_{k_{i}}}+\sum_{i=1}^{n_{g}}p_{g_{i}}\points{x_{g_{i}}}$
and $h=\sum_{i=1}^{n_{k}}p_{k}\points{x_{k_{i}}}+\sum_{i=1}^{n_{h}}p_{h_{i}}\points{x_{h_{i}}}$.
\end{lem}
\begin{proof}
Let us quickly make some elementary observations. Since $g\to h$
is EBM, there exist matrices $H\ge G$ and a vector $\left|\psi\right\rangle $
such that $\Prob[H,\left|\psi\right\rangle ]=h$ and $\Prob[G,\left|\psi\right\rangle ]=g$.
One can write 
\begin{align*}
 &  & H & \ge G\\
 & \iff & U_{h}X_{h}U_{h}^{\dagger} & \ge U_{g}X_{g}U_{g}^{\dagger} & \text{diagonalising}\\
 & \iff & X_{h} & \ge UX_{g}U & U:=U_{h}^{\dagger}U_{g}.
\end{align*}
It takes some work to argue about the dimension and multiplicities,
but once that is done, it is straightforward to conclude that $\left|\psi\right\rangle $
can be chosen to have the form claimed.
\end{proof}

\subsection{Proof of \Thmref{penEBMtopenWCF}}\label{subsec:penEBMtopenWCF}
\begin{figure}
\begin{centering}
\includegraphics[width=9cm]{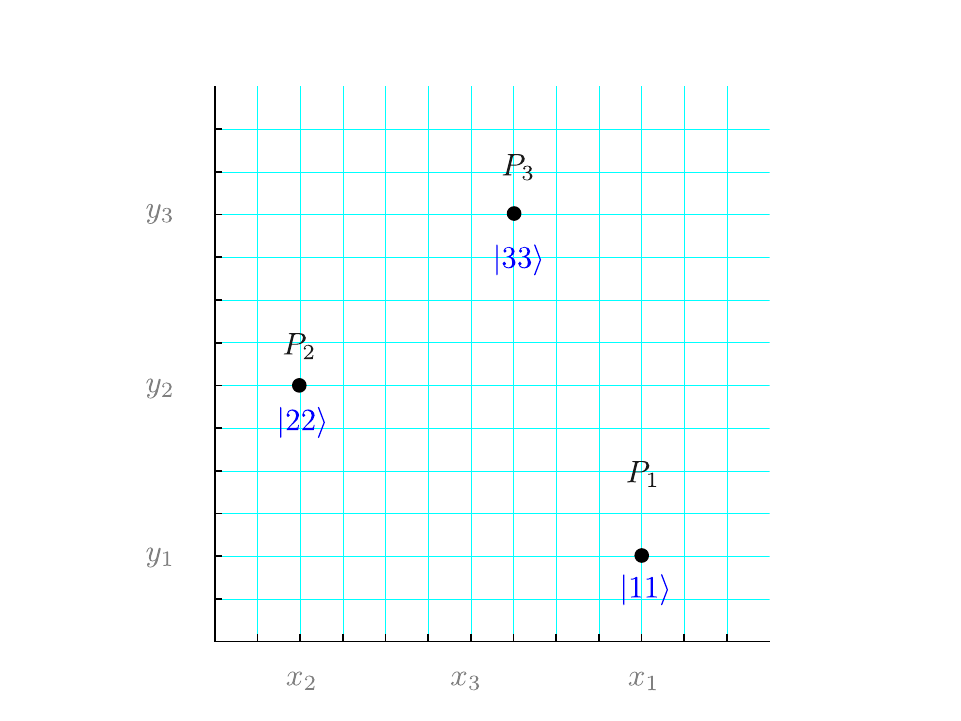}
\par\end{centering}
\caption{A frame in the Canonical Form.\protect\label{fig:CanonicalForm}}

\end{figure}

The rest of this section is devoted to proving \Thmref{penEBMtopenWCF}.
Suppose a $\Lambda$-penEBM point game $(p_{0},p_{1},\dots, p_{n})$ is
given. To each ``frame'', i.e., to each $p_{i}$, we associate
a ``canonical form''. 
\begin{defn}[Canonical Form]
 Consider the tuple $(\left|\psi\right\rangle _{ABM},Z^{A},Z^{B})$
where ${\cal A},{\cal B},{\cal M}$ are (finite dimensional) Hilbert
spaces and $\left|\psi\right\rangle _{ABM}\in{\cal A}\otimes{\cal B}\otimes{\cal M}$,
$Z^{A}\in{\cal A}$, $Z^{B}\in{\cal B}$. We say this tuple is in
the \emph{Canonical Form} with respect to a frame $p=\sum_{i}P_{i}\points{x_{i},y_{i}}$
of a TDPG if (see \Figref{CanonicalForm}) $\left|\psi\right\rangle =\sum_{i}\sqrt{P_{i}}\left|ii\right\rangle _{AB}\otimes\left|\varphi\right\rangle _{M}$,
$Z^{A}=\sum_{i}x_{i}\left|i\right\rangle \left\langle i\right|_{A}$
and $Z^{B}=\sum_{i}y_{i}\left|i\right\rangle \left\langle i\right|_{B}$
where $\left|\varphi\right\rangle _{M}$ is an arbitrary state (representing
the state of extra uncoupled registers that might be present).
\end{defn}

Our approach is to assign a canonical form to each frame of the $\Lambda$-penEBM point game point game and construct unitaries (and projectors) that define a $\Lambda$-penWCF which admits the $Z$ variables in the canonical form as dual feasible solutions. This ensures that the resulting $\Lambda$-penWCF protocol
has the same cheating reward as the final point of the $\Lambda$-penEBM point game. 

We use the reverse time convention (as we did in the proof
of \Propref{penWCF_implies_penEBM}): The variables $(\left|\psi_{(i)}\right\rangle ,Z_{(i)}^{A},Z_{(i)}^{B})$
describing the $i$th frame $p_{i}$ of $\Lambda$-EBPM point game,
and the unitary $U_{(i)}$ transforming $p_{i}$ to $p_{i+1}$(which
we define shortly), are going to be related to those of the $\Lambda$-WCF
protocol by 
\begin{equation}
\left|\psi_{i}\right\rangle :=\left|\psi_{(n-i)}\right\rangle ,Z_{i}=Z_{(n-i)},\text{ and }U_{i}=U_{(n-i)}^{\dagger}.\label{eq:reverse}
\end{equation}

We emphasise that the proof is based on that for the standard WCF
case (as in \cite{Arora2019})---and this is by design because we
chose the scoring convention carefully when defining $\Lambda$-penWCF
protocols in \Defref{penWCF}. We now proceed to give an informal description of the unitaries to provide some intuition and afterwards a formal description of them.

\subsection*{Informal Description of the Unitaries}

Consider frames $p_{j}=:g$ and $p_{j+1}=:h$ of a $\Lambda$-penEBM
point game $(p_{0},\dots, p_{n})$. For simplicity, suppose $g\to h$
involves changing points along a horizontal line (i.e., corresponding
to Alice's move) as illustrated in \Figref{TransitionForTEF}. We
use $\{g_{i}\}$ to index the initial points and $\{h_{i}\}$ to index
the final points and $\{k_{i}\}$ to index points that do not change.
More precisely, we write $g=\sum_{i}p_{g_{i}}\points{x_{g_{i}},y_{g_{i}}}+\sum_{i}p_{k_{i}}\points{x_{k_{i}},y_{k_{i}}}$
and $h=\sum_{i}p_{h_{i}}\points{x_{h_{i}},y_{h_{i}}}+\sum_{i}p_{k_{i}}\points{x_{k_{i}},y_{k_{i}}}$
where $y_{g_{i}}=y_{h_{i}}=y$ for all $i$. Let $\left\{ \left\{ \left|g_{i}\right\rangle \right\} _{i},\left\{ \left|h_{i}\right\rangle \right\} _{i},\left\{ \left|k_{i}\right\rangle \right\} _{i},\left|m\right\rangle \right\} $
denote orthonormal vectors (for instance, $\left\langle h_{i}|g_{i'}\right\rangle =0$).
We assume that a canonical form is given for $g$ using $\left|g_{i}\right\rangle $
instead $\left|i\right\rangle $ and that for $h$ is given using
$\left|h_{i}\right\rangle $ instead of $\left|i\right\rangle $.
Using \Lemref{EBM-line-characterisation}, it is not hard to see that
there is a unitary $U$ such that $\sum_{i}x_{h_{i}}\left|h_{i}\right\rangle \left\langle h_{i}\right|\ge E_{h}U\left(\sum_{i}x_{g_{i}}\left|g_{i}\right\rangle \left\langle g_{i}\right|\right)U^{\dagger}E_{h}$
where $E_{h}=\sum_{i}\left|h_{i}\right\rangle \left\langle h_{i}\right|$. 

Consider the following state which is in the canonical form for $g$,
\[
\left|\psi_{(1)}\right\rangle =\left(\sum_{i}\sqrt{p_{g_{i}}}\left|g_{i}g_{i}\right\rangle _{AB}+\sum_{i}\sqrt{p_{k_{i}}}\left|k_{i}k_{i}\right\rangle _{AB}\right)\otimes\left|m\right\rangle _{M}.
\]
We want Bob to send his part of $\left|g_{i}\right\rangle $ states
to Alice via the message register. One way of doing this is to have
Bob conditionally swap to obtain the following 
\[
\left|\psi_{(2)}\right\rangle =\sum_{i}\sqrt{p_{g_{i}}}\left|g_{i}g_{i}\right\rangle _{AM}\otimes\left|m\right\rangle _{B}+\sum_{i}\sqrt{p_{k_{i}}}\left|k_{i}k_{i}\right\rangle _{AB}\otimes\left|m\right\rangle _{M}.
\]
The intuition here is that since $y_{g_{i}}=y_{h_{i}}=y$, and since
Bob performs this step, it will not correspond to a non-trivial constraint
in the dual SDP. Alice can now update the probabilities locally by
applying $U'$---which is the same as $U$ but uses $\left|gg_{i}\right\rangle $
and $\left|hh_{i}\right\rangle $ instead of $\left|g_{i}\right\rangle $
and $\left|h_{i}\right\rangle $ while leaving the remaining space
unchanged---to obtain 
\[
\left|\psi_{(3)}\right\rangle =\sum_{i}\sqrt{p_{h_{i}}}\left|h_{i}h_{i}\right\rangle _{AM}\otimes\left|m\right\rangle _{B}+\sum_{i}\sqrt{p_{k_{i}}}\left|k_{i}k_{i}\right\rangle \otimes\left|m\right\rangle _{M}.
\]
This is the step that must satisfy a non-trivial constraint in the
dual SDP. Finally, Bob now accepts the new state by ``unswapping''
to obtain 
\[
\left|\psi_{(4)}\right\rangle =\left(\sum_{i}\sqrt{p_{h_{i}}}\left|h_{i}h_{i}\right\rangle _{AB}+\sum_{i}\sqrt{p_{k_{i}}}\left|k_{i}k_{i}\right\rangle _{AB}\right)\otimes\left|m\right\rangle _{M}.
\]
We will see that this step also satisfies the constraint in the dual
SDP. 

A few brief remarks before we move on to the formal proofs. (1) Note
the the actual protocol runs backwards relative to how we described
the operations here (see \Eqref{reverse}). (2) Even though we are
going from $p_{i}$ to $p_{i+1}$, instead of using a single unitary,
we are using three unitaries (of which only one is actually non-trivial).
This basically allows us to decouple the message register after each
$p_{i}$ to $p_{i+1}$ transition. (3) In our description above, we
assumed that points along a single horizontal line are affected. However,
we can handle multiple horizontal lines at once. 

\subsection*{Formal Description of the unitaries}

\begin{figure}
\begin{centering}
\subfloat[\label{fig:TransitionForTEF}In \cite{Arora2019}, they could only handle point game transitions along a single horizontal line at a time.]{\begin{centering}
\hspace{1.7em} \includegraphics[width=17.3cm]{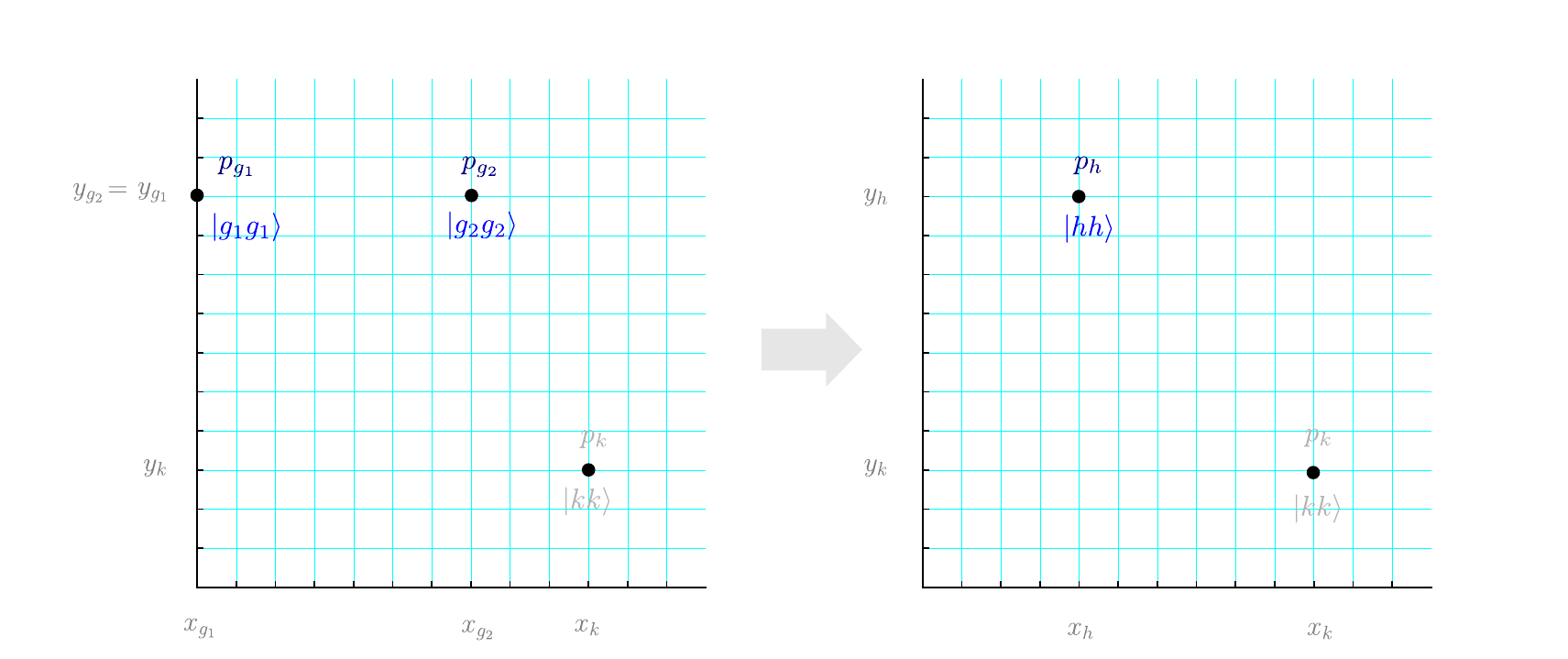}
\par\end{centering}
}
\par\end{centering}
\begin{centering}
\subfloat[\label{fig:multipleHorizontalPoints}Point game transition along multiple horizontal lines.]{\begin{centering}
\includegraphics[width=18cm]{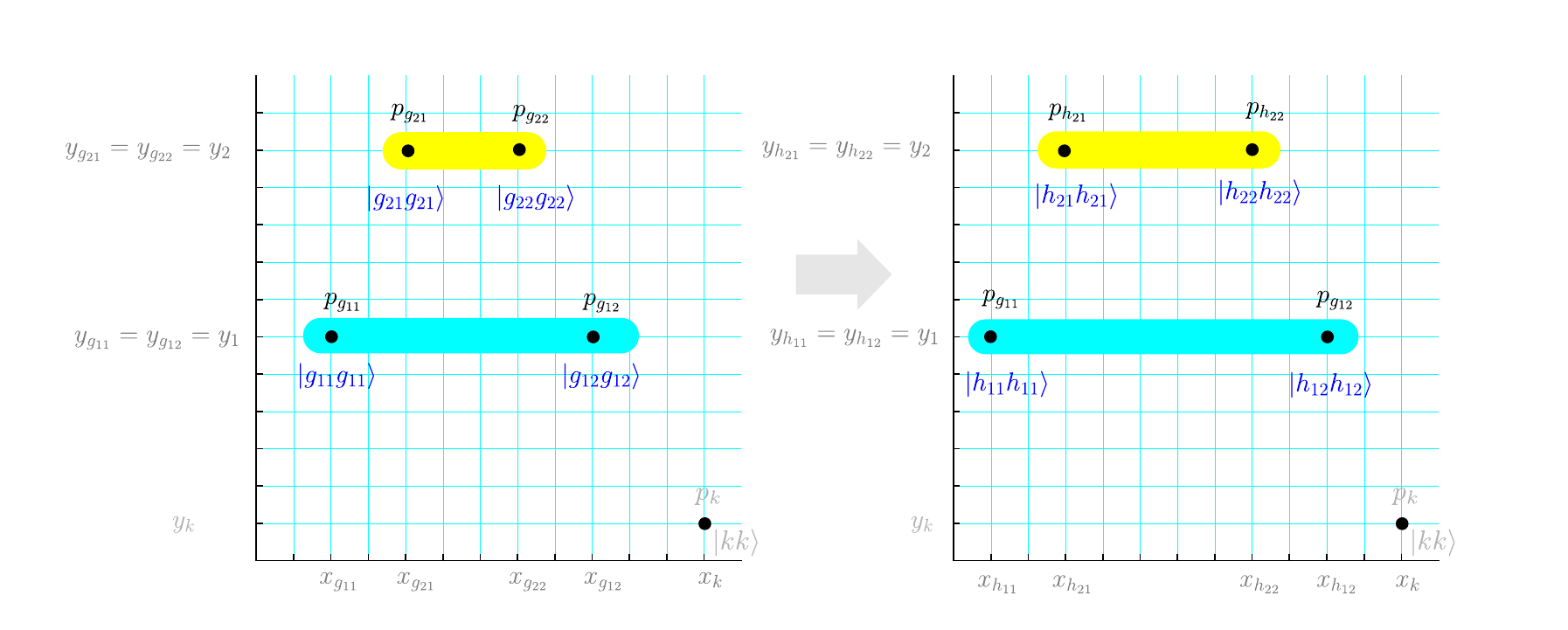}
\par\end{centering}

}
\par\end{centering}
\centering{}\caption{The parallelising step to improve round-efficiency. Comparison with prior work. Allowing multiple horizontal transitions to take place at once.}
\end{figure}

Consider frames $p_{j}=:g$ and $p_{j+1}=:h$ of a $\Lambda$-EBM
point game $(p_{0},\dots, p_{n})$ as illustrated in \Figref{multipleHorizontalPoints}
where multiple horizontal line transitions take place. We write 
\[
g=:\sum_{\ell}\left(\sum_{i}p_{g_{\ell i}}\points{x_{g_{\ell i}},y_{g_{\ell}}}\right)+\sum_{i}p_{k_{i}}\points{x_{k_{i}},y_{k_{i}}}
\]
and 
\[
h=:\sum_{\ell}\left(\sum_{i}p_{h_{\ell i}}\points{x_{h_{\ell i}},y_{h_{\ell}}}\right)+\sum_{i}p_{k_{i}}\points{x_{k_{i}},y_{k_{i}}}
\]
where we use labels $g_{\ell i}$ (resp. $h_{\ell i}$) to denote
the properties of the $i$th point, involved in the $\ell$th horizontal
transition, in frame $g$ (resp. frame $h$) and since $y_{g_{\ell i}}=y_{g_{\ell i'}}$
for all $i,i'$, we simply use $y_{g_{\ell}}$ (similarly for $y_{h_{\ell}}$).
We further simplify the notation and write
\[
y_{g_{\ell}}=y_{h_{\ell}}=:y_{\ell}
\]
as $\ell$ indexes horizontal transitions which means the $y$ coordinates
for the corresponding points in $g$ and $h$ are the same. We are
going to define a dual SDP \Thmref{dualSDP} and also construct a
dual feasible solution to it, iteratively. Let $\left\{ \left\{ \left|g_{\ell i}\right\rangle \right\} _{\ell i},\left\{ \left|h_{\ell i}\right\rangle \right\} _{\ell i},\left\{ \left|k_{i}\right\rangle \right\} _{i},\left|m\right\rangle \right\} $
denote orthonormal vectors (as in the previous subsection, except
for the change in indexing). Consider the following definitions.
\begin{enumerate}
\item \textbf{The $j$th frame.\label{enu:The-jth-frame.}}

Consider the following definitions:
\begin{align*}
\left|\psi_{(j,1)}\right\rangle  & :=\left(\sum_{\ell i}\sqrt{p_{\ell i}}\left|g_{\ell i}g_{\ell i}\right\rangle _{AB}+\sum_{i}\sqrt{p_{k_{i}}}\left|k_{i}k_{i}\right\rangle _{AB}\right)\otimes\left|m\right\rangle _{M},\\
Z_{(j,1)}^{A} & :=\sum_{\ell i}x_{g_{\ell i}}\left|g_{\ell i}\right\rangle \left\langle g_{\ell i}\right|_{A}+\sum_{i}x_{k_{i}}\left|k_{i}\right\rangle \left\langle k_{i}\right|_{A},\\
Z_{(j,1)}^{B} & :=\sum_{\ell i}y_{\ell}\left|g_{\ell i}\right\rangle \left\langle g_{\ell i}\right|_{B}+\sum_{i}y_{k_{i}}\left|k_{i}\right\rangle \left\langle k_{i}\right|_{B}.
\end{align*}

\begin{claim}
The tuple $\left(\left|\psi_{(j)}\right\rangle ,Z_{(j)}^{A},Z_{(j)}^{B}\right)$
is in the canonical form for the frame $p_{j}=:g$. \label{claim:canonical-j}
\end{claim}

\item \textbf{Bob sends to Alice. }

Consider the following definitions:
\begin{align*}
\left|\psi_{(j,2)}\right\rangle  & :=\sum_{\ell i}\sqrt{p_{g_{\ell i}}}\left|g_{\ell i}g_{\ell i}\right\rangle _{AM}\otimes\left|m\right\rangle _{B}+\sum_{i}\sqrt{p_{k_{i}}}\left|k_{i}k_{i}\right\rangle _{AB}\otimes\left|m\right\rangle _{M}\\
U_{(j,1)} & :=\sum_{\ell i}\left|g_{\ell i}m\right\rangle \left\langle mg_{\ell i}\right|_{BM}+\hc+\underbrace{\left[\mathbb{I}_{BM}-\left(\sum_{\ell i}\left|g_{\ell i}m\right\rangle \left\langle g_{\ell i}m\right|+\sum_{\ell i}\left|mg_{\ell i}\right\rangle \left\langle mg_{\ell i}\right|\right)_{BM}\right]}_{=\mathbb{I}_{{\rm rest}}}\\
E_{(j,2)} & :=\sum_{\ell i}\left|g_{\ell}m\right\rangle \left\langle g_{\ell}m\right|_{BM}+\sum_{i}\left|k_{i}\right\rangle \left\langle k_{i}\right|_{B}\otimes\mathbb{I}_{M}\\
Z_{(j,2)}^{A} & =Z_{(j,1)}^{A}\\
Z_{(j,2)}^{B} & =\sum_{\ell}y_{\ell}\mathbb{I}_{B}^{\left\{ \left\{ g_{\ell i}\right\} _{i},m\right\} }+\sum_{i}y_{k_{i}}\left|k_{i}\right\rangle \left\langle k_{i}\right|_{B}
\end{align*}
where $U_{(j,1)}$ is basically permuting $\left|g_{\ell i}m\right\rangle _{BM}$
with $\left|mg_{\ell i}\right\rangle _{BM}$ .
\begin{claim}
\label{claim:BobSendsToAlice}It holds that (a) $\left|\psi_{(j,2)}\right\rangle =E_{(j,2)}U_{(j,1)}\left|\psi_{(j,1)}\right\rangle $
and (b) $Z_{(j,2)}\otimes\mathbb{I}_{M}\ge E_{(j,2)}U_{(j,1)}\left(Z_{(j,1)}^{B}\otimes\mathbb{I}_{M}\right)U_{(j,1)}^{\dagger}E_{(j,2)}$.
\end{claim}

\begin{proof}
Item (a) holds trivially by definition of $U_{(j,1)}$ and $E_{(j,2)}$.
In this proof, let $U$ denote $U_{(j,1)}$, let $E$ denote $E_{(j,2)}$.
Note that $U^{\dagger}=U$. To establish item (b), we write 
\begin{align*}
 & EU\left(Z_{(j,1)}^{B}\otimes\mathbb{I}_{M}\right)UE\\
= & EU\left(\sum_{\ell i}y_{\ell}\left|g_{\ell i}\right\rangle \left\langle g_{\ell i}\right|_{B}\otimes\mathbb{I}_{M}+\sum_{i}y_{k_{i}}\left|k_{i}\right\rangle \left\langle k_{i}\right|_{B}\otimes\mathbb{I}_{M}\right)UE\\
= & EU\Bigg(\sum_{\ell i}y_{\ell}\left|g_{\ell i}\right\rangle \left\langle g_{\ell i}\right|_{B}\otimes\left|m\right\rangle \left\langle m\right|_{M}+\underbrace{\sum_{\ell i}y_{\ell}\left|g_{\ell i}\right\rangle \left\langle g_{\ell i}\right|\otimes\left(\mathbb{I}_{M}-\left|m\right\rangle \left\langle m\right|_{M}\right)}\\
 & \quad\quad\quad+\underbrace{\sum_{i}y_{k_{i}}\left|k_{i}\right\rangle \left\langle k_{i}\right|_{B}\otimes\mathbb{I}_{M}}\Bigg)UE\quad\quad\quad\text{highlighted terms are unaffected by \ensuremath{U}}\\
= & E\Bigg(\sum_{\ell i}y_{\ell}\left|mg_{\ell i}\right\rangle \left\langle mg_{\ell i}\right|_{BM}+\sum_{\ell i}y_{\ell}\left|g_{\ell i}\right\rangle \left\langle g_{\ell i}\right|\otimes\left(\mathbb{I}_{M}-\left|m\right\rangle \left\langle m\right|_{M}\right)\\
 & \quad\quad\quad+\sum_{i}y_{k_{i}}\left|k_{i}\right\rangle \left\langle k_{i}\right|_{B}\otimes\mathbb{I}_{M}\Bigg)E\\
= & \left(\sum_{\ell i}y_{\ell}\left|g_{\ell i}\right\rangle \left\langle g_{\ell i}\right|\otimes\left(\mathbb{I}_{M}-\left|m\right\rangle \left\langle m\right|_{M}\right)+\sum_{i}y_{k_{i}}\left|k_{i}\right\rangle \left\langle k_{i}\right|_{B}\otimes\mathbb{I}_{M}\right)\\
\le & \sum_{\ell}y_{\ell}\mathbb{I}_{B}^{\left\{ \left\{ g_{\ell i}\right\} ,m\right\} }\otimes\mathbb{I}_{M}+\sum_{i}y_{k_{i}}\left|k_{i}\right\rangle \left\langle k_{i}\right|_{B}\otimes\mathbb{I}_{M}=Z_{(j,2)}^{B}\otimes\mathbb{I}_{M}.
\end{align*}
This means $EU\left(Z_{(j,1)}^{B}\otimes\mathbb{I}_{M}\right)UE\le Z_{(j,2)}^{B}\otimes\mathbb{I}_{M}$.
\end{proof}

\item \textbf{Alice applies the non-trivial unitary.\label{enu:Alice-applies-the-nontrivial-unitary}}

Consider the following definitions: 
\begin{align*}
\left|\psi_{(j,3)}\right\rangle  & :=\sum_{\ell i}\sqrt{p_{h_{\ell i}}}\left|h_{\ell i}h_{\ell i}\right\rangle _{AM}\otimes\left|m\right\rangle _{B}+\sum_{i}\sqrt{p_{k_{i}}}\left|k_{i}k_{i}\right\rangle _{AB}\otimes\left|m\right\rangle _{M}\\
U_{(j,2)} & :=\left|w\right\rangle \left\langle v\right|+\text{other terms acting non-trivally on \ensuremath{{\cal H}:=\spn}\ensuremath{\left\{  \left\{  \left|h_{\ell i}h_{\ell i}\right\rangle \right\}  _{\ell i},\left\{  \left|g_{\ell i}g_{\ell i}\right\rangle \right\}  _{\ell i}\right\} } }\\
E_{(j,3)} & :=\left(\sum_{\ell i}\left|h_{\ell i}\right\rangle \left\langle h_{\ell i}\right|+\sum_{i}\left|k_{i}\right\rangle \left\langle k_{i}\right|\right)_{A}\otimes\mathbb{I}_{M}\\
Z_{(j,3)}^{A} & =\sum_{\ell i}x_{h_{\ell i}}\left|h_{\ell i}\right\rangle \left\langle h_{\ell i}\right|_{A}+\sum_{i}x_{k_{i}}\left|k_{i}\right\rangle \left\langle k_{i}\right|_{A}\\
Z_{(j,3)}^{B} & =Z_{(j,2)}^{B}
\end{align*}
where 
\[
\left|v\right\rangle =\frac{\sum_{\ell i}\sqrt{p_{g_{\ell i}}}\left|g_{\ell i}g_{\ell i}\right\rangle }{\sqrt{\sum_{\ell i}p_{g_{\ell i}}}},\quad\left|w\right\rangle =\frac{\sum_{\ell i}\sqrt{p_{h_{\ell i}}}\left|h_{\ell i}h_{\ell i}\right\rangle }{\sqrt{\sum_{\ell i}p_{h_{\ell i}}}},
\]
and $U_{(j,2)}$ is satisfies 
\begin{equation}
\sum_{\ell i}x_{h_{\ell i}}\left|h_{\ell i}h_{\ell i}\right\rangle \left\langle h_{\ell i}h_{\ell i}\right|\ge E_{(j,2)}U_{(j,2)}\left(\sum_{\ell i}x_{g_{\ell i}}\left|g_{\ell i}g_{\ell i}\right\rangle \left\langle g_{\ell i}g_{\ell i}\right|\right)U_{(j,2)}^{\dagger}E_{(j,2)}.\label{eq:TEFconstraint}
\end{equation}

\begin{claim}
\label{claim:EBM_implies_to_TEF}Let $g\to h$ be an EBM (horizontal)
transition. Then, there is a unitary $U_{(j,2)}$ of the form described
above that satisfies \Eqref{TEFconstraint}. 
\end{claim}

\begin{proof}
It follows from \Lemref{EBM-line-characterisation} by an appropriate
change of basis.
\end{proof}

\begin{claim}
\label{claim:TEF-again}It holds that (a) $\left|\psi_{(j,3)}\right\rangle =E_{(j,3)}U_{(j,2)}\left|\psi_{(j,2)}\right\rangle $
and (b) $Z_{(j,3)}^{A}\otimes\mathbb{I}_{M}\ge E_{(j,3)}U_{(j,2)}\left(Z_{(j,2)}^{A}\otimes\mathbb{I}_{M}\right)U_{(j,2)}^{\dagger}E_{(j,3)}$.
\end{claim}

{\begin{proof}
Again, item (a) follows directly by definitions of $U_{(j,2)}$ and
$E_{(j,3)}$. We focus on establishing item (b) and we use $U$ and
$E$ to denote $U_{(j,2)}$ and $E_{(j,3)}$ respectively. We write

\begin{align*}
EU\left(Z_{(j,2)}^{A}\otimes\mathbb{I}_{M}\right)U^{\dagger}E & =EU\Bigg[\sum_{\ell i}x_{g_{\ell i}}\left|g_{\ell i}g_{\ell i}\right\rangle \left\langle g_{\ell i}g_{\ell i}\right|_{AM}+\sum x_{g_{\ell i}}\left|g_{\ell i}\right\rangle \left\langle g_{\ell i}\right|_{A}\otimes\left(\mathbb{I}-\left|g_{\ell i}\right\rangle \left\langle g_{\ell i}\right|\right)_{M}\\
 & \quad\quad+\sum_{i}x_{k_{i}}\left|k_{i}\right\rangle \left\langle k_{i}\right|_{A}\otimes\mathbb{I}_{M}\Bigg]U^{\dagger}E
\end{align*}
and observe that $Z_{(j,2)}^{A}\otimes\mathbb{I}_{M}$ is block diagonal
in ${\cal H}$ and ${\cal H}^{\perp}$. In fact, $Z_{(j,3)}^{A}\otimes\mathbb{I}_{M}$
and $U$ are also block diagonal in ${\cal H}$ and ${\cal H}^{\perp}$.
In the expression above, terms in $\mathsf{I}$ belong to the ${\cal H}\times{\cal H}$
block and $U$ acts non-trivially only on this block. Restricting
to this block, it is immediate from \Eqref{TEFconstraint} that in
the ${\cal H}\times{\cal H}$ block, the following holds: 
\[
\sum x_{h_{\ell i}}\left|h_{\ell i}h_{\ell i}\right\rangle \left\langle h_{\ell i}h_{\ell i}\right|\ge EU\left(\sum x_{g_{\ell i}}\left|g_{\ell i}\right\rangle \left\langle g_{\ell i}\right|\right)U^{\dagger}E.
\]
In the other block, $U$ acts as identity. On this block, $Z_{(j,3)}-EU\left(Z_{(j,2)}\otimes\mathbb{I}\right)U^{\dagger}E$
is 
\begin{align*}
 & \left(\sum_{\ell i}x_{h_{\ell i}}\left|h_{\ell i}\right\rangle \left\langle h_{\ell i}\right|_{A}+\sum_{i}x_{k_{i}}\left|k_{i}\right\rangle \left\langle k_{i}\right|\right)\otimes\mathbb{I}_{M}\\
 & \quad\quad-\underbrace{E\left(\sum x_{g_{\ell i}}\left|g_{\ell i}\right\rangle \left\langle g_{\ell i}\right|_{A}\otimes\left(\mathbb{I}-\left|g_{\ell i}\right\rangle \left\langle g_{\ell i}\right|\right)_{M}\right)E}_{=0}-\cancel{E}\left(\sum_{i}x_{k_{i}}\left|k_{i}\right\rangle \left\langle k_{i}\right|_{A}\otimes\mathbb{I}_{M}\right)\cancel{E}\\
= & \sum_{\ell i}x_{h_{\ell i}}\left|h_{\ell i}\right\rangle \left\langle h_{\ell i}\right|_{A}\ge0
\end{align*}
because $x_{h_{\ell i}}\ge0$ for all $\ell,i$. 
\end{proof}
}
\item \textbf{Bob accepts Alice's change.\label{enu:Bob-accepts-Alice's}}

Consider the following definitions: 
\begin{align*}
\left|\psi_{(j,4)}\right\rangle  & :=\left(\sum_{\ell i}\sqrt{p_{h_{\ell i}}}\left|h_{\ell i}h_{\ell i}\right\rangle _{AB}+\sum_{i}\sqrt{p_{k_{i}}}\left|k_{i}k_{i}\right\rangle _{AB}\right)\otimes\left|m\right\rangle _{M}\\
U_{(j,3)} & :=U_{(j,1)}=\sum_{\ell i}\left|h_{\ell i}m\right\rangle \left\langle mh_{\ell i}\right|_{BM}+\hc+\underbrace{\left[\mathbb{I}_{BM}-\left(\sum_{\ell i}\left|h_{\ell i}m\right\rangle \left\langle h_{\ell i}m\right|+\sum_{\ell i}\left|mh_{\ell i}\right\rangle \left\langle mh_{\ell i}\right|\right)_{BM}\right]}_{=\mathbb{I}_{{\rm rest}}}\\
E_{(j,3)} & :=\left(\sum_{\ell i}\left|h_{\ell i}\right\rangle \left\langle h_{\ell i}\right|+\sum_{i}\left|k_{i}\right\rangle \left\langle k_{i}\right|\right)_{B}\otimes\mathbb{I}_{M}\\
Z_{(j,4)}^{A} & =Z_{(j,3)}^{A}\\
Z_{(j,4)}^{B} & =\sum_{\ell i}y_{\ell i}\left|h_{\ell i}\right\rangle \left\langle h_{\ell i}\right|_{B}+\sum_{i}y_{k_{i}}\left|k_{i}\right\rangle \left\langle k_{i}\right|_{B}.
\end{align*}

\begin{claim}
The tuple $\left(\left|\psi_{(j,4)}\right\rangle ,Z_{(j,4)}^{A},Z_{(j,4)}^{B}\right)$
is in canonical form for the frame $p_{j+1}$. \label{claim:canonical4}
\end{claim}

{\begin{proof}
Follows by inspection of the appropriate definitions. 
\end{proof}
}
\begin{claim}
\label{claim:TEF_j4}It holds that (a) $\left|\psi_{(j,4)}\right\rangle =E_{(j,4)}U_{(j,3)}\left|\psi_{(j,3)}\right\rangle $
and (b) 
\[
Z_{(j,4)}^{B}\otimes\mathbb{I}_{M}\ge E_{(j,3)}U_{(j,3)}\left(Z_{(j,3)}^{B}\otimes\mathbb{I}_{M}\right)U_{(j,3)}^{\dagger}E_{(j,3)}.
\]
\end{claim}

{\begin{proof}
As before, item (a) follows from the definitions of $\left|\psi_{(j,4)}\right\rangle ,\left|\psi_{(j,3)}\right\rangle $
and $E_{(j,4)},U_{(j,3)}$. For item (b), denote by $E,U$ the objects
$E_{(j,4)},U_{(j,3)}$. Recall $Z_{(j,3)}^{B}=Z_{(j,2)}^{B}=\sum_{\ell}y_{\ell}\mathbb{I}_{B}^{\left\{ \left\{ g_{\ell i}\right\} ,m\right\} }+\sum_{i}y_{k_{i}}\left|k_{i}\right\rangle \left\langle k_{i}\right|$.
We write 
\begin{align*}
EU\left(Z_{(j,3)}^{B}\otimes\mathbb{I}_{M}\right)U^{\dagger}E & =EU\Bigg(\sum_{\ell}y_{\ell}\mathbb{I}_{B}^{\left\{ \left\{ g_{\ell i}\right\} ,m\right\} }\otimes\mathbb{I}_{M}\\
 & \quad\quad+\sum_{i}y_{k_{i}}\left|k_{i}\right\rangle \left\langle k_{i}\right|_{B}\otimes\mathbb{I}_{M}\Bigg)U^{\dagger}E\\
 & =\underbrace{E\left(\sum_{\ell}y_{\ell}\mathbb{I}_{B}^{\left\{ g_{\ell i}\right\} }\otimes\mathbb{I}_{M}\right)E}_{=0}+ & \text{\ensuremath{\because} \ensuremath{U} acts trivially on this}\\
 & \quad EU\left(\sum_{\ell}y_{\ell}\mathbb{I}_{B}^{\{m\}}\otimes\mathbb{I}_{M}\right)U^{\dagger}E+\\
 & \quad\sum_{i}y_{k_{i}}\left|k_{i}\right\rangle \left\langle k_{i}\right|_{B}\otimes\mathbb{I}_{M}\\
 & =\sum_{\ell}y_{\ell}\left|h_{\ell i}m\right\rangle \left\langle h_{\ell i}m\right|_{B}+\sum_{i}y_{k_{i}}\left|k_{i}\right\rangle \left\langle k_{i}\right|_{B}\otimes\mathbb{I}_{M}\\
 & \le\left(\sum_{\ell i}y_{\ell i}\left|h_{\ell i}\right\rangle \left\langle h_{\ell i}\right|_{B}+\sum_{i}y_{k_{i}}\left|k_{i}\right\rangle \left\langle k_{i}\right|_{B}\right)\otimes\mathbb{I}_{M}\\
 & =Z_{(j,4)}^{B}\otimes\mathbb{I}_{M}.
\end{align*}
This establishes item (b) as well, completing the proof. 
\end{proof}
}

\end{enumerate}
\begin{thm}[Strengthening of \Thmref{penEBMtopenWCF}; $\Lambda$-EBM point game
$\implies$ $\Lambda$-penWCF protocol]
Let $(p_{0}\dots p_{n})$ be a $\Lambda$-EBM point game with final
point $\points{\beta,\alpha}$. Then, there exists a $\Lambda$-penWCF
protocol with cheating rewards $R_{A}^{*}\le\alpha$ and $R_{B}^{*}\le\beta$
which uses $2n$ rounds of communication and $\log\left[3\cdot\left(\max_{j}\left(\mathsf{PointCount}(p_{j})+\mathsf{PointCount}(p_{j+1})\right)+1\right)\right]$
qubits where $\mathsf{PointCount}(p_{j})=\left|\left\{ (x,y):p_{j}(x,y)\neq0\right\} \right|$
is the number of points with non-zero weight. \label{thm:LambdaEBMpointgame-implies-LambdaPenWCF}
\end{thm}

{\begin{proof}
We have done most of the work already. Consider the definitions of
$\left|\psi_{(j,d)}\right\rangle ,Z_{(j,d)}^{A},Z_{(j,d)}^{B},E_{(j,d)},U_{(j,d)}$
for $j\in\{0,\dots n\}$ and $d\in\{1,2,3,4\}$ ($E_{j,1}:=\mathbb{I}$
and $U_{(j,4)}$ was not defined). Now, for the four steps of say
frame $j$, we used vectors $\left\{ \left\{ \left|g_{\ell i}\right\rangle \right\} _{\ell i},\left\{ \left|h_{\ell i}\right\rangle \right\} _{\ell i},\left\{ \left|k_{i}\right\rangle \right\} _{i},\left|m\right\rangle \right\} $
where $g$s were used for initial points, $h$ for final points and
$k$ for points that do not change (relative to the transition $p_{j}\to p_{j+1}$).
At the first step of the frame $j+1$, we re-use the vectors $\left\{ \left\{ \left|h_{\ell i}\right\rangle \right\} ,\left\{ \left|k_{i}\right\rangle \right\} ,\left|m\right\rangle \right\} $
but we relabel them as to again be consistent with the same convention---$g$s
are used for initial points, $k$ for points that do not change (relative
to the transition $p_{j+1}\to p_{j+2}$) while $\left|m\right\rangle $
stays unchanged. And the vector $\left\{ \left|g_{\ell i}\right\rangle \right\} $
can be re-used or extended as needed, in the remaining steps. 

Using this convention for the basis vectors of the Hilbert space,
one can consider the following definition for the protocol 
\begin{align*}
(U_{m},U_{m-1},\dots) & :=\left(U_{(j,1)},U_{(j,2)},U_{(j,3)},U_{(j+1,1)},U_{(j+1,2)},U_{(j+1,3)}\right)_{j=(0,2,\dots n)}\\
(E_{m},E_{m-1},\dots) & :=\left(E_{(j,1)},E_{(j,2)},E_{(j,3)},E_{(j+1,1)},E_{(j+1,2)},E_{(j+1,3)}\right)_{j=(0,2\dots n)}
\end{align*}
where $m=3n$. Crucially, communication is only required immediately
after $U_{(j,1)}$ and $U_{(j,2)}$ but not after $U_{(j,3)}$ because
the message register gets decoupled and can be discarded. Thus, even
though there are $3n$ unitaries that are applied, only $2n$ messages
are exchanged. The number of qubits can be seen to be as asserted
by inspection. 

Further, let 
\begin{align*}
(Z_{m}^{A},Z_{m-1}^{A},\dots) & :=\left(Z_{(j,1)}^{A},Z_{(j,2)}^{A},Z_{(j,3)}^{A},\underbrace{Z_{(j,4)}^{A}=Z_{(j+1,1)}^{A}},Z_{(j+1,2)}^{A},Z_{(j+1,3)}^{A}\right)_{j=(0,2\dots n)}\\
(Z_{m}^{B},Z_{m-1}^{B},\dots) & :=\left(Z_{(j,1)}^{B},Z_{(j,2)}^{B},Z_{(j,3)}^{B},\overbrace{Z_{(j,4)}^{B}=Z_{(j+1,1)}^{B}},Z_{(j+1,2)}^{B},Z_{(j+1,3)}^{B}\right)_{j=(0,2\dots n)}
\end{align*}
where our choice of basis ensures the highlighted terms are the same
(which can be seen by their respective definitions and by \Claimref{canonical-j}
and \Claimref{canonical4}). Further, one can define the measurements
$\Pi_{A}^{(0/1/\bot)}$ and $\Pi_{B}^{(0/1/\bot)}$ implicitly from
$Z_{A,m}$ and $Z_{B,m}$ as stated in point 4 of \Thmref{dualSDP},
and also $\left|\psi_{A,0}\right\rangle ,\left|\psi_{B,0}\right\rangle $
is defined as the eigenvector of $Z_{A,0}$ and $Z_{B,0}$. With these,
one can define a $\Lambda$-penWCF protocol and check that the corresponding
dual SDP as stated in \Thmref{dualSDP}, using \Claimref{BobSendsToAlice},
\Claimref{EBM_implies_to_TEF}, \Claimref{TEF-again}, \Claimref{canonical4}
and \Claimref{TEF_j4}, admits a solution specified by the dual variables
(i.e. the $Z$s) above. 
\end{proof}
}

\section{$\Lambda$-Penalty Time Dependent Point Games ($\Lambda$-pen TDPG)}\label{sec:Lambda-TDPG}

{In the previous section, we looked at $\Lambda$-pen EBM point games
that were built using EBM transitions. Prior works \cite{Mochon07,ACG+14}
observed that it is helpful to consider EBM functions instead of transitions.
This is because the set of EBM functions form a convex cone and one
can use conic duality to obtain a better characterisation. We, therefore,
first define EBM functions, recall its alternate characterisation---called
valid functions---using operator monotone functions. We end the section
by defining $\Lambda$-(penalty) point games using valid functions
instead of EBM functions, together with the analogue of \Thmref{LambdaEBMpointgame-implies-LambdaPenWCF}
that uses valid functions instead of EBM transitions.}

\subsection{EBM and valid functions (results from conic duality)}
\begin{defn}[$K$, EBM functions]
 A function $a:\Rnonneg\to\mathbb{R}$ with finite support is an
\emph{EBM function} if the line transition $a^{-}\to a^{+}$ is EBM
(see \Defref{EBMlinetransition}), where $a^{+}:\Rnonneg\to\Rnonneg$
and $a^{-}:\Rnonneg\to\Rnonneg$ denote, respectively, the positive
and negative part of $a$ (i.e. $a=a^{+}-a^{-}$ with $\supp(a^{+})\cap\supp(a^{-})=\emptyset$
and $a^{\pm}\ge0$).
\end{defn}

{It turns out that the closure of EBM functions equals the set of valid
functions, which are defined below. This follows from the fact that
the bi-dual of a convex cone equals the closure of a cone. The dual
of EBM functions is the set of operator monotone functions and, in
turn, the dual of operator monotone functions is defined to be the
set of valid functions. What helps here is that the set of operator
monotone functions are known to have a very nice characterisation. }
\begin{defn}[Valid functions and transitions.]
\label{def:validFunctions}We first define \emph{valid functions}
with one and two variables. 
\begin{enumerate}
\item Let $a:\Rnonneg\to\mathbb{R}$ be a function with finite support such
that $\sum_{x}a(x)=0$. Then $a$ is a \emph{valid function} if for
all $\lambda>0$, $\sum_{x}\frac{-a(x)}{\lambda+x}\ge0$ and $\sum_{x}x\cdot a(x)\ge0$.
\item Let $t:\Rnonneg\times\Rnonneg\to\mathbb{R}$ is a 
\begin{enumerate}
\item \emph{horizontally valid function} if for all $y\ge0$, $t(\cdot,y)$
is a valid function;
\item \emph{vertically valid function} if for all $x\ge0$, $t(x,\cdot)$
is a valid function.
\end{enumerate}
\end{enumerate}
Given finitely supported functions $g,h:\Rnonneg\times\Rnonneg\to\Rnonneg$
(resp. $g,h:\Rnonneg\to\Rnonneg$), we say $g\to h$ is a \emph{horizontally/vertically
valid transition} \emph{(resp. valid line transition)} if $t:=h-g$
is a horizontally/vertically valid function (resp. valid function). 

When clear from the context, we drop \emph{horizontally/vertically}
and simply use the term valid function (even for bivariate functions).
\end{defn}

\subsection{$\Lambda$-pen TDPG $\protect\implies$ $\Lambda$-penWCF protocols}\label{subsec:penTDPGimpliespenWCF}

{Just as we defined $\Lambda$-EBM point games that used the definition
of an EBM transition, one can define $\Lambda$-point games with EBM
functions. The basic idea being that instead of requiring the point
game $(p_{0},\dots, p_{n})$ to be such that for each odd $i$ (resp.
even $i$), $p_{i}\to p_{i+1}$ is horizontally (resp. vertically)
valid, we require that $p_{i+1}-p_{i}$ is a horizontal EBM function
(resp. vertical EBM function)---where horizontal/vertical EBM functions
are defined in the same spirit as we defined horizontal/vertical EBM
transitions, i.e., by fixing one coordinate and requiring the remaining
function to be an EBM function. 

Since the closure of EBM functions turns out to be the same as the
set valid functions, one can similarly define $\Lambda$-point games
with valid functions. Interestingly, the analogue of \Lemref{EBM-line-characterisation}
holds also for valid functions, once projectors are allowed as in
\Claimref{EBM_implies_to_TEF}. Using this, one can use essentially
the same proof that $\Lambda$-EBM point games implies $\Lambda$-pen
WCF protocol to conclude that $\Lambda$-point games with valid functions
also imply $\Lambda$-pen WCF protocols (with $2n$ rounds of communication).}
\begin{defn}[$\Lambda$-pen TDPG]
\label{def:LambdaPointGameWithValidFunctions} A $\Lambda$-penalty
(time dependent) point game with valid functions is a sequence $(t_{1},\dots t_{n})$
of $n$ functions such that $\{t_{1},t_{3}\dots\}$ (resp. $\{t_{0},t_{2}\dots\}$)
are vertically valid (resp. horizontally valid) such that 
\begin{itemize}
\item $\frac{1}{2}\points{\Lambda,\Lambda+1}+\frac{1}{2}\points{\Lambda+1,\Lambda}+\sum_{i=1}^{n}t_{i}=\points{\beta,\alpha}$
(for some $\alpha,\beta\in[\Lambda,\Lambda+1${]})
\item $\forall j\in\{1,\dots n\}$, $\frac{1}{2}\points{\Lambda,\Lambda+1}+\frac{1}{2}\points{\Lambda+1,\Lambda}+\sum_{i=1}^{n}t_{i}\ge0$. 
\end{itemize}
By the $j\text{th}$ frame (of the point game), we mean $p_{j}:=\frac{1}{2}\points{\Lambda,\Lambda+1}+\frac{1}{2}\points{\Lambda+1,\Lambda}+\sum_{i=1}^{n}t_{i}$
(for $j\in\{0,1,\dots, n\}$ with $p_{0}:=\frac{1}{2}\points{\Lambda,\Lambda+1}+\frac{1}{2}\points{\Lambda+1,\Lambda}$).
The point $\points{\beta,\alpha}$ is referred to as the final point
(of the point game).

{The first condition above simply ensures that one reaches a single
final point while the second condition ensures that each intermediate
frame is valid. }
\end{defn}

\begin{claim}
\label{claim:valid_implies_TEF}Let $t$ be a horizontally valid function.
Then, there is a unitary $U_{(j,2)}$ of the form described in Step 3 ``Alice applies the non-trivial unitary'' (right above \Claimref{EBM_implies_to_TEF}) that satisfies \Eqref{TEFconstraint}.
\end{claim}

{\begin{proof}[Proof sketch.]
 Follows from prior works. See, for instance, Lemma 19 in \cite{arora2024protocolsquantumweakcoin}.\footnote{\cite{arora2024protocolsquantumweakcoin} is a concise, self-contained version, describing results from \cite{Arora2019} and \cite{Arora2019b} while \cite{cryptoeprint:2022/1101} is the comprehensive self-contained version that includes all the details.} %
\end{proof}
}

{We can now state the analogue of \Thmref{LambdaEBMpointgame-implies-LambdaPenWCF}
for $\Lambda$-point games with valid functions.}
\begin{thm}[$\Lambda$-penTDPG $\implies$ $\Lambda$-penWCF protocol]
 Let $(p_{0},\dots, p_{n})$ be the frames of a $\Lambda$-penalty
(time dependent) point game, $(t_{1},\dots t_{n})$, with valid functions
(see \Defref{LambdaPointGameWithValidFunctions}) that has $\points{\beta,\alpha}$
as the final point. Then, there exists a $\Lambda$-penWCF protocol
with cheating rewards $R_{A}^{*}\le\alpha$ and $R_{B}^{*}\le\beta$
which uses $2n$ rounds of communication and $3\cdot\log\left[\left(\max_{j}\left(\mathsf{PointCount}(p_{j})+\mathsf{PointCount}(p_{j+1})\right)+1\right)\right]$
qubits where $\mathsf{PointCount}(p_{j})=\left|\left\{ (x,y):p_{j}(x,y)\neq0\right\} \right|$
is the number of points with non-zero weight. \label{thm:LambdaPenTDPG-implies-Lambda-penWCF}
\end{thm}

{\begin{proof}[Proof sketch.]
Proceed as in \Subsecref{Lambda-penEBM-point-game}, and use \Claimref{valid_implies_TEF}
instead of \Claimref{EBM_implies_to_TEF}. 
\end{proof}
}

\section{$\Lambda$-penalty Time Independent Point Games}\label{sec:Lambda-TIPG}

{We now define $\Lambda$-pen Time Independent Point Games ($\Lambda$-pen
TDPGs).}
\begin{defn}[$\Lambda$-pen TIPG, $\varepsilon$-approx $\Lambda$-pen TIPG] \label{def:penTIPGformal}
 A $\Lambda$-penalty time independent point game ($\Lambda$-pen
TIPG) is specified by two functions $(h,v)$ where $h$ is horizontally
valid and $v$ is vertically satisfying the constraint 
\begin{equation}
h+v=1\cdot\points{\beta,\alpha}-\frac{1}{2}\points{\Lambda,\Lambda+1}-\frac{1}{2}\points{\Lambda+1,\Lambda}\label{eq:Lambda-TIPG-condition}
\end{equation}
for some $\beta,\alpha\in[\Lambda,\Lambda+1]$. We call the point
$\points{\beta,\alpha}$ the final point of the game and call the
set ${\cal S}=\supp(h)\cup\supp(v)\backslash\supp(h+v)$, the set
of intermediate points. 

An $\varepsilon$-approximate $\Lambda$-penalty time independent
point game ($\varepsilon$-approx $\Lambda$-pen TIPG) is defined
exactly as above except that instead of \Eqref{Lambda-TIPG-condition},
we require
\[
\left\Vert h+v-\left(1\cdot\points{\beta,\alpha}-\frac{1}{2}\points{\Lambda,\Lambda+1}-\frac{1}{2}\points{\Lambda+1,\Lambda}\right)\right\Vert _{1}\le\varepsilon
\]
where $\left\Vert f\right\Vert _{1}:=\sum_{x\in\supp(f)}\left|f(x)\right|$.
\end{defn}

{It is easy to see that a $\Lambda$-pen TDPG results in a $\Lambda$-pen
TIPG with the same final point. Now we prove the converse. Note that in the theorem below we use slightly different notation to that in \Thmref{LambdaPenTIPGtoLambdaPenTDPG}. }
\begin{thm}[$\varepsilon$-approx $\Lambda$-pen TIPG $\implies$ $\Lambda$-pen
TDPG]
\label{thm:LambdaPenTIPGtoLambdaPenTDPG} Given an $\varepsilon$-approximate
$\Lambda$-penalty TIPG $(h,v)$ with $h=v^{T}$, final point $(\beta,\alpha)$,
with $m_{1}=\min\left\{ \mincoordinate(h),\mincoordinate(v)\right\} $,
choose any $c_{1}\in\Big(0,\frac{m_{1}^{2}}{(\Lambda+1)\cdot\Lambda}\Big).$
Then, for every $\delta\in(\delta_{\min},1)$ there is a $\Lambda$-penalty
TDPG $(p_{0},\dots p_{n})$ with final point $(\beta+\mathsf{err},\alpha+\mathsf{err})$
where $\mathsf{err}=\sqrt{\delta\cdot(m_{2}-\alpha)(m_{2}-\beta)}$,
$n=10+2/\eta_{2}$. These, in turn are specified by
\begin{itemize}
\item $\delta_{\min}:=(1-\varepsilon_{2})\cdot\frac{c_{3}\varepsilon_{1}}{1+c_{3}\varepsilon_{1}}+\varepsilon_{2}$,
with $c_{3}:=c_{1}^{-1}-1$,
\item $m_{2}:=\max\left\{ \maxcoordinate(h),\tilde{m}_{2}\right\} $ with 
\item $\tilde{m}_{2}:=\min\left\{ (1-w_{1})\left(\frac{1}{(\Lambda+1)}-\frac{w_{1}}{m_{1}}\right)^{-1}:w_{1}\in\{w_{1}^{\pm}\}\right\} $
for 
\[
w_{1}^{\pm}=\frac{\sqrt{8c_{1}\Lambda^{2}(\Lambda+1)^{2}+m_{1}^{2}(8c_{1}\Lambda(\Lambda+1)+1)-8c_{1}\Lambda\left(2\Lambda^{2}+3\Lambda+1\right)m_{1}}\pm m_{1}}{2(\Lambda+1)(m_{1}-\Lambda)},
\]
and finally,
\item $\eta_{2}:=\frac{\deltacatalyst}{\left\Vert h^{-}\right\Vert }\cdot\frac{c_{1}(1-\varepsilon_{1})+\varepsilon_{1}}{(1-\deltacatalyst)}$
with $\deltacatalyst=1-\left((1-\varepsilon_{1})+\varepsilon_{1}/c_{1}\right)\cdot\frac{1-\delta}{1-\varepsilon_{2}}.$
\end{itemize}
Above, $\varepsilon_{1},\varepsilon_{2}\le\varepsilon$ are such that
$s=\left(1-\varepsilon_{1}\right)\sideal+\varepsilon_{1}\cdot\serror$
and $e=(1-\varepsilon_{2})\eideal+\varepsilon_{2}\cdot\eerror$ where
$\left\Vert \serror\right\Vert _{1},\left\Vert \eerror\right\Vert _{1}=1$. The number of points in $p_{i}$ is at most the number of points in $(h,v)$, i.e. $|\supp(p_{i})|\le |\supp\{ h \} \cup \supp\{ v \} |$ for all $i\in \{0,\cdots , n\}$.
\end{thm}

{Compared to the analogous result for WCF, there are two key differences. 
\begin{enumerate}
\item In WCF, the point games could not involve points with any coordinate below
$0$. And the original proofs relied on this fact. This fact itself
translates, in the cheat penalty setting, to the constraint that the
points cannot have coordinates lower than $\Lambda$. However, this
is an unnecessary constraint and indeed, the point games we consider,
do in fact involve points with coordinates lower than $\Lambda$.
We thus generalise the proof slightly to allow points with coordinates
below $\Lambda$.
\item The other difference is that our numerical algorithm produces approximate
penalty TIPGs. For readers familiar with the proof for WCF, this involves
the use of catalyst points and absorbing them in the end. This is
very similar to having residual points from the approximation that
need to be absorbed in the end into the final point. We therefore
combine steps to obtain the final bound on error and communication
complexity. 
\end{enumerate}
Before we start the proof, we note that combining \Thmref{LambdaPenTDPG-implies-Lambda-penWCF}
and \Thmref{LambdaPenTIPGtoLambdaPenTDPG}, one immediately obtains
a $\Lambda$-penalty WCF protocol starting from an $\varepsilon$-approximate
$\Lambda$-penalty TDPG, together with bounds on the number of qubits
and the number of rounds of communication. }

\subsection{Establishing that $\Lambda$-pen TIPG $\protect\implies$ $\Lambda$-pen
TDPGs assuming intermediate lemmas}

{We later consider sequences of valid transitions and to this end,
it helps to introduce the notion of transitively valid functions. }
\begin{defn}[transitively valid]
 Consider two functions $g,h:\Rnonneg\times\Rnonneg\to\Rnonneg$
with finite support. The transition $g\to h$ is \emph{transitively
valid (with $n$ steps)} if there is a sequence of finitely supported
functions $p_{i}:\Rnonneg\times\Rnonneg\to\Rnonneg$ such that $g\to p_{1},p_{1}\to p_{2},\cdots , p_{n-1}\to p_{n},p_{n}\to h$
are all valid transitions (see \Defref{validFunctions}). 
\end{defn}

{To prove \Thmref{LambdaPenTIPGtoLambdaPenTDPG}, we observe that the
following statements are enough---proofs of which we get back to
later. We use the following notation for the remainder of this section.}
\begin{notation}
\label{nota:TIPG_maxcoordinate_etc}Let $(h,v)$ be a $\varepsilon$-approx
$\Lambda$-pen TIPG where $h=v^{T}$ (i.e., $h(x,y)=v(y,x)$ for all
$x,y$). Let $h+v=e-s$ where $e$ and $s$ have disjoint support
and both $e,s\ge0$. 

For any valid function, say $h$, we use $h=h^{+}-h^{-}$ to denote
the positive and negative parts of $h$, i.e., $h^{+}$ and $h^{-}$
have disjoint support and $h^{+},h^{-}\ge0$ are both non-negative
at all coordinates. 
Define $\sideal:=\frac{1}{2}\points{\Lambda,\Lambda+1}+\frac{1}{2}\points{\Lambda+1,\Lambda}$
and $\eideal:=\points{\beta,\alpha}$ for some $\beta,\alpha\in[\Lambda,\Lambda+1]$.
A ``configuration'' or a ``frame'' is any bivariate function $f:\Rnonneg\times\Rnonneg\to\Rnonneg$
with finite support. 
Finally, $\maxcoordinate(f)$ of a frame is the highest $x$ or $y$
coordinate that any point with non-zero weight has, i.e. $\maxcoordinate(f):=\max\left\{ \{x:\exists y'\text{ s.t. }f(x,y')>0\}\cup\{y:\exists x'\text{ s.t.}f(x',y)>0\}\right\} $.
Similarly, $\mincoordinate(f)$ is the lowest $\max\{x,y\}$ coordinate\footnote{Consider the $\Lambda=0$ case. Without the $\max$, the answer for
$h$ or $v$ would be $0$ because they involve points on the axes.} with non-zero weight in $f$, i.e. $\mincoordinate(f):=\min\{\max\{x,y\}:f(x,y)>0\}$. 
\end{notation}

{The motivation for introducing $\maxcoordinate$ and $\mincoordinate$
notation will become clear shortly. We start claiming that $s\to e$
becomes transitively valid if $h^{-}$ is already present in the ``background''. }
\begin{lem}
\label{lem:TIPG_lem1_s_plus_hminus}It holds that $s+h^{-}\to s+h^{+}$
and $s+h^{+}\to e+h^{-}$ are both valid transitions. 
\end{lem}

{Obviously, one cannot simply assume the starting frame is $s+h^{-}$.
The following basically allows one to ensure that $s+\gamma h^{-}\to e+\gamma h^{-}$
is also transitively valid, where the number of steps depends on the
scalar $0<\gamma<1$. }
\begin{lem}
\label{lem:TIPG_lem2_addZeta_or_add_deltaZeta_in_many_steps}Let $0<\gamma<1$
and $\xi$ be any frame. If $s+\xi\to e+\xi$ is transitively valid
with $n_{1}$ steps then $s+\gamma\xi\to\delta_{1}s+(1-\delta_{1})e+\gamma\xi$
is transitively valid with $n_{2}:=n_{1}\cdot\left\lfloor \frac{1}{\gamma}\right\rfloor $
steps where $\delta_{1}=1-\gamma\cdot\left\lfloor 1/\gamma\right\rfloor $.
When $1/\gamma$ is some natural number, we get $s+\gamma\xi\to e+\gamma\xi$
is transitively valid with $n_{2}=n_{1}/\gamma$ steps. 
\end{lem}

{That handles a big part of showing that a $\Lambda$-pen TIPG can
be converted into a $\Lambda$-pen TDPG. We need to address a few
things: (1) how to produce and absorb the $\gamma h^{-}$ frame and
(2) how to handle the approximation $\varepsilon$. 

First, we rewrite the approximate starting and ending frames.}

\begin{claim}
\label{claim:h=00003DvT}If $h=v^{T}$ and $\left\Vert h+v-(\eideal-\sideal)\right\Vert _{1}\le\varepsilon<1$,
then $s=\left(1-\varepsilon_{1}\right)\sideal+\varepsilon_{1}\cdot\serror$
and $e=(1-\varepsilon_{2})\eideal+\varepsilon_{2}\cdot\eerror$ where
$\left\Vert \serror\right\Vert _{1},\left\Vert \eerror\right\Vert _{1}=1$
and $\varepsilon_{1}$ and $\varepsilon_{2}$ are both at most $\varepsilon$. 
\end{claim}

{We now show that one can go from 
\begin{equation}
\sideal\to(1-\delta)\eideal+\delta\cdot\points{m_{2},m_{2}}\label{eq:s_ideal_to_eideal_plus_rest}
\end{equation}
 but with some constraints. First, a parameter $c_{1}$ must be selected
which must be in some range specified by the minimum coordinate of
interest $m_{1}$ and the cheat penalty $\Lambda$. Now, this $c_{1}$
parameter specifies $\delta_{\min}$ which is the smallest $\delta$
can be and also $m_{2}$ which specifies a large coordinate (explained
shortly). The proof that \Eqref{s_ideal_to_eideal_plus_rest} is transitively
valid essentially starts by using a small weight from $\sideal$ and
producing (a) $\serror$ together with (b) a small amount of $h^{-}$.
In this process, a residual point at $\points{m_{2},m_{2}}$ with
a small weight is also produced. Then using \lemref{TIPG_lem1_s_plus_hminus}
and \Lemref{TIPG_lem2_addZeta_or_add_deltaZeta_in_many_steps}, the
proof allows one to convert $s+\gamma h^{-}\to e+\gamma h^{-}$. Finally,
with some raises, one obtains the desired final configuration/frame,
i.e. $(1-\delta)\eideal+\delta\cdot\points{m_{2},m_{2}}$. The subtlety
in all this is, of course, making everything quantitative and keeping
track of the number of intermediate transitions. }
\begin{lem}
\label{lem:from_s_ideal_to_e}Suppose the premise of \Claimref{h=00003DvT}
holds and let 
\[
m_{1}=\min\{\mincoordinate(\serror),\mincoordinate(h^{-})\},\quad\text{and}\quad c_{1}\in\Big[0,\frac{m_{1}^{2}}{(\Lambda+1)\cdot\Lambda}\Big).
\]
It holds that for every $\delta\in(\delta_{\min},1)$, the transition
\[
\sideal\to(1-\delta)\eideal+\delta\cdot\points{m_{2},m_{2}}
\]
is transitively valid with $6+2/\eta_{2}$ steps (assuming $1/\eta_{2}$
is an integer), where we define
\begin{itemize}
\item $\delta_{\min}:=(1-\varepsilon_{2})\cdot\frac{c_{3}\varepsilon_{1}}{1+c_{3}\varepsilon_{1}}+\varepsilon_{2}$,
with $c_{3}:=c_{1}^{-1}-1$,
\item $m_{2}:=\max\left\{ \maxcoordinate(h),\tilde{m}_{2}\right\} $ with 
\item $\tilde{m}_{2}:=\min\left\{ (1-w_{1})\left(\frac{1}{(\Lambda+1)}-\frac{w_{1}}{m_{1}}\right)^{-1}:w_{1}\in\{w_{1}^{\pm}\}\right\} $
for 
\[
w_{1}^{\pm}=\frac{\sqrt{8c_{1}\Lambda^{2}(\Lambda+1)^{2}+m_{1}^{2}(8c_{1}\Lambda(\Lambda+1)+1)-8c_{1}\Lambda\left(2\Lambda^{2}+3\Lambda+1\right)m_{1}}\pm m_{1}}{2(\Lambda+1)(m_{1}-\Lambda)},
\]
and finally,
\item $\eta_{2}:=\frac{\deltacatalyst}{\left\Vert h^{-}\right\Vert }\cdot\frac{c_{1}(1-\varepsilon_{1})+\varepsilon_{1}}{(1-\deltacatalyst)}$
with $\deltacatalyst=1-\left((1-\varepsilon_{1})+\varepsilon_{1}/c_{1}\right)\cdot\frac{1-\delta}{1-\varepsilon_{2}}.$
\end{itemize}
\end{lem}

{It remains to show that the point $\points{m_{2},m_{2}}$ can be absorbed.
The following lemma says that for any desired deviation $\epsilon>0$
from the ideal final frame $\points{\beta,\alpha}$, there is a $\delta_{\max}$
below which for any $\delta$, $(1-\delta)\eideal+\delta\cdot\points{m_{2},m_{2}}\to\points{\beta+\epsilon,\alpha+\epsilon}$
is transitively valid. In the case where $\delta_{\min}=0$ (i.e.
we are in the exact case where $\varepsilon_{1}=\varepsilon_{2}=0$),
it means that one can always find a $\delta<\delta_{\max}$ and thus
obtain protocols with end points getting arbitrarily close to $\points{\beta,\alpha}$---at
the expense of increased communication (scaling roughly as $1/\delta$). }
\begin{lem}
\label{lem:FinalMerge}Recall that $\eideal=\points{\beta,\alpha}$
and fix any $\points{m_{2},m_{2}}$ (such that $m_{2}\ge\max\{\beta,\alpha\}$).
For every $\epsilon>0$, there is a $\delta_{\max}:=\frac{\epsilon^{2}}{(m_{2}-\alpha)(m_{2}-\beta)}$
such that for all $\delta<\delta_{\max}$, the transition 
\[
(1-\delta)\eideal+\delta\cdot\points{m_{2},m_{2}}\to\points{\beta+\epsilon,\alpha+\epsilon}
\]
is transitively valid with 4 steps. 
\end{lem}

{However, when $\delta_{\min}>0$, one cannot get arbitrarily close
to $\points{\beta,\alpha}$. In that case, for every $c_{1}$, one
can compute $\delta_{\min}(c_{1})$ and $m_{2}(c_{1})$. Given $m_{2}$,
and $\epsilon$, using the lemma above, one can also compute $\delta_{\max}(\epsilon,c_{1})$.
The smallest $\epsilon$ for which $\delta_{\min}(c_{1})\le\delta_{\max}(\epsilon,c_{1})$
allows one to choose $\delta_{\min}(c_{1})\le\delta\le\delta_{\max}(\epsilon,c_{1})$
which specifies a protocol with $6+2/\eta_{2}(c_{1},\delta)$ rounds
of interaction and $\epsilon=\delta_{\max}\cdot(m_{2}-\beta)$ bias.}

{\begin{proof}[Proof of \Thmref{LambdaPenTIPGtoLambdaPenTDPG}.]
 The proof of \Lemref{from_s_ideal_to_e} uses \Lemref{TIPG_lem1_s_plus_hminus},
\Lemref{TIPG_lem2_addZeta_or_add_deltaZeta_in_many_steps} and \Claimref{h=00003DvT}
as we alluded to above and prove later. The proof of \Lemref{FinalMerge}
is not too involved and is also deferred (follows from prior works).
Assuming \Lemref{from_s_ideal_to_e} and \Lemref{FinalMerge}, the
proof of \Thmref{LambdaPenTIPGtoLambdaPenTDPG} is immediate. 
\end{proof}
}

\subsection{Proving the intermediate lemmas}

{Before we get to the more interesting proofs, let us establish \Claimref{h=00003DvT}
which basically says that $\left\Vert h+v-(\eideal-\sideal)\right\Vert _{1}\le\varepsilon$
implies that one can write $s=(1-\varepsilon_{1})\sideal+\varepsilon_{1}\cdot\serror$
and $e=(1-\varepsilon_{2})\eideal+\varepsilon_{2}\cdot\eerror$ where
the functions on the right are non-negative and normalised, and $\varepsilon_{1}$
and $\varepsilon_{2}$ are at most $\varepsilon$.}

{\begin{proof}[Proof of \Claimref{h=00003DvT}]
 We start by recalling that $\supp(e)\cap\supp(s)=\emptyset$. Further,
it also holds that $\supp(e)\supseteq\supp(\eideal)$ because otherwise,
$\left\Vert e-s-(\eideal-\sideal)\right\Vert _{1}\ge\left\Vert \eideal\right\Vert =1$
which violates our premise that $\varepsilon<1$. Similarly, one can
argue that $\supp(s)\supseteq\supp(\sideal)$. Thus, one can write
$\left\Vert e-s-(\eerror-\sideal)\right\Vert =\left\Vert e-\eideal\right\Vert _{1}+\left\Vert s-\sideal\right\Vert _{1}\le\varepsilon$.
This means that both $\left\Vert e-\eideal\right\Vert _{1}\le\varepsilon$
and $\left\Vert s-\sideal\right\Vert _{1}\le\varepsilon$. Finally,
using these inequalities and since $h=v^{T}$, we can write $s=(1-\varepsilon_{1})\sideal+\varepsilon_{1}\cdot\serror$
where $\varepsilon_{1}\le\varepsilon$ and $\supp(\sideal)\cap\supp(\serror)=\emptyset$
(if no such $\varepsilon_{1}$ existed, then the inequality could
not hold). Since $\left\Vert s\right\Vert _{1}=1$, and $\left\Vert \sideal\right\Vert =1$,
we have that $\left\Vert \serror\right\Vert =1$. One can similarly
establish that $e=(1-\varepsilon_{2})\eideal+\varepsilon_{2}\eerror$
where $\supp(\eideal)\cap\supp(\eerror)=\emptyset$ and $\varepsilon_{2}\le\varepsilon$. 
\end{proof}
}

{As a warm-up, we state and prove a simple observation that will be
useful later.}
\begin{lem}
If the transition $g'\to h'$ is transitively valid with $n$ steps
and $\zeta:\Rnonneg\times\Rnonneg\to\Rnonneg$ is a non-negative function
with finite support, then the transition $\delta\cdot g'+\zeta\to\delta\cdot h'+\zeta$
is also transitively valid with $n$ steps for all $\delta>0$. \label{lem:addToValidStillValid}
\end{lem}

{\begin{proof}
It suffices to show that the statement holds for valid line transitions
(because a valid transition is a valid line transition along each
$x$ (or $y$) coordinate). Given that $\ell\to r$ is a valid line
transition, it is immediate that $\delta\cdot\ell+\xi\to\delta\cdot r+\xi$
for any $\xi:\Rnonneg\to\Rnonneg$ finitely supported function. This
is because $\delta\cdot(r-\ell)$ is a valid function if $(r-\ell)$
is a valid function, which is the case because $\ell\to r$ is a valid
transition. The number of steps are preserved because one can apply
this reasoning to each step of $g'\to p_{1},p_{1}\to p_{2}\dots p_{n}\to h'$
to obtain $\delta\cdot g'+\zeta\to\delta\cdot p_{1}+\zeta,\dots\delta\cdot p_{n}+\zeta\to\delta\cdot h'+\zeta$. 
\end{proof}
}

{We now establish \Lemref{TIPG_lem1_s_plus_hminus} which says that
$s+h^{-}\to e+h^{-}$ is transitively valid with $2$ steps.}

{\begin{proof}[Proof of \Lemref{TIPG_lem1_s_plus_hminus}]
 It is clear that $s+h^{-}\to s+h^{+}$ is valid because $h$ is
a valid function and we can add $s$ to the valid transition $h^{-}\to h^{+}$
while preserving validity as in \Lemref{addToValidStillValid}. Showing
$s+h^{+}\to e+h^{+}$ requires some work. One can rewrite $h+v=s-e$
as $h^{+}-h^{-}+v^{+}-v^{-}=e-s$ which in turn can be rewritten as
\[
h^{+}+s=\underbrace{e+h^{-}-v^{+}}_{\xi}+v^{-}.
\]
We prove shortly that $\xi\ge0$ but assume this is the case for the
moment. Then, since $v^{-}\to v^{+}$ is a valid transition, using
\Lemref{addToValidStillValid}, we have that $\xi+v^{-}\to\xi+v^{+}=e+h^{-}\cancel{-v^{+}+v^{+}}$
is also a valid transition. Thes establishes that $s+h^{+}\to e+h^{-}$
is a valid transition, assuming $\xi\ge0$ which we now prove. Observe
that $\xi$ can only be negative at $\supp(v^{+})$. Since $\supp(v^{-})\cap\supp(v^{+})=\emptyset$,
it follows that $\xi\ge0\iff\xi+v^{-}\ge0$. But recall that $\xi+v^{-}=h^{+}+s$
which is manifestly non-negative. 
\end{proof}
}

{Next, we establish \Lemref{TIPG_lem2_addZeta_or_add_deltaZeta_in_many_steps}
which says, in particular, that if $s+h^{-}\to e+h^{-}$ is transitively
valid with $n_{1}$ steps, then $s+\gamma h^{-}\to e+\gamma h^{-}$
is also transitively valid with $n_{2}:=n_{1}/\gamma$ steps. }

{\begin{proof}[Proof of \Lemref{TIPG_lem2_addZeta_or_add_deltaZeta_in_many_steps}]
 We proceed as follows 
\begin{align*}
s+\gamma\xi & =(1-\gamma)s+\gamma(s+\xi)\\
 & \to(1-\gamma)s+\quad\quad\quad+\gamma(e+\xi)\\
 & =(1-2\gamma)s+\gamma(s+\xi)+\gamma e\\
 & \to(1-2\gamma)s+\gamma(s+\xi)+2\gamma e\\
 & \to(1-3\gamma)s+\gamma(s+\xi)+3\gamma e\\
 & \vdots\\
 & \to(1-m\gamma)s+\quad\quad\quad+m\gamma e+\gamma\xi
\end{align*}
where $m=\left\lfloor 1/\gamma\right\rfloor $ so that $0\le1-m\gamma\le\gamma$
and $m\gamma\ge1-\gamma$. This corresponds to $n_{1}\cdot m$ valid
transitions thus $s+\gamma\xi\to\delta_{1}s+(1-\delta_{1})e+\gamma\xi$
is transitively valid. 
\end{proof}
}

{The statements so far, did not really depend on the actual definitions
of $h,v$ (that also specify $s,e$). To prove the subsequent statements,
we assume we are starting with $\sideal$ and want to end up with
a single point. Let us start with a simple claim. }
\begin{claim}
\label{claim:splitToGetTo_m1}For every coordinate $m_{1}\in(0,\Lambda]$,
and any weight $c_{1}\in\Big[0,\frac{m_{1}^{2}}{(\Lambda+1)\cdot\Lambda}\Big)$,
the transition 
\[
\sideal\to(1-\delta_{1})\sideal+\delta_{1}\Big(c_{1}\points{m_{1},m_{1}}+\underbrace{(1-c_{1})}_{=:c_{2}}\points{m_{2},m_{2}}\Big)
\]
is transitively valid in $2$ steps and $m_{2}:=\min\left\{ (1-w_{1})\left(\frac{1}{(\Lambda+1)}-\frac{w_{1}}{m_{1}}\right)^{-1}:w_{1}\in\{w_{1}^{\pm}\}\right\} $
where $w_{1}^{\pm}=\frac{\sqrt{8c_{1}\Lambda^{2}(\Lambda+1)^{2}+m_{1}^{2}(8c_{1}\Lambda(\Lambda+1)+1)-8c_{1}\Lambda\left(2\Lambda^{2}+3\Lambda+1\right)m_{1}}\pm m_{1}}{2(\Lambda+1)(m_{1}-\Lambda)}.$
\end{claim}

{\begin{proof}[Proof of \Claimref{splitToGetTo_m1}]
 The idea is elementary. Simply perform a horizontal split and then
a vertical split starting from, say, $\points{\Lambda+1,\Lambda}$.
The horizontal split 
\begin{align*}
\sideal & =\frac{1}{2}\points{\Lambda+1,\Lambda}+\frac{1}{2}\points{\Lambda,\Lambda+1}\\
 & \to\frac{1}{2}w_{1}\points{m_{1},\Lambda}+\frac{1}{2}\left(1-w_{1}\right)\points{m_{2}',\Lambda}+\frac{1}{2}\points{\Lambda,\Lambda+1}
\end{align*}
 gives the following constraint 
\[
\frac{w_{1}}{m_{1}}+\frac{1-w_{1}}{m_{2}'}\le\frac{1}{(\Lambda+1)}.
\]
Here we treat $0\le w_{1}\le1$ as a free variable but the constraint
above requires, $w_{1}\le\frac{m_{1}}{(\Lambda+1)}$ and we have 
\[
m_{2}':=(1-w_{1})\left(\frac{1}{(\Lambda+1)}-\frac{w_{1}}{m_{1}}\right)^{-1}.
\]
(Note that as $w_{1}$ approaches $m_{1}/(\Lambda+1)$, $m_{2}'$
approaches infinty).

Now, we split vertically 
\begin{align*}
 & \frac{1}{2}w_{1}\points{m_{1},\Lambda}+\frac{1}{2}\left(1-w_{1}\right)\points{m_{2}',\Lambda}+\frac{1}{2}\points{\Lambda,\Lambda+1}\\
\to & \frac{1}{2}w_{1}\left(w_{2}\points{m_{1},m_{1}}+(1-w_{2})\points{m_{1},m_{2}''}\right)+\frac{1}{2}\left(1-w_{1}\right)\points{m_{2}',\Lambda}+\frac{1}{2}\points{\Lambda,\Lambda+1}
\end{align*}
which gives the following constraint
\[
\frac{w_{2}}{m_{1}}+\frac{(1-w_{2})}{m_{2}''}\le\frac{1}{\Lambda}.
\]
Treating $0\le w_{2}\le1$ also as a free variable, we get $w_{2}\le\frac{m_{1}}{\Lambda}$
with 
\[
m_{2}'':=(1-w_{2})\left(\frac{1}{\Lambda}-\frac{w_{2}}{m_{1}}\right)^{-1}.
\]

We demand that $m_{2}''=m_{2}'=:m_{2}$ to obtain a relation between
$w_{1}$ and $w_{2}$, i.e. 
\begin{align*}
 & \frac{(1-w_{2})}{(1-w_{1})}=\frac{\left(\frac{1}{\Lambda}-\frac{w_{2}}{m_{1}}\right)}{\left(\frac{1}{\Lambda+1}-\frac{w_{1}}{m_{1}}\right)}\\
\iff & \frac{(1-w_{2})}{(1-w_{1})}=\frac{\left(\frac{m_{1}-\Lambda w_{2}}{\Lambda}\right)}{\left(\frac{m_{1}-(\Lambda+1)w_{1}}{(\Lambda+1)}\right)}\\
\iff & \frac{\Lambda(1-w_{2})}{(m_{1}-\Lambda w_{2})}=\frac{(\Lambda+1)(1-w_{1})}{(m_{1}-(\Lambda+1)w_{1})}.
\end{align*}
We also have another relation between $w_{1}$ and $w_{2}$ which
is $c_{1}=\frac{1}{2}w_{1}w_{2}$. Substituting $w_{2}=\frac{2c_{1}}{w_{1}}$
in the expression above, we can solve for $w_{1}$in terms of $m_{1},\Lambda$
and $c_{1}$ 
\begin{align*}
 & \frac{\Lambda(1-2c_{1}/w_{1})}{(m_{1}-2\Lambda c_{1}/w_{1})}=\frac{(\Lambda+1)(1-w_{1})}{(m_{1}-(\Lambda+1)w_{1})}\\
\iff & \Lambda(1-2c_{1}/w_{1})(m_{1}-(\Lambda+1)w_{1})=(\Lambda+1)(1-w_{1})(m_{1}-2\Lambda c_{1}/w_{1})
\end{align*}
which yields a quadratic that can be solved to obtain 
\[
w_{1}^{\pm}=\frac{\sqrt{8c_{1}\Lambda^{2}(\Lambda+1)^{2}+m_{1}^{2}(8c_{1}\Lambda(\Lambda+1)+1)-8c_{1}\Lambda\left(2\Lambda^{2}+3\Lambda+1\right)m_{1}}\pm m_{1}}{2(\Lambda+1)(m_{1}-\Lambda)}
\]
and that in turn yields $m_{2}=\min\left\{ (1-w_{1})\left(\frac{1}{(\Lambda+1)}-\frac{w_{1}}{m_{1}}\right)^{-1}:w_{1}\in\{w_{1}^{\pm}\}\right\} .$
\end{proof}
}

{We now establish \Lemref{from_s_ideal_to_e} which shows that one
can start from $\sideal$ and get to $(1-\delta)\eideal+\delta\cdot\points{m_{2},m_{2}}$
for $\delta\ge\delta_{\min}$ where $\delta_{\min}$ vanishes as $\varepsilon_{1},\varepsilon_{2}$
vanish (i.e. if $s=\sideal$ and $e=\eideal$).}

{\begin{proof}[Proof of \Lemref{from_s_ideal_to_e}]
Our goal is to establish that, for every $1>\delta>0$, $\sideal\to(1-\delta)\eideal+\delta\cdot\erest$
is transitively valid and to give a bound on the number steps it takes. 

Let $1>\deltacatalyst>0$ be a free parameter. Let $\deltasfix>0$
be another parameter which is fixed by $\deltacatalyst$ (and
other parameters). We can now easily see the following are transitively
valid transitions.

\begin{align}
\sideal & \to\quad(1-\deltasfix-\deltacatalyst)\sideal+\deltasfix c_{1}\points{m_{1},m_{1}}+\deltacatalyst c_{1}\points{m_{1},m_{1}}\nonumber \\
 & \quad+(\deltasfix+\deltacatalyst)c_{2}\points{m_{2},m_{2}} & \text{follows from \Claimref{splitToGetTo_m1}}\nonumber \\
 & \to\quad(1-\deltasfix-\deltacatalyst)\sideal+\deltasfix c_{1}\serror+\deltacatalyst c_{1}\frac{h^{-}}{\left\Vert h^{-}\right\Vert }\nonumber \\
 & \quad+(\deltasfix+\deltacatalyst)c_{2}\points{m_{2},m_{2}}\label{eq:createdAllTheStates}
\end{align}
where the second transition follows from the fact that $m_1$ is the minimum coordinate appearing in the point game. We want the coefficient of $\sideal$ and $\serror$ to have
the ratio $(1-\deltasfix-\deltacatalyst)/(\deltasfix c_{1})=(1-\varepsilon_{1})/\varepsilon_{1}$
which yields 
\[
\varepsilon_{1}-\varepsilon_{1}\deltasfix-\varepsilon_{1}\deltacatalyst=\deltasfix c_{1}(1-\varepsilon_{1})\iff\deltasfix=\frac{\varepsilon_{1}(1-\deltacatalyst)}{c_{1}(1-\varepsilon_{1})+\varepsilon_{1}}
\]
we therefore have 
\[
\deltasfix=\begin{cases}
\frac{\left(1-\deltacatalyst\right)}{c_{1}\left(\frac{1}{\varepsilon_{1}}-1\right)+1} & \varepsilon_{1}>0\\
0 & \varepsilon_{1}=0.
\end{cases}
\]
To proceed, it would be easier to rewrite \Eqref{createdAllTheStates}
as follows, where we solve for $\eta_{1},\eta_{2},\eta_{3}$ below.
We therefore impose that

\begin{align*}
\text{\Eqref{createdAllTheStates}} & =\quad(1-\eta_{1})\left((1-\varepsilon_{1})\sideal+\varepsilon_{1}\serror+\eta_{2}h^{-}\right)+\eta_{3}\points{m_{2},m_{2}} & \text{}
\end{align*}
where $(1-\eta_{1})\varepsilon_{1}=\deltasfix c_{1}$ which yields
\[
\eta_{1}:=1-\frac{\deltasfix c_{1}}{\varepsilon_{1}}=1-\frac{c_{1}(1-\deltacatalyst)}{c_{1}(1-\varepsilon_{1})+\varepsilon_{1}}
\]
and $(1-\eta_{1})\eta_{2}=\frac{\deltacatalyst c_{1}}{\left\Vert h^{-}\right\Vert }$
which yields 
\begin{equation}
\eta_{2}:=\frac{\deltacatalyst\cancel{c_{1}}}{\left\Vert h^{-}\right\Vert }\cdot\frac{\varepsilon_{1}}{\deltasfix\cancel{c_{1}}}=\frac{\deltacatalyst}{\left\Vert h^{-}\right\Vert }\cdot\frac{c_{1}(1-\varepsilon_{1})+\varepsilon_{1}}{(1-\deltacatalyst)}\label{eq:eta2_using_delta_ctl}
\end{equation}
and finally $\eta_{3}:=1-(1-\eta_{1})\left(1+\eta_{2}\left\Vert h^{-}\right\Vert \right)$
from normalisation. Note that all of these are fixed by $\deltacatalyst$.

Proceeding with this change of variables, we have that the following
transitions are transitively valid (we count the steps at the end).
\begin{align*}
\sideal & \to(1-\eta_{1})\left((1-\varepsilon_{1})\sideal+\varepsilon_{1}\serror+\eta_{2}h^{-}\right)+\eta_{3}\points{m_{2},m_{2}}\\
 & =(1-\eta_{1})\left(s+\eta_{2}h^{-}\right)+\eta_{3}\points{m_{2},m_{2}}\\
 & \to(1-\eta_{1})\left(e+\eta_{2}h^{-}\right)+\eta_{3}\points{m_{2},m_{2}} & \text{in \ensuremath{1/\eta_{2}} steps}\\
 & =(1-\eta_{1})\left((1-\varepsilon_{2})\eideal+\varepsilon_{2}\eerror+\eta_{2}h^{-}\right)+\eta_{3}\points{m_{2},m_{2}}\\
 & \to(1-\eta_{1})\left((1-\varepsilon_{2})\eideal+(\varepsilon_{2}+\eta_{2}\left\Vert h^{-}\right\Vert )\points{m_{2},m_{2}}\right)+\eta_{3}\points{m_{2},m_{2}}\\
 & =(1-\delta)\eideal+\delta\points{m_{2},m_{2}}
\end{align*}
where $\delta=\eta_{1}+\varepsilon_{2}-\varepsilon_{2}\eta_{1}=(1-\varepsilon_{2})\eta_{1}+\varepsilon_{2}$.
Now, 
\[
\eta_{1}\ge1-\frac{1}{(1-\varepsilon_{1})+\varepsilon_{1}/c_{1}}=\frac{\left(1-\varepsilon_{1}\right)+\varepsilon_{1}/c_{1}-1}{(1-\varepsilon_{1})+\varepsilon_{1}/c_{1}}=\frac{\left(\frac{1}{c_{1}}-1\right)\varepsilon_{1}}{1+\left(\frac{1}{c_{1}}-1\right)\varepsilon_{1}}=\frac{c_{3}\varepsilon_{1}}{1+c_{3}\varepsilon_{1}}
\]
(where $c_{3}:=c_{1}^{-1}-1$) which is saturated as $\deltacatalyst$
goes to zero, and so 
\[
\delta\ge(1-\varepsilon_{2})\cdot\frac{c_{3}\varepsilon_{1}}{1+c_{3}\varepsilon_{1}}+\varepsilon_{2}.
\]
As a sanity check, note that when $\varepsilon_{1}=\varepsilon_{2}=0$,
the bound just requires $\delta\ge0$ which basically means that we
can choose the weight of the catalyst to be arbitrarily small to ensure
$\delta$ is arbitrarily small, in the exact case. For approximate
penalty TIPGs, we cannot go lower than the bound above. For small
enough $\varepsilon_{1}$ and $\varepsilon_{2}$, however, one should
still be able to obtain low bias. We end by writing an expression
for $\deltacatalyst$ in terms of $\delta$ as 
\begin{align*}
 &  & \delta= & (1-\varepsilon_{2})\left(1-\frac{(1-\deltacatalyst)}{(1-\varepsilon_{1})+\varepsilon_{1}/c_{1}}\right)+\varepsilon_{2}\\
\iff &  & \frac{(1-\deltacatalyst)}{(1-\varepsilon_{1})+\varepsilon_{1}/c_{1}}= & 1-\frac{\delta-\varepsilon_{2}}{1-\varepsilon_{2}}\\
\iff &  & \frac{(1-\deltacatalyst)}{(1-\varepsilon_{1})+\varepsilon_{1}/c_{1}}= & \frac{1\cancel{-\varepsilon_{2}}-\delta\cancel{+\varepsilon_{2}}}{1-\varepsilon_{2}}\\
\iff &  & (1-\deltacatalyst)= & \left((1-\varepsilon_{1})+\varepsilon_{1}/c_{1}\right)\cdot\frac{1-\delta}{1-\varepsilon_{2}}\\
\iff &  & \deltacatalyst & =1-\left((1-\varepsilon_{1})+\varepsilon_{1}/c_{1}\right)\cdot\frac{1-\delta}{1-\varepsilon_{2}}.
\end{align*}
Finally, $\eta_{2}$ can be computed using $\deltacatalyst$ using
\Eqref{eta2_using_delta_ctl} and $1/\eta_{2}$ gives (roughly) the
number of steps needed in the catalyst step.

It remains to count the total number of steps.
\begin{itemize}
\item $4$ steps: 2 for for setting up $\points{m_{1},m_{1}}$, and $2$
for turning them into $\serror$ and $h^{-}$.
\item $2/\eta_{2}$ steps: the catalyst step (for simplicity, we assume
$1/\eta$ is an integer)
\item 2 steps: final two raises to have all remaining points be at $\points{m_{2},m_{2}}$.

Thus, the total number of steps turns out to be $6+2/\eta_{2}$. 
\end{itemize}
\end{proof}
}

{It remains to establish \Lemref{FinalMerge} and its proof is essentially
the same as that in the standard WCF setting. We restate it here for
convenience.}

{\begin{proof}[Proof of \Lemref{FinalMerge}]
Define $\delta$ and $\delta'$ to be such that the following two
merges are valid (we use $m$ instead of $m_{2}$ for simplicity)
\begin{align*}
\delta'\points{m,\alpha}+\delta\points{m,m} & \to(\delta+\delta')\points{m,\alpha+\epsilon}\\
(1-\delta-\delta')\points{\beta,\alpha+\epsilon}+(\delta+\delta')\points{m,\alpha+\epsilon} & \to\points{\beta+\epsilon,\alpha+\epsilon}.
\end{align*}
The merge conditions are 
\begin{align*}
\delta'\alpha+\delta m & =(\delta+\delta')(\alpha+\epsilon)\\
(1-\delta-\delta')\beta+(\delta+\delta')m & =\beta+\epsilon
\end{align*}
using which one can solve for $\delta$ and $\delta'$ as 
\begin{align*}
\delta & =\frac{\epsilon^{2}}{(m-\beta)(m-\alpha)}\\
\delta' & =\frac{\epsilon}{(m-\beta)}\left(1-\frac{\epsilon}{m-\alpha}\right).
\end{align*}
These can now be combined to see that the following transitions are
valid 
\begin{align*}
(1-\delta)\points{\beta,\alpha}+\delta\points{m,m} & \to(1-\delta-\delta')\points{\beta,\alpha}+\delta'\points{m,\alpha}+\delta\points{m,m} & \text{raise}\\
 & \to(1-\delta-\delta')\points{\beta,\alpha}+(\delta'+\delta)\points{m,\alpha+\epsilon} & \text{using the first merge}\\
 & \to(1-\delta-\delta')\points{\beta,\alpha+\epsilon}+(\delta'+\delta)\points{m,\alpha+\epsilon} & \text{raise}\\
 & \to\points{\beta+\epsilon,\alpha+\epsilon} & \text{using the second merge}.
\end{align*}
Note that this takes four valid transitions. Now, note that one can
set $\delta_{\max}=\frac{\epsilon^{2}}{(m-\beta)(m-\alpha)}$ and
run the argument above for any $\delta\le\delta_{\max}$ by first
raising to ensure the starting frame is $(1-\delta_{\max})\points{\beta,\alpha}+\delta_{\max}\points{m,m}$
and one can transitively go to $\points{\beta+\epsilon,\alpha+\epsilon}$
in four steps. 
\end{proof}
}

\section{An Algorithm for Finding Approximate TIPGs}

\label{sec:alg}

Despite the crucial role of point games for solving the coin-flipping problem, there have been notably few examples of point games constructed in the literature.  Indeed, the family of TIPGs in \cite{Mochon07} and \cite{pelchat13} are the only known family of TIPGs for the weak coin-flipping problem in which the bias tends to zero. To the best of our knowledge there are no general algorithms for constructing point games.

Theorem~\ref{thm:LambdaPenTIPGtoLambdaPenTDPG} showed that to solve the point game problems arising from the cheat-penalised weak coin-flipping problem, it suffices to find a time-independent point game which \textit{approximately} computes the move on the right-hand side of equation (\ref{eq:Lambda-TIPG-condition}).  We therefore concern ourselves in this section with the following general problem.

\vskip0.2in

\textit{Given $\epsilon > 0$, a start-configuration $s$ and an end-configuration $e$, find a horizontally valid move $h$ and a vertically valid move $v$ such that
\begin{eqnarray}
    \label{eq:shvf}
    \left\| h + v - (e - s) \right\|_1 & \leq & \epsilon.
\end{eqnarray}}

\vskip0.2in

The difficulty in solving this problem is that it involves optimising a function over a high-dimensional space (namely, the set of all TIPGs) defined by an infinite number of linear constraints (conditions (\ref{valid1})--(\ref{valid2})).  The insight behind our algorithm is that we can break this search problem into parts, where the first is a convex optimisation over a low-dimensional space, and the second is merely linear algebra.  The algorithm is based on the two-dimensional profile function, a concept introduced in \cite{miller2020impossibility}.

We describe the algorithm in this section.  Mathematica code for the algorithm is available at \cite{penWCFcode}. We provide an example calculation in Section~\ref{sec:tipgsforpenwcf}.  We leave a formal analysis of the performance of the algorithm to future work.

\subsection{Setup}

\label{subsec:setup}

Assume that configurations $s, e \colon \mathbb{R}_{\geq 0}^2 \to \mathbb{R}$ are given, with
\begin{eqnarray}
    \sum_{x,y} s(x,y) & = & \sum_{x,y} e(x,y).
\end{eqnarray}
We build towards the definition of a two-dimensional profile function of a move.  Our definition is analogous that in \cite{miller2020impossibility}, although based on a different (simpler) mathematical formula for the one-dimensional profile function.

\begin{defn}[Profile function]
For any $x > 0$, let $P_x: \mathbb{R}_{\geq 0} \to \mathbb{R}$ be defined as
\begin{align}
    P_x(\lambda)=\left\{ \begin{array}{lc}
    \lambda x/(\lambda + x) &
    \textnormal{ if } \lambda > 0 \\ \\
    1 & \textnormal{ if } \lambda \leq 0. \end{array} \right.
\end{align}
Let
\begin{align}
    P_0 (\lambda)=\left\{ \begin{array}{lc}
    0 &
    \textnormal{ if } \lambda \geq 0 \\ \\
    1 & \textnormal{ if } \lambda < 0. \end{array} \right.
\end{align}
\end{defn}

For any one-dimensional move $f$, let the profile function $\hat{f} \colon \mathbb{R} \to \mathbb{R}$  of $f$ be defined by
\begin{eqnarray}
    \hat{f} ( \lambda ) & = & \sum_x f(x) P_x (  \lambda ).
\end{eqnarray}
By definition, $f$ is valid if and only if $\hat{f}$ is nonnegative and $\hat{f}
(-1) = 0$.  

For any two-dimensional move $r$, define the profile function $\hat{r} \colon \mathbb{R}^2 \to \mathbb{R}$ by
    \begin{equation}
        \hat{r}(\alpha,\beta)=\sum_{x,y\in\mathbb{R}}r(x,y)P_x(\alpha)P_y(\beta).
    \end{equation}
It is easy to see that both vertically valid and horizontally valid moves must have non-negative profiles. Therefore, the end-configuration $e$ can be reached from the start-configuration $s$ only if $\hat{s} \leq \hat{e}$.  

If $W$ is a set, then we write $\mathbb{R}^W$ to mean the set of all functions from $W$ to $\mathbb{R}$.  If $U \subseteq W$ is a finite subset of $W$, then we can define an inner product on $\mathbb{R}^W$ by
\begin{eqnarray}
    \left< p, q \right>_U & = & \sum_{\lambda \in U} p ( \lambda ) q ( \lambda) \\
    \left\| p \right\|_U & = & \sqrt{ \sum_{x \in U} p(x)^2 }
\end{eqnarray}
for any $p,q \colon W \to \mathbb{R}$.  If $W$ is itself a finite set, then we may simply write $\left< \cdot , \cdot \right>$ and $\left\| \cdot \right\|$ (without any subscript) to denote $\left< \cdot , \cdot \right>_W$ and $\left\| \cdot \right\|_W$.

\subsection{Step 1: Choose threshhold constants and search sets}

We  first choose a positive real number $\delta$. Roughly speaking, $\delta$ determines the precision of one of the initial steps in our search for a point game solution, and $M$ places a limit on the size of the weights in the point games included in the search.  
These choices require striking a balance: smaller choices of $\delta$ and larger choices of $M$ will yield a search that takes longer to execute, but which is more likely to yield a point game solution with smaller bias and less communication complexity.

Choose a finite set $S \subseteq \mathbb{R}_{\geq 0}$ and a finite set $T \subseteq \mathbb{R}$ such that $-1 \in T$.  These sets are used to simplify our search for a solution to the point game problem $(s,e)$: we only consider two-dimensional moves that are supported in the set $S \times S$, and when we are assessing whether two moves have a similar profile, we look only at the profile values on the set $T \times T$.

\subsection{Step 2: Search for (approximately) valid moves with (approximately) correct profiles}

\label{subsec:approxprofile}

The next step is to search for moves $h$ and $v$ such that $h+v$ has nearly the same profile as $e - s$, and such that $h$ and $v$ are (at least approximately) horizontally valid and vertically valid, respectively.  We do this search in a way that is intended to minimise the complexity of the search space.

Let $S = \{ s_1, s_2, \ldots, s_\ell \}$, where $s_i < s_{i+1}$ for all $i \in \{ 1, 2, \ldots, \ell \}$.  Let $D \colon \mathbb{R}^{  S \smallsetminus \{ s_\ell \} } \to \mathbb{R}^S$. For $f\in\mathbb{R}^{  S \smallsetminus \{ s_\ell \} }$, we write $D(f)=D_f\in\mathbb{R}^S$. We define $D$ as 
\begin{eqnarray}
D_f(s_1) & = & f( s_1 ) \\
D_f(s_j) & = & f(s_j) - f(s_{j-1}) \hskip0.2in
\textnormal{ for } 2 \leq j \leq \ell -1 \\
D_f (s_\ell) & = & - f(s_{\ell - 1 }).
\end{eqnarray}
The image of $D$ is exactly the set of functions in $\mathbb{R}^S$ that sum to zero. Define also the linear map $H:\mathbb{R}^S\to\mathbb{R}^T$ by
\begin{eqnarray}
    H_f (t) & = & \hat{f}(t)
\end{eqnarray}
where $\hat{f}$ denotes the profile associated with $f$. In other words, the operator $H$ maps a function $f$ defined on $S$ to its corresponding profile, restricted to the set $T$.

Let $H' = H \circ D$, 
and let the image of $H'$ (which is a set of functions from $T$ to $\mathbb{R}$) be denoted by $\Im H'$.
The image $\Im H'$ has an inner product  given by $\left< \cdot , \cdot \right>_T$, and the domain space $\mathbb{R}^{S \smallsetminus \{ s_\ell \}}$ has a native inner product.  Compute the singular values
\begin{eqnarray}
    c_1 \geq c_2 \geq \ldots \geq c_{\ell-1}
\end{eqnarray}
for $H'$ and corresponding (unit-length, right) singular vectors
\begin{eqnarray}
    v_1, v_2, \ldots, v_{\ell -1 } \in \mathbb{R}^{S \smallsetminus \{ s_\ell \}}.
\end{eqnarray}
Choose $k \in \{ 1, 2, \ldots, \ell-1 \}$ such that $c_j \geq \delta$ for all $j \leq k$ and $c_j < \delta$ for all $j > k$.  Let $V \subseteq \mathbb{R}^{S \smallsetminus \{s_\ell\}}$ be the span of $\{ v_1, v_2, \ldots, v_k \}$.   (Informally, the moves in $D(V^\perp)$ tend to have profiles that are close to zero, and therefore we will essentially ignore the space $D(V^\perp)$ during this step of the algorithm, and focus our search within $D(V)$.)

Let us say that a one-dimensional move 
$f \colon S \to \mathbb{R}$ is \textbf{$T$-valid} if $\sum_{x \in S} f(x) = 0$ and $\hat{f}(t) \geq 0$ for all $t \in T$. Compute an $h \colon S \times S \to \mathbb{R}$
which minimises
\begin{eqnarray}
    \left\|  (\hat{h} + \hat{v} ) - (\hat{e} - \hat{s}) \right\|_{T \times T}
\end{eqnarray}
where $v:=h^{\top}$, and $h$ is subject to the following constraint: each row of the move $h$ is contained in $D ( V )$ and is $T$-valid.
\subsection{Step 3: Compute approximately valid moves with the correct start- and end-configuration}

\label{subsec:correct}

At this point we expect to have moves $h$ and $v$ that are approximately horizontally valid and approximately vertically valid, respectively, and are such that the profile of $h + v$ is close to that of $e -s $. However, this does not mean that $h+v$ is close to $e-s$. To remedy this, we now introduce additional moves %
$p$ and $q$ (correction factors) such that $p$ is approximately horizontally valid, $q$ is approximately vertically valid, and $(h+p) + (v+q) )$ is equal to $e - s$.

Again we use singular value analysis, although we focus directly on the linear transformation $H$ rather than on $H'$.  Let
\begin{eqnarray}
    d_1 \geq d_2 \geq \ldots \geq d_\ell
\end{eqnarray}
be the singular values of $H$, and let
\begin{eqnarray}
    w_1, w_2, \ldots, w_\ell \in \mathbb{R}^S
\end{eqnarray}
be the corresponding singular vectors in the domain $H$.  The vectors $\{ w_i \}$ form an othornomal basis for $\mathbb{R}^S$, and $\hat{w}_1, \ldots, \hat{w}_\ell$ form an orthogonal set (under $\left< \cdot , \cdot \right>_T$) such that
\begin{eqnarray}
    \left\| \hat{w_j} \right\|_T & = & d_j.
\end{eqnarray}
For any $j,k \in \{ 1, 2, \ldots, \ell \}$, define $w_j \otimes w_k \colon S \times S \to \mathbb{R}$ by
\begin{eqnarray}
(w_j \otimes w_k) ( x,y) & = & w_j (x ) w_k ( y ).
\end{eqnarray}
The functions $\{ w_j \otimes w_k \}$ likewise form an orthonormal basis for $\mathbb{R}^{S \times S}$, and the profiles of $\{ w_j \otimes w_k \}$ are all orthogonal to one another under $\left< \cdot , \cdot \right>_T$.   The $T$-norm of the profile of $w_j \otimes w_k$ is $d_j d_k$.   

We can therefore compute an orthogonal decomposition of the difference function $(h+v) - (e-s)$.  Find real values $\{ t_{jk} \}_{1 \leq j,k \leq \ell}$ such that
\begin{eqnarray}
\label{orthodec}
    (e-s) - (h+v)  & = & \sum_{j,k } t_{jk} (w_j \otimes w_k).
\end{eqnarray}
Let
\begin{eqnarray}
  \label{exp:p}  p & = & \sum_{j > k} t_{jk} ( w_j \otimes w_k)   +  \frac{1}{2} \sum_k  t_{kk} ( w_k \otimes w_k) \\
    \label{exp:q} q & = & \sum_{j < k} t_{jk} ( w_j \otimes w_k ) + \frac{1}{2} \sum_k  t_{kk} (w_k \otimes w_k) ,
\end{eqnarray}
and let $h' = h + p$ and $v' = v + q$.  Note that $p=q^{\top}$ and given that $v=h^{\top}$, we have $h' = v^{\prime \top}$. Finally,
\begin{eqnarray}
    h' + v' = e - s.
\end{eqnarray}

The intuition behind this construction is as follows. Since the profile of $(e-s) - (h + v)$ has small $T$-norm, the $T$-norms of the profiles of $t_{jk} (w_j \otimes w_k)$ (which are equal to $t_{jk} d_j d_k$, respectively) must likewise be small.  Thus, for any $j,k \in \{ 1, 2, \ldots, \ell \}$, either $t_{jk}$ is small, the $T$-norm of $\hat{w}_j$ is small, or the $T$-norm of $\hat{w}_k$ is small.  In the case of the terms $t_{jk} (w_j \otimes w_k)$ that appear in the expression  for $p$ (see Equation~\ref{exp:p}), we always have $d_j \leq d_k$, and so either $t_{jk}$ or $d_j$ must be small, which makes $p$ approximately horizontally valid.  A similar heuristic suggests that $q$ should be approximately vertically valid.  Therefore we expect this step to yield a pair $(h', v')$ such that $h'$ is approximately horizontally valid and $v'$ is approximately vertically valid.

\subsection{Step 4: Project approximately valid moves to valid moves}\label{sec:project_valid}

The final step is to replace the approximately valid moves $h'$ and $v'$ with valid moves $h_*$ and $v_*$ such that $\norm{h_* + v_* - e + s}_{S\times S}$ is minimised. To achieve this, we project each approximate move onto the corresponding set of valid moves. The projection procedure, outlined below, ensures that the resulting $h_*$ and $v_*$ remain as close as possible to $h'$ and $v'$ while satisfying the validity constraints.

To implement the projection, we consider the following constrained minimization problem:
\begin{equation}
\begin{aligned}
\min_{v_* \in \mathbb{R}^{S\times S}} \quad & \norm{v'-v_*}_{S\times S}^2 \\
\text{s.t.} \quad & H(v_*)\geq 0 \\
\end{aligned}
\end{equation}
This problem seeks the valid move $v_*$ closest to the approximate move $v'$. It can be reformulated as a quadratic program of the form
\begin{equation}
\begin{aligned}
\min_{x} \quad & \tfrac{1}{2} x^\top Q x + c^\top x \\
\text{s.t.} \quad & A x \leq b, \\
\end{aligned}
\end{equation}
where $Q$ is the identity matrix, $c = -v'$, $A = -H$, and $b = 0$. We assume that the move $v_*$ and $v'$ have been mapped to a vector and that $H$ has been mapped to a matrix. In this formulation, the quadratic objective captures the squared distance to $v'$, while the linear inequality encodes the validity constraint.

Since $Q = I \geq 0$, the objective function is strictly convex. Combined with the linear inequality constraint, this makes the problem a convex quadratic program. Convexity guarantees both the existence of a global optimum and the availability of efficient numerical algorithms to compute it. In practice, this ensures that the projection step can be carried out robustly, yielding the closest valid move $v_*$ to the given approximation $v'$.

Note that the constraint ensures that $\hat{v}_*(\lambda)$ is nonnegative on the set $T$. Consequently, $T$ should be chosen to cover as broad a range of values as possible in order to guarantee the validity of the move. Recall that we also required $-1\in T$, but this constraint is automatically satisfied by imposing that the sum of the weights of a move is $0$. In practice, however, we find that a relatively small choice of $T$ is already sufficient to enforce the validity condition. We observe that, in general, applying this projection procedure to obtain $v_*$ and $h_*$ also ensures that $v_* = h_*^\top$. This symmetry arises because the corresponding condition is already enforced when constructing the approximate moves $v'$ and $h'$.

\section{penTIPGs for Cheat-penalised  WCF}\label{sec:tipgsforpenwcf}

We now present several of the penTIPGs obtained numerically, together with the protocols derived from them in terms of bias, number of messages, and qubits used. The penTIPG data are provided as matrices corresponding to the valid moves $h_*$ and $v_*$. In addition, we include the matrix representation of the profile function $H$, along with the number of non-truncated elements, as described in Section~\ref{subsec:approxprofile}. The sets $S$ and $T$ are also made available. All of these values are provided in a supplementary text file, which can be accessed at \cite{penWCFcode}. In this text file we present the matrices for the moves as $h'$ and $v'$ together with the matrix corresponding to $H$ to project to valid moves. For completeness, we also present several representative penTIPGs explicitly as matrices within this section.

The bias and number of messages in the protocols derived from the penTIPGs can be determined using Theorem~\ref{thm:LambdaPenTIPGtoLambdaPenTDPG}, while the number of qubits required follows from Theorem~\ref{thm:LambdaPenTDPG-implies-Lambda-penWCF}. The qubit cost depends on the size of the set $S$ chosen in the construction. Notably, the number of elements in $S$ does not need to be large in order to achieve solutions with low bias. Even more strikingly, our numerical results suggest that arbitrarily low bias can be attained with a constant number of qubits, in sharp contrast to Mochon's family of protocols, where vanishing bias requires taking the limit of infinitely many qubits.

As stated in Theorem~\ref{thm:LambdaPenTIPGtoLambdaPenTDPG}, the number of points in the penTDPG---and therefore the number of messages---depends on the parameter $\eta_2$. This parameter in turn depends on $\delta$ and $c_1$. Numerically, we set $c_1$ as close as possible to $\frac{m_1^2}{(\Lambda+1)\Lambda}$ and then find that the closer we set $\delta$ to $\delta_{\mathrm{min}}$, the lower the bias. There is of course a trade-off between the number of messages and the bias obtained in this way.

\subsection{penTIPGs}\label{sec:list_TIPGs}
In this subsection we list some of the penTIPGs we found numerically. We provide  the matrix encoding $v*$. The move $h_*$ can be found by noting $h_*=v_*^\top$. We also provide the sets $S$ and $T$ together with a truncation parameter which defines how many singular values we keep for the optimisation. Note that we don't add $-1$ to the set $T$ as that constraint is automatically satisfied in the numerics. For penTIPG 1 we present some detail on the bias and round complexity.

\subsubsection{penTIPG $1$ with $\Lambda = 1$}
The parameters are defined as:
\[
S = \{0.3, 0.7, 1, 1.25, 1.500005, 1.75, 2, 3\},
\]
\[
\text{truncation} = 6,
\]
\[
T = \{0.1, 0.3, 0.5, 1, 1.5, 2, 3, 4, 10, 1000\},
\]

\textbf{Matrix \( v_* \):}
\[
\begin{bmatrix}
\begin{tabular}{cccccccc}
 0.000000 & -0.012549 & 0.000867 & 0.002990 & -0.005655 & 0.000098 & 0.001934 & 0.002993 \\
 0.012549 & 0.000000 & -0.027156 & -0.021163 & 0.007004 & 0.005266 & 0.025788 & -0.007110 \\
 -0.000867 & 0.027156 & 0.000000 & -0.035587 & -0.014383 & -0.147784 & -0.304895 & -0.028092 \\
 -0.002990 & 0.021163 & 0.035587 & 0.000000 & -0.196252 & 0.006301 & 0.165097 & -0.031862 \\
 0.005655 & -0.007004 & 0.014383 & 0.196252 & 0.500000 & 0.227074 & 0.092193 & -0.028695 \\
 -0.000098 & -0.005266 & 0.147784 & -0.006301 & -0.227074 & 0.000000 & 0.077654 & 0.016646 \\
 -0.001934 & -0.025788 & -0.195105 & -0.165097 & -0.092193 & -0.077654 & 0.000000 & 0.064828 \\
 -0.002993 & 0.007110 & 0.028092 & 0.031862 & 0.028695 & -0.016646 & -0.064828 & 0.000000 \\
\end{tabular}
\end{bmatrix}
\]

\begin{figure}[H]
\centering
\includegraphics[width=0.6\paperwidth]{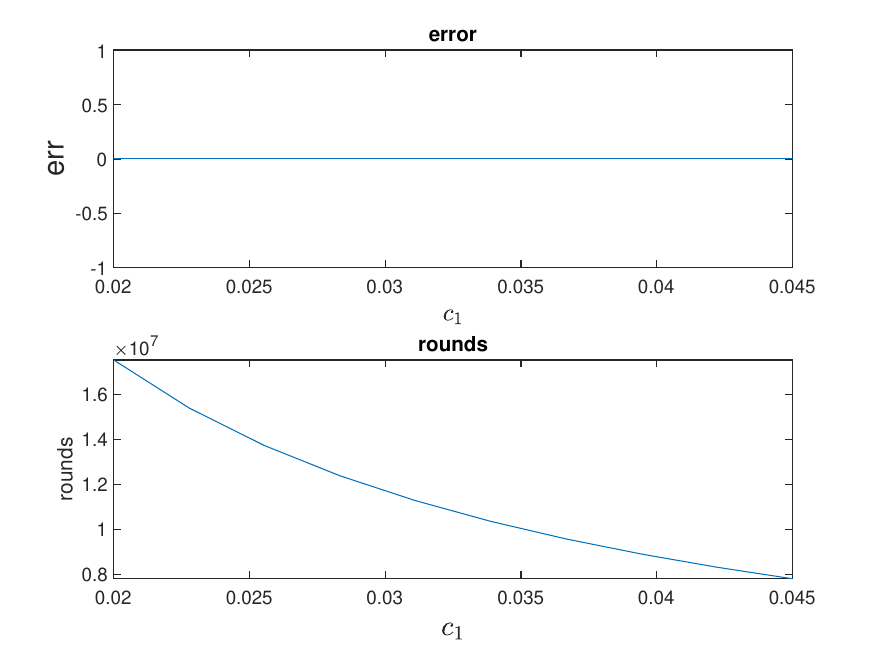}
\caption{The error and rounds of communication for penTIPG 1 when choosing $\delta=\delta_{\mathrm{min}}+10^{-5}$.}\label{fig:TIPG1-1}
\end{figure}

\begin{figure}[H]
\centering
\includegraphics[width=0.6\paperwidth]{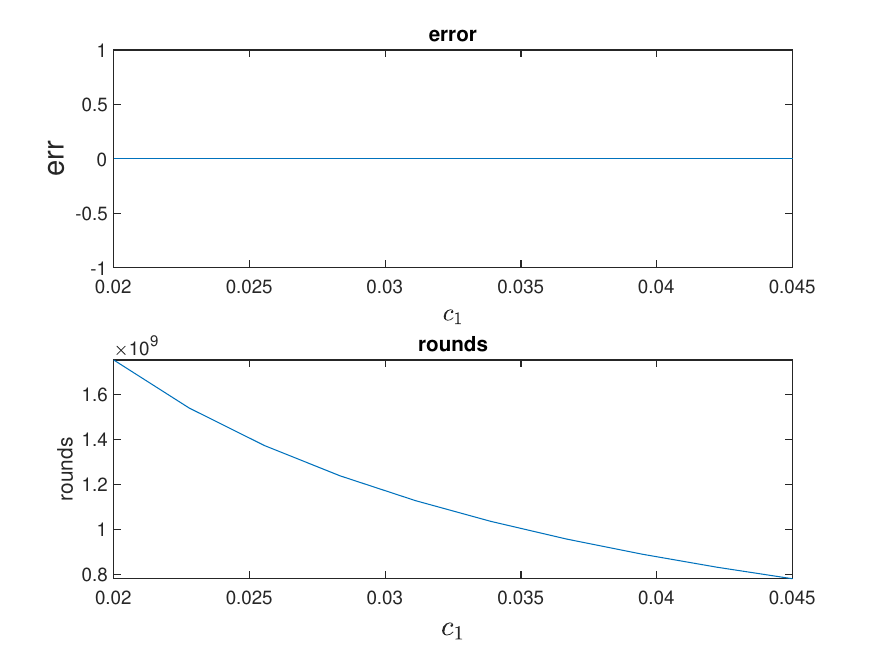}
\caption{The error and rounds of communication for penTIPG 1 when choosing $\delta=\delta_{\mathrm{min}}+10^{-7}$.}\label{fig:TIPG1-2}
\end{figure}

\paragraph{Trade-off between error and rounds.}
The error $\mathsf{err}$ and rounds of communication obtained in the protocol are controlled by the choice of $\delta$ and $c_1$ in Theorem~\ref{thm:LambdaPenTIPGtoLambdaPenTDPG}. For example, when choosing $\delta=\delta_{\mathrm{min}}+ 10^{-5}$ and $c_1$ close to $\frac{m_1^2}{(\Lambda+1)\Lambda}$, we can obtain $\mathsf{err}=0.004$ and $8\times 10^6$ rounds of communication as depicted in Figure~\ref{fig:TIPG1-1}. Note that the final bias of this protocol is dominated by $\mathsf{err}$ and therefore to decrease the bias we need to decrease $\delta$. By choosing $\delta = \delta_{\mathrm{min}}+10^{-7}$, we find that $\mathsf=0.0004$ and the number of rounds is $7\times 10^9$ as shown in Figure~\ref{fig:TIPG1-2}. Even though the number of rounds has increased by two orders of magnitude, the number of qubits used is the same. Therefore, reducing the error is just a matter of choosing $\delta$ as close as possible to $\delta_\mathrm{min}$.

\subsubsection{penTIPG $2$ with $\Lambda=1$}

\[
S = \{0.6, 0.8, 1, 1.225, 1.505, 1.75, 2, 2.5\},
\]
\[
\text{truncation} = 6,
\]
\[
T = \{0.1, 0.3, 0.5, 1, 1.5, 2, 3, 4, 10, 1000\},
\]

\textbf{Matrix \( v_* \):}
\[
\begin{bmatrix}
\begin{tabular}{cccccccc}
 0.000000 & -0.022240 & -0.010330 & 0.001572 & 0.017053 & 0.012344 & -0.008076 & -0.002015 \\
 0.022240 & 0.000000 & -0.032154 & -0.023705 & -0.041653 & 0.014959 & 0.091371 & -0.039714 \\
 0.010330 & 0.032154 & 0.000000 & -0.059100 & -0.005204 & -0.150847 & -0.293117 & -0.041182 \\
 -0.001572 & 0.023705 & 0.059100 & 0.000000 & -0.210070 & -0.008831 & 0.155849 & -0.023644 \\
 -0.017053 & 0.041653 & 0.005204 & 0.210070 & 0.500000 & 0.212440 & 0.038713 & 0.006219 \\
 -0.012344 & -0.014959 & 0.150847 & 0.008831 & -0.212440 & 0.000000 & 0.055970 & 0.025358 \\
 0.008076 & -0.091371 & -0.206883 & -0.155849 & -0.038713 & -0.055970 & 0.000000 & 0.048148 \\
 0.002015 & 0.039714 & 0.041182 & 0.023644 & -0.006219 & -0.025358 & -0.048148 & 0.000000 \\
\end{tabular}
\end{bmatrix}
\]

\subsubsection{penTIPG $3$ with $\Lambda=0.01$}
\[
S = \{0.005, 0.007, 0.01, 0.125, 0.51000000000001, 0.75, 1.01, 1.02\},
\]
\[
\text{truncation} = 6,
\]
\[
T = \{0.1, 0.3, 0.5, 1, 1.5, 2, 3, 4, 10, 1000\},
\]

\textbf{Matrix \( v_* \):}

\[
\begin{bmatrix}
\begin{tabular}{cccccccc}
 0.000000 & -0.011971 & -0.006441 & 0.006342 & -0.062810 & 0.058173 & 0.025592 & -0.029527 \\
 0.011971 & 0.000000 & -0.010990 & -0.001980 & -0.012568 & -0.046571 & 0.105927 & -0.059991 \\
 0.006441 & 0.010990 & 0.000000 & -0.017410 & 0.036784 & -0.126314 & -0.342226 & -0.079367 \\
 -0.006342 & 0.001980 & 0.017410 & 0.000000 & -0.104147 & 0.024187 & 0.066522 & -0.025626 \\
 0.062810 & 0.012568 & -0.036784 & 0.104147 & 0.500000 & 0.243379 & 0.068062 & 0.013066 \\
 -0.058173 & 0.046571 & 0.126314 & -0.024187 & -0.243379 & 0.000000 & 0.145096 & -0.017206 \\
 -0.025592 & -0.105927 & -0.157774 & -0.066522 & -0.068062 & -0.145096 & 0.000000 & 0.118873 \\
 0.029527 & 0.059991 & 0.079367 & 0.025626 & -0.013066 & 0.017206 & -0.118873 & 0.000000 \\
\end{tabular}
\end{bmatrix}
\]

In Figure~\ref{fig:TIPG4_plot}, we give a graphical depiction of $v_*$. %

\section{Other Cheat-penalised WCF Protocols}\label{sec:other_protocols}
In this section, we give an overview of a few other cheat-penalised protocols for WCF with the purpose of comparing them to the protocols found through our method. 
\subsection{Spekkens-Rudolph}\label{sec:SR-protocol}
 In \cite{SR02} a cheat-sensitive protocol was proposed with a bias close to $0.207$. The main idea is that Alice prepares a quantum state on two registers and sends one of them to Bob. Then, based on a measurement by Bob on the state, a party is chosen to verify the state on both registers. If the verification succeeds, the party who did not perform the verification is declared the winner. This construction can be naturally extended to the cheat-penalised setting. We summarise the protocol below.

\begin{itemize} 
\item Alice prepares $\ket{\psi} = \sqrt{x}\ket{00}+\sqrt{1-x}\ket{11} \in D(\mathcal{A}\mathcal{B})$ and sends $\mathcal{B}$ to Bob.  
\item Bob measures $\mathcal{B}$ with the measurement $(E_0,E_1)$ to get $b$, where $E_0 = \frac{1}{2x} \kb{0}$, for $x \in (1/2, 1]$.    
    He sends $b$ to Alice. 
\item If $b = 0$, he sends $\mathcal{B}$ back to Alice. 
    Alice measures the POVM $\{\ketbra{\psi_0}, I-\ketbra{\psi_0}\}$. 
    If the result is $\ketbra{\psi_0}$, Bob wins. If not, she declares Bob a cheater.  
\item If $b = 1$, Alice sends $\mathcal{A}$ to Bob. 
    He measures the POVM $\{\ketbra{\psi_1}, I-\ketbra{\psi_1}\}$.  
   If the result is $\ketbra{\psi_1}$, Alice wins. If not, he declares Alice a cheater.  
\end{itemize} 
Where we define $\ket{\psi_b}=\frac{(I\otimes \sqrt{E_b})\ket{\psi}}{\sqrt{\bra{\psi_b}I\otimes E_b \ket{\psi}}}$.

In the weak coin-flipping setting studied by \cite{SR02}, the optimal bias of $\frac{1}{\sqrt{2}} - \frac{1}{2} \approx 0.207$ is achieved by choosing $x = \frac{1}{\sqrt{2}}$. In the cheat-penalised setting, the semidefinite programs characterising the optimal strategies for cheating Alice and cheating Bob are given as follows.

\textit{SDP cheating Alice}. 

$$ \max_{\rho_{AB}} \quad (\Lambda+1)\cdot \Tr\left[(I^A \otimes \sqrt{E_1})\rho_{AB}(I^A \otimes \sqrt{E_1})\ketbra{\psi_1}{\psi_1}\right] + \Lambda\cdot \Tr\left[\rho_{AB}(I^A \otimes E_0)\right]   $$

subject to $$ Tr_{AB}\left[\rho_{AB}\right]=1 $$
          $$\rho_{AB}\geq 0. $$

\textit{SDP cheating Bob}

$$ \max_{\rho_{ABC}} \quad (\Lambda+1)\cdot \Tr\left[(\ketbra{\psi_0}\otimes \ketbra{0}_C)\rho_{ABC}\right] + \Lambda\cdot \Tr\left[(I^{AB} \otimes \ketbra{1}_C)\rho_{ABC}\right]   $$

subject to $$ \Tr_{BC}\left[\rho_{ABC}\right]=\Tr_B\left[\ketbra{\psi}\right] $$
          $$\rho_{ABC}\geq 0. $$

\subsubsection{Point game}
\begin{figure}[H]
\centering
\includegraphics[width=0.6\paperwidth]{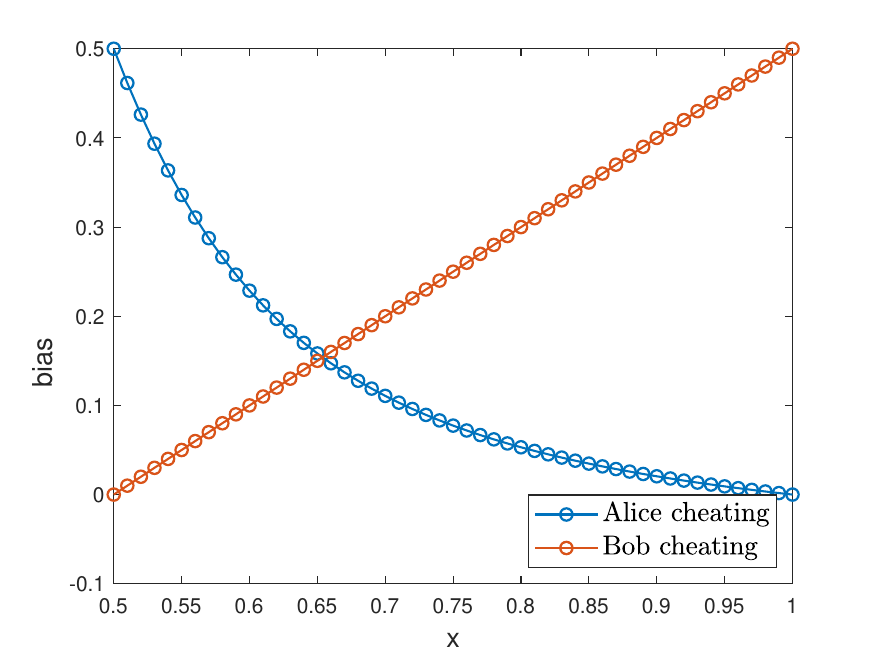}
\caption{Bias for Spekkens-Rudolph protocol when $\Lambda=6$. The intersection of the curves for the bias of Alice and Bob give the optimal $x$ for the protocol.}\label{fig:SR-cheating}
\end{figure}

We consider the Spekkens-Rudolph point game as in \cite{Mochon07} but with the starting point as defined for cheat-penalised WCF, i.e., with coordinates $w$ and $v$ such that $w>v>0$.
\begin{align} 
\frac{1}{2} \points{v,w} + \frac{1}{2} \points{w,v} 
& \quad \nobarfrac{\text{h. split}}{\to} \quad  
\frac{1}{2} \points{v,w} + p \points{z_1,v} + (\frac{1}{2}-p) \points{z_2,v} \\ 
& \quad \nobarfrac{\text{v. raise}}{\to} \quad
\frac{1}{2} \points{v,w} + p \points{z_1,v} + (\frac{1}{2}-p) \points{z_2,w} \\ 
& \quad \nobarfrac{\text{h. merge}}{\to} \quad
(1-p) \pg{\frac{\frac{v}{2}+(1/2-p)z_2}{1-p}}{w} + p \points{z_1,v}  \\ 
& \quad \nobarfrac{\text{v. merge}}{\to} \quad
 \points{z_1,(1-p)w+pv}  
\end{align} 
where due to the splits and merge we have the conditions 
\begin{align}
    \frac{1}{2w}&\geq \frac{p}{z_1} + \frac{\frac{1}{2}-p}{z_2},\\
    z_1 &= \frac{\frac{v}{2}+(1/2-p)z_2}{1-p},\\
    0&<z_1<1,\\
    z_2&>1,\\
    0&<p<1.
\end{align}
If we solve for $z_1 = (1-p)w+pv$ under these constraints, we can find the cheating reward. By solving these equations using a non-linear optimisation routine we find that $x=0.653$ gives the optimal measurement in the SDPs. This can be also checked by solving the SDPs and finding the $x$ at which both Alice's and Bob's cheating bias are the same as shown in Figure~\ref{fig:SR-cheating}. The value for optimal cheating bias is $0.152$.  The number of messages in this protocol is $8$ and based on Theorem \ref{thm:LambdaPenTDPG-implies-Lambda-penWCF} the number of qubits is $6$.

\subsection{Dip-Dip-Boom}\label{sec:DipBoom}
The point game for the Spekkens-Rudolph protocol can be refined, as described in Section 3.2.2 of \cite{Mochon07}, by increasing the number of points generated in the initial splits (see Figure~5 in \cite{Mochon07}). This point game characterises what is known as the Dip-Dip-Boom protocol. Mochon analyses the cheating reward in the cheat-penalised setting in the limit where the number of initial points tends to infinity. From his analysis, he derives a bound on the cheating reward. While we do not reproduce the full derivation here, we provide additional details relevant to our discussion.

 Mochon works in the setting where winning the coin flip gives $1$ point, loosing gives $0$ points and if you get caught cheating you lose $\Lambda$ points. From the starting frame $\frac{1}{2}\points{1,0}+\frac{1}{2}\points{0,1}$ we want to find the minimal $R^*$ such that $\points{R^*,R^*}$ can be reached by this protocol. The initial splits generate a continuous probability distribution of points of the form
\begin{align}\label{eq:cont-points}
    \frac{1}{2}\int_{R^*}^\infty p(z) \points{z,0}\dd z + \frac{1}{2}\int_{R^*}^\infty p(z) \points{0,z}\dd z .
\end{align}
 By conservation of probability, it is shown that $p(z)=\frac{2(R^*)^2}{z^3}$ to obtain a valid transition. Finally, to go from the initial frame to the frame in Equation \ref{eq:cont-points}, the following inequality must be satisfied (Equation 101 in \cite{Mochon07})
 \begin{align}\label{eq:ineq-split}
     -\frac{1}{\Lambda+1}\leq -\left(\frac{\Lambda-2R^*}{\Lambda^2} + \frac{2(R^*)^2}{\Lambda^3}\log(\frac{R^* + \Lambda}{R^*})\right).
 \end{align}
 Mochon~\cite{Mochon07} argues that this is the correct constraint corresponding to a split in the cheat-penalised setting, in accordance with his rewards convention.\footnote{Note that our convention, while equivalent in terms of constructing protocols, is different from Mochon's: we award $\Lambda+1$ points on winning the coin flip, $\Lambda$ on losing the coin flip and $0$ on getting caught cheating.}
 Finally, we show the following claim giving the scaling of $R^*$ when $\Lambda\to \infty$. In~\cite{Mochon07}, this was stated without proof.
\begin{claim}[Section 3.3.2 in \cite{Mochon07}]
    As $\Lambda \to \infty$, the optimal cheating reward for the Dip-Dip-Boom protocol is $R^* = \frac{1}{2}+\frac{\log \Lambda}{4\Lambda} + O(\frac{1}{\Lambda})$.
\end{claim}
\begin{proof}
    Throughout the proof we assume the equality in Equation \ref{eq:ineq-split} holds. To see that equality is enough, note that we can write Equation \ref{eq:ineq-split} as
    \begin{align}
        H_{\Lambda}(z) :=\Lambda(\Lambda-2z)+2z^2\log\!\left(1+\frac{\Lambda}{z}\right)
\;\le\;\frac{\Lambda^3}{\Lambda+1}.
    \end{align}
    For each fixed $\Lambda>0$, the function $H_\Lambda(z)$ is continuous and strictly decreasing in $z>0$, with
\[
H_\Lambda(0^+)=\Lambda^2,\qquad 
H_\Lambda(z)\to 0 \quad (z\to\infty).
\]
Thus, there exists a unique $R^*(\Lambda)>0$ such that
\[
H_\Lambda(R^*(\Lambda)) = \frac{\Lambda^3}{\Lambda+1}.
\]
At this value the inequality holds with equality, and strictly for $z>R^*(\Lambda)$. Therefore, we can consider Equation \ref{eq:ineq-split} as an equality for finding a lower bound to the cheating reward.

    Consider the expansion
    \begin{align}
        \frac{1}{\Lambda+1}=\frac{1}{\Lambda}-\frac{1}{\Lambda^2}+O\left(\frac{1}{\Lambda^3}\right).
    \end{align}
Then, multiplying both sides of Equation \ref{eq:ineq-split} by $\Lambda^2$ gives
\begin{align}
    \Lambda -1 + O\left(\frac{1}{\Lambda}\right)=\Lambda - 2R^* + \frac{2(R^*)^2\log(1+\frac{\Lambda}{R^*})}{\Lambda}.
\end{align}
We then have the equality
\begin{align}\label{eq:expanded_lambda}
     1 + O\left(\frac{1}{\Lambda}\right)= 2R^* - \frac{2(R^*)^2\log(1+\frac{\Lambda}{R^*})}{\Lambda}.
\end{align}
As $\Lambda\to \infty$, we can consider $R^*=\frac{1}{2}+\delta$ with $\delta=o(1)$. Then, for large $\Lambda$ we have $\log(1+\frac{\Lambda}{R^*})\approx\log(\frac{\Lambda}{R^*})=\log(\Lambda)-\log(R^*)$ and moreover
\begin{align}
    \log R^* &= \log(\frac{1}{2}+\delta)\\
    &= \log(\frac{1}{2}) + \log(1+2\delta)\\
    &= \log(\frac{1}{2}) + 2\delta + O(\delta^2).
\end{align}
Therefore, we have that $\log(1+\frac{\Lambda}{R^*})\approx\log(\Lambda)+\log(2)-2\delta+O(\delta^2)$. Note that in this last expression any smaller orders in $1/\Lambda$ will scale as $O\left(\frac{1}{\Lambda}\right)$. Furthermore, $2(R^*)^2 = \frac{1}{2}+2\delta + 2\delta^2$ and thus,
\begin{align}
    \frac{2(R^*)^2\log(1+\frac{\Lambda}{R^*})}{\Lambda}= \frac{(\frac{1}{2}+2\delta)(\log \Lambda + \log 2-2\delta)}{\Lambda}+O\left(\frac{\delta^2 \log \Lambda}{\Lambda}\right)
\end{align}
We shall neglect terms of order $\delta^2$ as these terms are small when $\Lambda\to \infty$. From Equation \ref{eq:expanded_lambda}, we have
\begin{align}
         1 + O\left(\frac{1}{\Lambda}\right)&= 1+2\delta- \frac{2(R^*)^2\log(1+\frac{\Lambda}{R^*})}{\Lambda}.\\
         &\approx 1 + 2\delta - \frac{\log\Lambda}{2\Lambda}-\frac{\log 2}{2\Lambda} + \frac{\delta}{\Lambda}-\frac{2\delta\log\Lambda}{\Lambda}-\frac{2\delta\log 2}{\Lambda}
\end{align}
In the last expression, the leading term is $\frac{\log \Lambda}{2\Lambda}$. Therefore, if we want this expression to match the order $O\left(\frac{1}{\Lambda}\right)$, we need to at least cancel this higher order term with $\delta$. We must then have
\begin{align}
2\delta = \frac{\log \Lambda}{2\Lambda} + O\left(\frac{1}{\Lambda}\right). 
\end{align}
This shows that $R^* = \frac{1}{2}+\frac{\log \Lambda}{4\Lambda} + O(\frac{1}{\Lambda})$.
\end{proof}
In fact, we can show the higher-order scaling of $R^*$. We provide this result next.
\begin{claim}
As $\Lambda\to \infty$, the optimal cheating reward for the Dip-Dip-Boom protocol is $R^* = \frac{1}{2}+\frac{\frac{1}{4}\log (2\Lambda)-\frac{1}{2}}{\Lambda} + \frac{\tfrac14\log^2(2\Lambda)-\tfrac58\log(2\Lambda)+\tfrac78}{\Lambda^2}
+O\!\left(\frac{\log^3\Lambda}{\Lambda^3}\right)$.
\end{claim}
\begin{proof}
    Consider the following expression for $R^*$,
    \begin{align}
        R^* = \frac12 + \frac{A\log \Lambda + B}{\Lambda} + \frac{C\log^2\Lambda + D \log\Lambda+E}{\Lambda^2} +O\left(\frac{\log^3\Lambda}{\Lambda^3}\right).
    \end{align}
    We need to find $A,B,C,D$ and $E$ such that equality in Equation \ref{eq:ineq-split} is solved up to the relevant order. For simplicity, define $u_1 = A\log \Lambda + B$, $u_2 =C\log^2 \Lambda + D \log \Lambda + E$, $\delta=R^*-\tfrac12$. Therefore, we have the following expansions.
    \begin{align}
        \frac{1}{\Lambda+1}&=\frac{1}{\Lambda} - \frac{1}{\Lambda^2}+\frac{1}{\Lambda^3}-\frac{1}{\Lambda^4} + O(\Lambda^{-5})\\
        R^* &= \frac{1}{2}+ \frac{u_1}{\Lambda}+\frac{u_2}{\Lambda^2} + O(\Lambda^{-3})\\
        (R^*)^2 &= \frac{1}{4} + \frac{1}{4} + \frac{u_1}{\Lambda} + \frac{u_2 + u_1^2}{\Lambda^2} + O(\Lambda^{-3})\\
        \log R^* &= \log \frac{1}{2} + 2\delta - 2\delta^2 + O(\delta^3)\\
        &= \log \frac{1}{2} + \frac{2u_1}{\Lambda} + \frac{2u_2 - 2u_1^2}{\Lambda^2} + O(\Lambda^{-3}).
    \end{align}
    By replacing these expansions in Equation \ref{eq:ineq-split} and imposing equality of left and right hand side, we can match the orders and obtain coefficients $A,B,C,D,E$. The result is the expression 
    $$R^* = \frac{1}{2}+\frac{\frac{1}{4}\log (2\Lambda)-\frac{1}{2}}{\Lambda} + \frac{\tfrac14\log^2(2\Lambda)-\tfrac58\log(2\Lambda)+\tfrac78}{\Lambda^2}
+O\!\left(\frac{\log^3\Lambda}{\Lambda^3}\right).$$
\end{proof}
\subsection{ABDR04 protocol}
The Dip-Dip-Boom protocol analysed in Section~\ref{sec:DipBoom} demonstrates that, when both the penalty and the number of rounds are taken to the infinite limit, the cheating reward converges to $\frac{1}{2}$. In contrast, the cheat-penalised protocol introduced in~\cite{ambainis2004multiparty} establishes that it is sufficient to take only the penalty to the infinite limit in order to achieve this bound. This protocol, which we describe below, requires only $3$ messages between Alice and Bob. Consider $\Lambda\geq 4$ and $\delta=\frac{2}{\sqrt{\Lambda}}$. Define the state $\ket{\psi_a}=\sqrt{\delta}\ket{a}\ket{a}+\sqrt{1-\delta}\ket{2}\ket{2}\in \mathbb{C}^3\otimes \mathbb{C}^3$ with $a\in\{0,1\}$.
\begin{itemize} 
\item Alice picks $a\in \{0,1\}$ uniformly at random and prepares $\ket{\psi_a}\in \mathcal{A}\mathcal{B}$ and sends $\mathcal{B}$ to Bob.
\item Bob picks $b\in\{0,1\}$ uniformly at random and sends $b$ to Alice.
\item Alice sends $a$ and the register $\mathcal{A}$ to Bob. Then Bob measures $\mathcal{A}\mathcal{B}$ with the POVM $\{\ketbra{\psi_a},I-\ketbra{\psi_a}\}$. If Bob measures the state $\ket{\psi_a}$, the verification succeeds and the coin flip is $a\oplus b$, otherwise Bob declares that Alice cheated.
\end{itemize} 
As demonstrated in \cite{ambainis2004multiparty}, the cheating reward for both Alice and Bob is given by $\frac{1}{2} + \frac{1}{\sqrt{\Lambda}}$. %
We do not describe the point games in this case as they are not relevant here but they can be constructed using known techniques, if needed. As discussed previously, the protocol necessitates the use of two qutrits or, alternatively, four qubits.

\section{Comparison of penWCF Protocols\label{sec:comparisonPenWCF}}

We compare previously known protocols with those obtained in our work in terms of the bias achieved relative to the number of communication rounds and the number of qubits required. We use for the comparison penTIPG 2 which yields protocol (iii) and penTIPG 3 which yields protocols (i) and (ii) depending on the choice of parameters (more specifically on the choice of $\delta$).

\begin{figure}[H]
\centering
 \includegraphics[scale=1.2, width=1\textwidth]{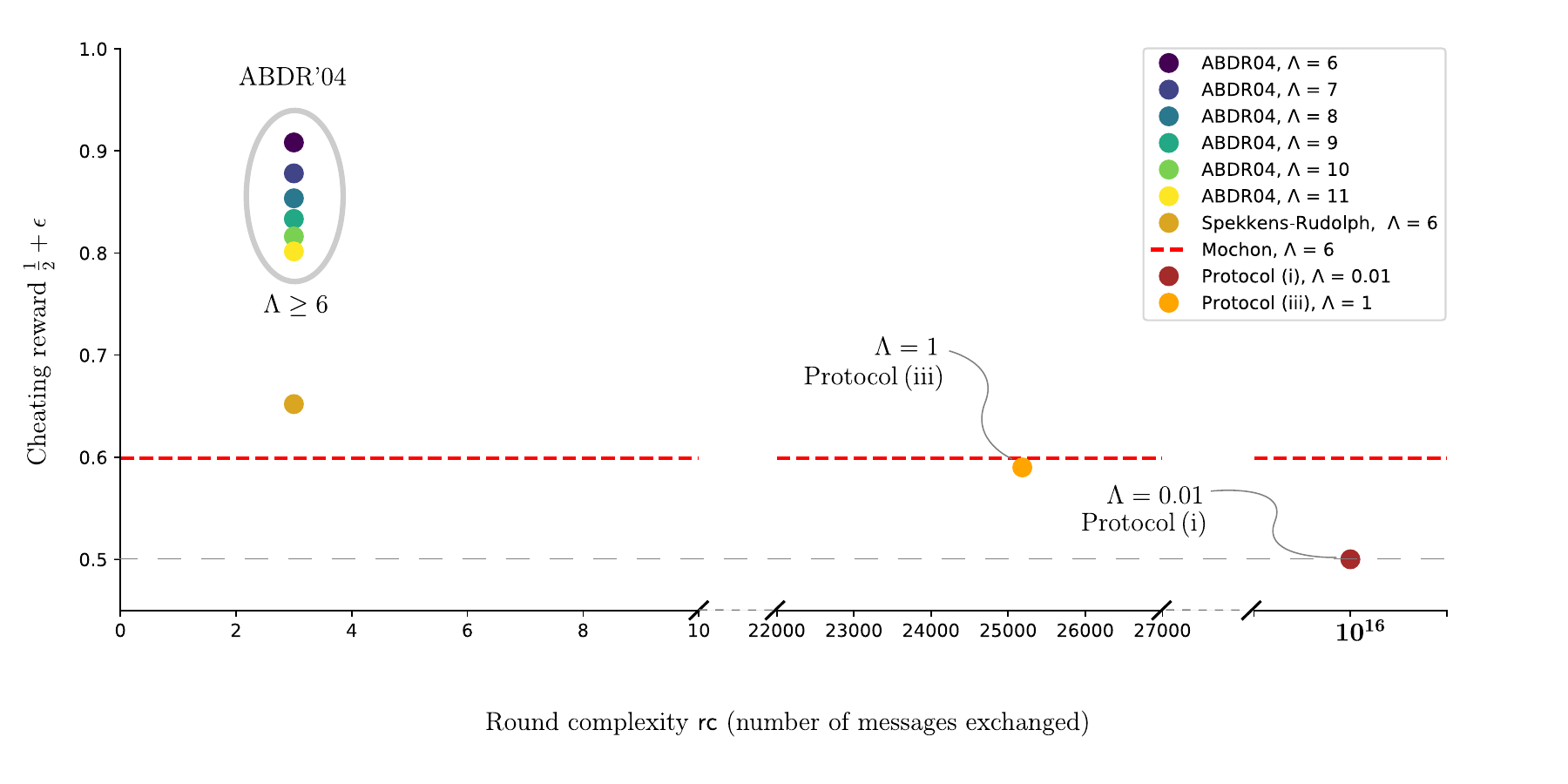}
\caption{Comparison of cheat-penalised weak coin-flipping protocols in terms of the reward and the number of messages. We compare our protocols penTIPG 2 and penTIPG 3 to ABDR04 \cite{ambainis2004multiparty}, Spekkens-Rudolph protocol \cite{SR02} and Mochon's bound \cite{Mochon07}. 
bias and number of qubits. 
The protocol penTIPG 2 corresponds to Protocol (iii)---our potentially amenable to experiment protocol, with $\Lambda=1$. The protocol penTIPG 3 corresponds to protocol (i). 
}\label{fig:reward-messages}
\end{figure}

\paragraph{Round complexity.} Figure~\ref{fig:reward-messages} illustrates this comparison for $\Lambda \in \{1, 0.01\}$. Our protocols achieve a bias below the bound of the cheat-penalised version of Dip-Dip-Boom protocol, whose performance is derived under the assumption of infinitely many communication rounds between Alice and Bob. While the ABDR04 protocol can in principle reach arbitrarily low bias, it does so only in the limit of infinite penalty. By contrast, our results demonstrate that comparably low bias can already be achieved with finite penalty, albeit at the cost of a large number of communication rounds. Despite this large resource cost, the number of rounds in our protocol is still significantly smaller than in Mochon's TIPGs for WCF achieving bias $\frac{1}{2}+O(\epsilon)$. In particular, the lower bound from \cite{miller2020impossibility} shows that the number of rounds required scales as $\exp(\Omega(1/\sqrt{\epsilon}))$. Previous estimations \cite{Alnawakhtha2025} for the constants in the lower bound indicate that to obtain a bias of $\epsilon=10^{-8}$ one would require at least $10^{23}$ rounds. With protocol (i), we can reach a bias of $10^{-8}$ with $10^{16}$ rounds. Moreover, to reach a bias of $10^{-10}$ we only need $10^{18}$ rounds, which is still $10^5$ times better in communication complexity compared to the lower bound with $\epsilon=10^{-8}$.

\paragraph{Space complexity.} Figure~\ref{fig:reward-qubits} compares bias against the number of qubits used in each protocol. Here, our numerics provide striking evidence that arbitrarily low bias can be obtained using only a constant number of qubits, even when the penalty is kept small. All the protocols in our work require $24$ qubits, protocol (ii) based on penTIPG 3 can reach down to bias $\epsilon=10^{-10}$. 
This reveals a sharp distinction from prior approaches: the Dip-Dip-Boom and ABDR04 protocol require infinite penalty. Note also that we are considering the Dip-Dip-Boom protocol in the ``infinite limit" of points in the point game, i.e., infinite rounds. Mochon~\cite{Mochon07} also constructs other families TIPGs for WCF that achieve a bias of $\tfrac{1}{2} + O(\epsilon)$ in the weak coin-flipping setting. However, these TIPGs require an unbounded number of qubits as the bias approaches $\tfrac{1}{2}$. Our findings therefore highlight that our penTIPG-based constructions can achieve low bias with genuinely finite resources, suggesting a new route to practical protocols.

Given the previous considerations, a natural question is whether our algorithm can produce protocols achieving arbitrarily low bias when $\Lambda = 0$. Such a result would establish the existence of weak coin-flipping protocols that are more space-efficient than Dip-Dip-Boom and Mochon's TIPGs reaching a bias $\tfrac12+\O(\epsilon)$. Our preliminary numerical evidence, however, suggests that this may not be the case. In particular, we were unable to find solutions close to the $\Lambda = 0$ regime when the chosen grid (i.e., the set $S$) excluded points below $\points{\Lambda, \Lambda}$. This indicates that the ability to approach arbitrarily low bias appears to require access to such points in the grid.

\section*{Acknowledgements} 

ASA acknowledges support from the U.S. Department of Defense through a QuICS Hartree Fellowship and from IIIT Hyderabad.
MESM  acknowledges support from the U.S. Department of Defense through a QuICS Hartree Fellowship. JS is funded in part by the Commonwealth of Virginia's Commonwealth Cyber Initiative (CCI) under grant number 469351.

\newpage{}

\bibliographystyle{alpha}
\bibliography{housekeeping}

\newpage{}

\appendix

\end{document}